\documentclass[aps,prb,twocolumn,grouppedaddress, nofootinbib, longbibliography]{revtex4-2}
\usepackage{amsmath,amssymb}
\usepackage{dsfont,yfonts}
\usepackage{tabularx}
\usepackage{multirow}
\usepackage{diagbox}
\newcolumntype{C}{>{\centering\arraybackslash}X}%
\usepackage{graphicx}
\usepackage{subfigure}
\usepackage[usenames,dvipsnames]{xcolor}
\usepackage{tikz}
\usepackage{pgffor}
\usepackage{verbatim}
\usepackage{bbm}
\usepackage{float}
\usepackage{mathrsfs}
\usepackage{xcolor,colortbl}
\usepackage[colorlinks, bookmarks=true,breaklinks=true,linkcolor=red, citecolor=blue, linktocpage=true, urlcolor=blue]{hyperref}

\usepackage{amsfonts}

\newcommand{\Z}{\mathbb{Z}}

\newcommand{\teta}{\tilde{\eta}}
\newcommand{\rev}{{\rm rev}}

\newcommand{\bfx}{\mathbf{x}}

\newcommand{\calM}{\mathcal{M}}

\newcommand{\modulo}[1]{\ ({\rm mod} \ #1)}
\makeatletter
\newcommand\xleftrightarrow[2][]{%
  \ext@arrow 9999{\longleftrightarrowfill@}{#1}{#2}}
\newcommand\longleftrightarrowfill@{%
  \arrowfill@\leftarrow\relbar\rightarrow}
\makeatother

\begin{document}

\title{Anomaly indicators and bulk-boundary correspondences for 3D interacting topological crystalline phases with mirror and continuous symmetries}

\author{Shang-Qiang Ning}
\affiliation{Department of Physics and HKU-UCAS Joint Institute for Theoretical and Computational Physics, The University of Hong Kong, Pokfulam Road, Hong Kong, China}

\author{Bin-Bin Mao}
\affiliation{Department of Physics and HKU-UCAS Joint Institute for Theoretical and Computational Physics, The University of Hong Kong, Pokfulam Road, Hong Kong, China}

\author{Zhengqiao Li}
\affiliation{Department of Physics and HKU-UCAS Joint Institute for Theoretical and Computational Physics, The University of Hong Kong, Pokfulam Road, Hong Kong, China}

\author{Chenjie Wang}
\email{cjwang@hku.hk}
\affiliation{Department of Physics and HKU-UCAS Joint Institute for Theoretical and Computational Physics, The University of Hong Kong, Pokfulam Road, Hong Kong, China}

\date{\today}

\begin{abstract}
We derive a series of quantitative bulk-boundary correspondences for 3D bosonic and fermionic symmetry-protected topological (SPT) phases under the assumption that the surface is gapped, symmetric and topologically ordered, i.e., a symmetry-enriched topological (SET) state. We consider those SPT phases that are protected by the mirror symmetry and continuous symmetries that form a group of $U(1)$, $SU(2)$ or $SO(3)$. In particular, the fermionic cases correspond to a crystalline version of 3D topological insulators and topological superconductors in the famous ten-fold-way classification, with the time-reversal symmetry replaced by the mirror symmetry and with strong interaction taken into account. For surface SETs, the most general interplay between symmetries and anyon excitations is considered. Based on the previously proposed dimension reduction and folding approaches, we re-derive the classification of bulk SPT phases and define a \emph{complete} set of bulk topological invariants for every symmetry group under consideration, and then derive explicit expressions of the bulk invariants in terms of surface topological properties (such as topological spin, quantum dimension) and symmetry properties (such as mirror fractionalization, fractional charge or spin). These expressions are our quantitative bulk-boundary correspondences. Meanwhile, the bulk topological invariants can be interpreted as \emph{anomaly indicators} for the surface SETs which carry 't Hooft anomalies of the associated symmetries whenever the bulk is topologically non-trivial. Hence, the quantitative bulk-boundary correspondences provide an easy way to compute the 't Hooft anomalies of the surface SETs. Moreover, our anomaly indicators are complete. Our derivations of the bulk-boundary correspondences and anomaly indicators are explicit and physically transparent. The anomaly indicators obtained in this work can be straightforwardly translated to their time-reversal counterparts that apply to the usual topological insulators and topological superconductors, due to a known correspondence between mirror and time-reversal topological phases.  
\end{abstract}

\maketitle

\tableofcontents

\section{Introduction}

Symmetry-protected topological (SPT) phases are a class of short-range entangled states of matter whose non-trivial topological properties require protection from symmetries\cite{chen2013,senthil2015}. Topological insulators and topological superconductors are among the famous examples. One of the most important features of SPT states is that the boundary cannot be trivially gapped. For example, the surface of a 3D topological insulator can either be a gapless Dirac fermion, or a spontaneous symmetry-breaking state, or certain time-reversal symmetric non-Abelian topologically ordered states\cite{MetlitskiPRB2015,Bonderson13d,wangc13b,chen14a,wangc14}. The non-triviality of these surface states lies in the fact that they carry a \emph{quantum anomaly}, more precisely the \emph{'t~Hooft anomaly}\cite{tHooftbook, KapustinPRL2014}, of the underlying symmetries. Generally speaking, an anomalous state cannot be regularized (e.g., by a lattice realization) in the same dimensions without violating the symmetries.\footnote{A caveat is that in a lattice realization, the symmetries should be implemented in an on-site fashion and should be in a linear representation of the symmetry group. If they are not on-site or in a projective representation of the symmetry group, anomalous states may still be realized on a lattice.} Instead, it has to live on the boundary of a bulk of one higher dimension. In fact, for given symmetries, there is a one-to-one correspondence between the bulk SPT phases and the types of 't~Hooft anomalies on the boundary. This correspondence is a manifestation of the famous \emph{bulk-boundary correspondence} in topological phases of matter\cite{Witten1988,Wen1995}.

In the past few years, special attention has been paid to 3D SPT systems with a gapped, symmetric, and topologically ordered surface state\cite{MetlitskiPRB2015,Bonderson13d,wangc13b,chen14a,wangc14, vishwanath13,wangc13,burnell14,WangJPRX2018, fidkowski13,metlitski14,you14,chen14,WangPRX2016,RyuPRB2016_BBC}, i.e., the surface is a \emph{symmetric-enriched topological} (SET) state\cite{EssinPRB2013, MesarosPRB2013, BarkeshliPRB2019, TarantinoNJP2016, TeoAP2015}.  On the one hand, surface SETs are a new scenario to terminate the 3D bulk state, which does not exist in lower dimensions. Also, they are relatively easier to deal with than gapless surface states in interacting  systems. So, they attract a lot of attention.  On the other hand, SETs themselves are of fundamental interests. They exhibit interesting physical properties, such as symmetry fractionalization on anyon excitations. It is important to determine the 't Hooft anomaly of a given SET state, a problem equivalent to establishing the bulk-boundary correspondence for 3D SPT systems.\cite{HarveyNPB1985,WittenRMP2016, chen14,BarkeshliPRB2019,WangPRX2016,RyuPRB2016_BBC}

For 3D SPT systems with an SET surface, the bulk-boundary correspondence can be established at a purely topological level, as both the bulk and surface are topological states. Ideally, one may perform the following steps: (i) define a set of bulk topological invariants to characterize the 3D SPT state; (ii) define a set of surface topological invariants to characterize the SET surface; and (iii) establish a set of equations of the bulk and surface topological invariants, which serve as a quantitative bulk-boundary correspondence. For example, 3D topological superconductors, with the time-reversal symmetry satisfying $\mathcal{T}^2=-1$, has a $\Z_{16}$ classification.\cite{wangc14,metlitski14,you14} The quantitative bulk-boundary correspondence can be expressed as follows\cite{WangPRL2017}:
\begin{align}
\eta_{\mathcal{T}} = \frac{1}{\sqrt{2}D} \sum_{a\in\mathcal{C}} d_a \theta_a\tilde{\mathcal{T}}_a^2,
\label{eq:Tindicator}
\end{align}
where $\eta_\mathcal{T}$ is a bulk invariant that takes a value in $1, e^{i\pi/8}, \dots, e^{i15\pi/8}$, and all quantities on the right-hand side describe the surface SET. More specifically, $\mathcal{C}$ denotes the surface topological order, $a$ denotes a surface anyon, $d_a$ and $\theta_a$ are quantum dimension and topological spin, $D=\sqrt{\sum_a d_a^2}$ is the total quantum dimension, and $\tilde{\mathcal{T}}_a^2=0,\pm 1$ describes time-reversal properties of the anyon $a$(see Ref.~\cite{WangPRL2017,TachikawaPRL2017} for details). Equation \eqref{eq:Tindicator} applies to general time-reversal symmetric fermionic SETs. It is not only a quantitative bulk-boundary relation, but also provides an easy way to compute the time-reversal 't Hooft anomaly of an arbitrary SET, which has the same $\Z_{16}$ classification as the bulk topological superconductor. In this respect, $\eta_\mathcal{T}$ is also referred to as an \emph{anomaly indicator}.

It is highly desired to derive quantitative bulk-boundary relations similar to Eq.~\eqref{eq:Tindicator} for 3D interacting SPT systems with other symmetries. Difficulties are expected, e.g., it is usually difficult to identify a complete set of topological invariants for the bulk or the surface, in particular if one aims for results that apply to the most general surface SETs. Nevertheless,  several successful attempts have been made. For example, Refs.~\onlinecite{folding} and \onlinecite{Mao2020} find a physical way to define anomaly indicators and derive similar equations to \eqref{eq:Tindicator} for mirror-symmetric bosonic and fermionic systems, respectively. Reference \onlinecite{LapaPRB2019} introduces several expressions of anomaly indicators for 3D topological insulators, i.e., those SPTs with time reversal symmetry and $U(1)$ charge conservation. Moreover, the work of Ref.~\onlinecite{BulmashPRR2020} introduces a general algorithm to compute anomalies by making use of a class of exactly solvable models. More discussions on prior works will be given in Sec.~\ref{sec:previous}. 

In this work, we aim to perform a systematic study on 3D interacting SPT phases protected by the mirror symmetry $\mathcal{M}$ and continuous symmetries of group $G$, where $G = U(1)$, $SU(2)$ or $SO(3)$, and establish a collection of quantitative bulk-boundary correspondences like \eqref{eq:Tindicator}. The overall symmetry group $\hat{G}$ formed by $\mathcal{M}$ and $G$ will be discussed more explicitly in Sec.~\ref{sec:symmetry}.  This work is a direct generalization of Refs.~\onlinecite{folding} and \onlinecite{Mao2020}. We will study both bosonic and fermion systems. The fermionic cases are the mirror version of the topological insulators and superconductors in the famous ten-fold way classification\cite{ff1,ff2}, namely topological crystalline insulators and superconductors. More specifically, we will derive or re-derive the bulk SPT classification, define anomaly indicators and surface topological invariants in a physically transparent way, and finally establish quantitative  bulk-boundary relations like the one in Eq.~\eqref{eq:Tindicator}. Our main results are summarized in Sec.~\ref{sec:main-results}.

Before moving on to the main results, here we briefly introduce our approach and make a few comments on its advantages.  We will use the folding approach introduced in Ref.~\onlinecite{folding} to obtain the bulk-boundary correspondences (see more details in Sec.~\ref{sec:general}). This approach is developed specifically to handle mirror symmetry, which is further based on the dimensional reduction approach for crystalline symmetries, first introduced in Ref.~\onlinecite{SongPRX2017}. Special treatment is needed for spatial symmetries (such as the mirror symmetry) and anti-unitary symmetries (such as time-reversal), because the usual method of gauging symmetries does not apply\cite{levin2012}. The folding approach transforms the original 3D problem to a 2D problem such that (i) the mirror symmetry $\mathcal{M}$ turns into an  onsite unitary $\Z_2$ symmetry and (ii) the original bulk-boundary relation becomes a problem of 2D gapped domain wall. See Fig.~\ref{fig:folding} for an illustration. Since $\mathcal{M}$ becomes an on-site unitary symmetry, we can now  gauge it to study topological properties. Another advantage is that the topological correspondence between the two sides of a gapped domain wall can be more readily established.

\begin{table*}
\caption{Classification of 3D bosonic SPT phases with different symmetries and values of the anomaly indicators $\eta_1, \eta_2, \eta_3$ and $\eta_4$. The symbol ``$\circ$'' means the corresponding anomaly indicator is not defined. Remark: in the case of $SU(2)\times\Z_2^\calM$, the alternative indicator $\teta_3$ can take $\pm 1$. However, it is forced to be 1 when the surface is an SET state, which indicates that the surface is either symmetry-breaking or gapless if $\teta_3=-1$. }
\label{tab:boson}
\begin{tabular}{cccccc}
\hline\hline
$\quad$ Symmetry $\quad$ & $\quad$ Classification $\quad$ & $\quad$ $\eta_1$ $\quad$ & $\quad$ $\eta_2$ $\quad$  & $\quad$ $\eta_3$ $\quad$  & $\quad$ $\eta_4$ $\quad$ \\
\hline
$U(1)\times \Z_2^\calM$ &   $\Z_2^4$ & $\pm 1$ & $\pm 1$ & $\pm 1$ & $\pm 1$ \\
$U(1)\rtimes\Z_2^\calM$ &  $\Z_2^3$ & $\pm 1$ & $\pm 1$ & $\pm 1$ & $\circ$ \\
$SU(2)\times \Z_2^\calM$ &  $\Z_2^3$ & $\pm 1$ & $\pm 1$ & $\eta_1{\teta_3}$& $\eta_2{\teta_3}$ \\
$SO(3)\times \Z_2^\calM$ & $\Z_2^4$ & $\pm 1$ & $\pm 1$ & $\pm 1$ & $\pm 1$\\
\hline
\end{tabular}
\end{table*}

\begin{table*}
\caption{Classification of 3D fermionic SPT phases with different symmetries and values of the anomaly indicators $\eta_{1f}, \eta_{2f}, \eta_{3f}$ and $\eta_{4f}$. We have adapted the Altland-Zirnbauer symmetry classes\cite{AZclass} by replacing the time-reversal with the mirror symmetry, with the understanding that ``$\mathcal{T}^2=\pm1$'' is mapped to ``$\mathcal{M}^2=\mp 1$''. Here, we list only the classes that have non-trivial 3D interacting SPT phases. Our classification is obtained by the dimensional reduction approach\cite{SongPRX2017}, and the results are in agreement with those from invertible topological field theory\cite{FreedarXiv2016}. Anomaly indicators for the DIII class was studied in Ref.~\onlinecite{Mao2020} and we list it here for completeness. The number $n$ in the forth column is an integer. The symbol ``$\circ$'' means the corresponding anomaly indicator is not defined. The red numbers cannot be realized by surface SETs, indicating that the surface is either symmetry-breaking or gapless.}
\label{tab:fermion}
\begin{tabular}{ccccccc}
\hline\hline
$\quad$  Class $\quad$ & $\quad$ Symmetry $\quad$ & $\quad$ Classification $\quad$ & $\quad$ $\eta_{1f}$ $\quad$ & $\quad$ $\eta_{2f}$ $\quad$ & $\quad$ $\eta_{3f}$ $\quad$ & $\quad$ $\eta_{4f}$ $\quad$ \\
\hline
DIII & $\Z_2^f\times \Z_2^\calM$ & $\Z_{16}$ & $e^{i n \pi/8}$
 & $\circ $  & $\circ $  & $\circ $\\
AIII & $U_f(1)\times\Z_2^\calM$ &  $\Z_{8}\times \Z_2$ & $e^{i n \pi/4}$ & $\pm 1$ & $\eta_{1f}^4$ & $\circ$  \\
AII & $U_f(1)\rtimes\Z_2^\calM$ &  $\Z_{2}^3$ & $\pm 1$ & $\pm 1$ & $\pm 1$ & $\circ$\\
AI & $\big[U_f(1)\rtimes \Z_4^{f\calM}\big]/\Z_2$ &  $\Z_{2}$ & $\circ$ & $\pm 1$ & $1$ & $\circ$ \\
CI & $SU_f(2)\times\Z_2^\calM$ &  $\Z_4\times \Z_2$ & $\pm 1$, \color{red}{$\pm i$} & $\pm 1$ & $1$ & $\circ$\\
CII & $\big[SU_f(2)\times\Z_4^{f\calM}\big]/\Z_2$ &  $\Z_2^3$ &   $\circ$ & $\pm 1$ & $1$, \color{red}{$-1$} & $\pm 1$\\
\hline
\end{tabular}
\end{table*}

\begin{table}[b]
\caption{Relations between $\teta_i$ and $\eta_i$ in bosonic systems. The indicators $\teta_i$ are defined in Sec.~\ref{sec:define-eta-u1-b}.}
\label{tab:eta-relation}
\begin{tabular}{l|l}
\hline 
\hline
$\quad U(1)\times\Z_2^\mathcal{M}$ & $\quad U(1)\rtimes \Z_2^\mathcal{M}\quad  $  \\
\hline
$\quad \teta_1 = \eta_1\quad $ & $\quad \teta_{1} = \eta_1\quad $\\
$\quad \teta_2 = \eta_2\quad $ & $\quad \teta_2 = \eta_2\quad $ \\
$\quad \teta_3 = \eta_1\eta_3\quad $ & $\quad \teta_3 = \eta_1\eta_3\quad $ \\
$\quad \teta_4 = \eta_1\eta_2\eta_3\eta_4\quad $ & $\quad \teta_4 = \eta_1\eta_3\quad $ \\
$\quad \teta_5 =  \eta_4 \quad $  & $\quad \teta_{5} =\eta_2 $ \\
\hline
\end{tabular}
\end{table}

\subsection{Main results}
\label{sec:main-results}

We study 3D bosonic and fermionic interacting SPT systems with both the mirror symmetry $\mathcal{M}$ and a continuous symmetry group $G=U(1)$, $SU(2)$ or $SO(3)$. Different cases of the total symmetry group $\hat{G}\supset G$ will be discussed in Sec.~\ref{sec:symmetry}. We give a systematic and physical characterization of the bulk and surface topological properties using the dimensional reduction and folding approaches\cite{SongPRX2017,folding}. In particular, we define a complete set of anomaly indicators for all the SPT phases under consideration and derive a series of quantitative bulk-boundary relations. The anomaly indicators are summarized in Table \ref{tab:boson} and \ref{tab:fermion}.  These indicators form a complete set of bulk topological invariants so that the bulk SPT classification can be inferred from the possible values that they can take. The inferred classifications are the same as those of interacting time-reversal topological insulators and superconductors\cite{wangc14,FreedarXiv2016}, in agreement with the crystalline equivalence principle\cite{ThorngrenPRX2018}.

More specifically, for bosonic systems with $G=U(1)$, we study both cases of $U(1)\times\Z_2^\calM$ and $U(1)\rtimes\Z_2^\calM$. The two cases have  $\Z_2^4$ and $\Z_2^3$ classification, respectively. We first define a set of anomaly indicators $\teta_1$, $\teta_2$, $\teta_3$, $\teta_4$ and $\teta_5$ to characterize the bulk SPT state (not all are independent). These indicators can be recombined into four other \emph{independent} indicators $\eta_1, \eta_2, \eta_3$ and $\eta_4$, which have the following expressions in terms of surface SET quantities:
\begin{subequations}
\label{eq:eta_b}
\begin{align}
\eta_1 &= \frac{1}{D}\sum_{a\in\mathcal{C}} d_a^2 \theta_a \label{eq:eta_b1}\\
\eta_2 &= \frac{1}{D}\sum_{a\in\mathcal{C}} d_a \theta_a\mu_a \label{eq:eta_b2}\\
\eta_3 &= \frac{1}{D}\sum_{a\in\mathcal{C}} d_a^2 \theta_a e^{i2\pi q_a} \label{eq:eta_b3}\\
\eta_4 &= \frac{1}{D}\sum_{a\in\mathcal{C}} d_a \theta_a \mu_a  e^{i2\pi q_a} \label{eq:eta_b4}
\end{align}
\end{subequations}
where all $\eta_i$ take a value $\pm 1$, $d_a$ and $\theta_a$ are quantum dimension and topological spin of anyon $a$ in the topological order $\mathcal{C}$, $\mu_a=0,\pm1$ is a quantity describing mirror fractionalization, $q_a$ (defined modulo $1$) is the fractional charge associated with $U(1)$ symmetry. We remark that the indicator $\eta_4$  only applies to $U(1)\times \Z_2^\calM$ but not  $U(1)\rtimes \Z_2^\calM$.   The two sets of indicators $\{\teta_i\}$ and $\{\eta_i\}$ are equivalent, with the relations shown in Table \ref{tab:eta-relation}. The indicators $\teta_i$ have transparent physical definitions (see Sec.~\ref{sec:define-eta-u1}) while $\eta_i$ have simpler expressions, so we keep both notations in this paper.  The bulk SPT is trivial if and only if $\eta_i=1$ for every $i$. Equivalently, the surface SET is anomaly-free if and only if $\eta_i=1$ for every $i$ when evaluated by Eqs.~\eqref{eq:eta_b1}-\eqref{eq:eta_b4}. The indicators $\eta_1$ and $\eta_2$ characterize pure mirror anomalies, while $\eta_3$ and $\eta_4$ are associated with mixed anomalies between $\mathcal{M}$ and $U(1)$. The indicators $\eta_1$ and $\eta_2$ and their time-reversal counterparts have already been discussed in Refs.~\onlinecite{WangPRL2017, BarkeshliCMP2019}, and the time-reversal counterparts of $\eta_3$ and $\eta_4$ have been proposed in Ref.~\onlinecite{LapaPRB2019}.

For fermion systems with $G=U_f(1)$, there are three possible symmetry groups, corresponding to the AI, AII and AIII classes in Table \ref{tab:fermion}.  We define three anomaly indicators $\eta_{1f}, \eta_{2f}$ and  $\eta_{3f}$ to characterize the bulk SPT state (where $\eta_{1f}$  is not applicable to AI class). Again, they form a complete set of bulk topological invariants so that the SPT classification can be inferred from the values they can take. With the folding approach, we are able to show that
\begin{subequations}
\label{eta_f}
\begin{align}
\eta_{1f} &= \frac{1}{\sqrt{2}D}\sum_{a\in\mathcal{C}} d_a \theta_a\mu_a \label{eq:eta-f1} \\
\eta_{2f} &= \frac{1}{\sqrt{2} D}\sum_{a\in\mathcal{C}} d_a^2 \theta_a e^{i\pi q_a}\label{eq:eta-f2} \\
\eta_{3f} &= \frac{1}{2D}\sum_{a,b\in\mathcal{C}}d_ad_be^{i2\pi q_a} e^{i2\pi q_b} S_{ab}\label{eq:eta-f3}
\end{align}
\end{subequations}
where $\mathcal{C}$ is now a fermionic topological order, $S_{ab}$ is the $S$ matrix,  $\mu_a=0,\pm 1$ describes mirror fractionalization of the surface SET,  $q_a$ (now defined modulo $2$) is the fractional charge associated with $U_f(1)$.  The bulk SPT is trivial if and only if $\eta_{if}=1$ for every $i$, and equivalently the surface SET is anomaly-free if and only if $\eta_{if}=1$ for every $i$ when evaluated by Eqs.~\eqref{eq:eta-f1}-\eqref{eq:eta-f3}. The indicator $\eta_{1f}$ describes pure mirror anomaly, which was discussed in Ref.~\cite{Mao2020}, and $\eta_{2f}$, $\eta_{3f}$ describe mixed anomalies between $\mathcal{M}$ and $U_f(1)$. We remark again that the time-reversal counterparts of the expressions in \eqref{eta_f} have already been proposed in Ref.~\onlinecite{LapaPRB2019} before.

\begin{subequations}
For bosonic systems with $G=SU(2)$ or $SO(3)$ in Table \ref{tab:boson}, the indicators $\eta_1$ and $\eta_2$ still apply as they characterize pure mirror anomalies. To characterize the mixed anomalies, we show that it is enough to consider their $U(1)$ subgroup, and thereby anomaly indicators inherit from those of $U(1)\times\Z_2^\calM$. For $G=SU(2)$, there is only one $\Z_2$ in the classification associated with the mixed anomaly, characterized by $\teta_3$. However, $\teta_3$ is enforced to be 1 if the surface is an SET, due to the fact that $SU(2)$ cannot be fractionalized by anyons. Therefore, any SPT state with $\teta_3=-1$ is enforced to support a gapless surface if no symmetry breaking occurs. This phenomena is known as ``symmetry-enforced gaplessness'' and was first discovered in Ref.~\onlinecite{wangc14}.  In the case of $SO(3)$, the bulk-boundary relations for $\eta_3$ and $\eta_4$ inherit from \eqref{eq:eta_b3} and \eqref{eq:eta_b4}: 
\begin{align}
\eta_{3,SO(3)} &= \frac{1}{D}\sum_{a\in\mathcal{C}} d_a^2 \theta_a e^{i2\pi s_a} \label{eq:eta_b3-SO3}\\
\eta_{4,SO(3)} &= \frac{1}{D}\sum_{a\in\mathcal{C}} d_a \theta_a \mu_a  e^{i2\pi s_a} \label{eq:eta_b4-SO3}
\end{align}
\end{subequations}
where $s_a$ is the spin carried by the anyon $a$ under $SO(3)$, which is either an integer or half-integer.  That is, the fractional charge $q_a$  in \eqref{eq:eta_b3} and \eqref{eq:eta_b4} are replaced by $s_a$ in \eqref{eq:eta_b3-SO3} and \eqref{eq:eta_b4-SO3}. Note that we have dropped the subscript ``$SO(3)$'' of $\eta_{3,SO(3)}$ and $\eta_{4,SO(3)}$ in Table \ref{tab:boson}.

\begin{subequations}
For fermionic systems with $G=SU_f(2)$, there are two symmetry classes,  corresponding to the CI and CII classes in Table \ref{tab:fermion}. Similarly to the bosonic case, many properties can be characterized by $\mathcal{M}$ and the $U_f(1)$ subgroup. However, compared to $G=U_f(1)$, we define a forth indicator $\eta_{4f}$ for CII class. The indicators $\eta_{1f}$, $\eta_{2f}$ and $\eta_{3f}$ are similarly defined, though they may take different values. We show that the two cases, that $\eta_{1f}=\pm i$ in CI class and that $\eta_{3f}=-1$ in CII class, cannot be realized from surface SETs and must lead to symmetry-enforced gaplessness. The expression of $\eta_{1f}$ is again given by \eqref{eq:eta-f1}. The others are
\begin{align}
\eta_{2f, SU_f(2) } &= \frac{1}{\sqrt{2} D}\sum_{a\in\mathcal{C}} d_a^2 \theta_a e^{i2\pi s_a}\label{eq:eta-f2-SU2} \\
\eta_{4f, SU_f(2)} &= \frac{1}{\sqrt{2} D}\sum_{a\in\mathcal{C}} d_a \theta_a \mu_a e^{i2\pi s_a}\label{eq:eta-f4-SU2}
\end{align}
\end{subequations}
Equation \eqref{eq:eta-f2-SU2} is related to \eqref{eq:eta-f2} by replacing $q_a$ with $2s_a$, where the factor of 2 is due to the convention that $SU_f(2)$ has an angle period $4\pi$ while $U_f(1)$ has an angle period $2\pi$. Also, \eqref{eq:eta-f4-SU2} is very similar to \eqref{eq:eta_b4-SO3}. Indeed, we derive the former from the latter in Sec.~\ref{sec:eta-CII}. Again, we have dropped the subscript ``$SU_f(2)$'' of $\eta_{3f,SU_f(2)}$ and $\eta_{4f,SU_f(2)}$ in Table \ref{tab:fermion}.
 
Finally, we remark that while the above quantitative bulk-boundary relations are our main results, the physical definitions of the anomaly indicators are also worth emphasized, which are given in Secs. \ref{sec:define-eta-u1}, \ref{sec:define-eta-su2-b} and \ref{sec:define-eta-SU2-f}. In addition, when re-deriving the classification of 3D SPT phases in Table \ref{tab:boson} and \ref{tab:fermion}, we also derive a few classifications of 2D SRE states, which are summarized in Table \ref{tab:2d-plane}. To our knowledge, some of these classifications are not known previously.

\subsection{Relation to prior works}
\label{sec:previous}

Here we discuss the prior works that are closely related to this work. 

First, regarding classification of strongly correlated topological crystalline phases, there have been many works in the literature\cite{SongPRX2017,ThorngrenPRX2018,HuangPRB2017, ElsePRB2019, SongNC2020, ShiozakiArxiv2018,JiangPRB2017, SongPRB2020, xiongJPA2018, GuoCMP2020_CobordCrystalline, FreedArxiv2019invertible, DebrayArxiv2021}. The dimensional reduction approach proposed in Ref.~\onlinecite{SongPRX2017} and the crystalline equivalence principle proposed in Ref.~\onlinecite{ThorngrenPRX2018} are two general classification schemes that emphasize more on physical properties. The two schemes are later shown to be equivalent\cite{HuangPRB2017, ElsePRB2019, SongNC2020, ShiozakiArxiv2018}. There are also more mathematical approaches such as the cobordism theory\cite{Kapustin2015b,GuoCMP2020_CobordCrystalline} and invertible topological field theory\cite{FreedarXiv2016, FreedArxiv2019invertible,DebrayArxiv2021}. For our purpose of defining bulk topological invariants using physical observables, we find the dimensional reduction approach more suitable and adopt it extensively in this work. More specifically, on classification of 3D topological crystalline phases protected by $\mathcal{M}$ and a Lie group $G=U(1)$, $SU(2)$ or $SO(3)$, not all cases have been worked out explicitly before (some $U(1)$ cases were done in Ref.~\cite{SongPRX2017}). So, we derive or re-derive these classifications using the dimensional reduction approach, in particular for some of the fermionic cases.  Nevertheless, classifications of the time-reversal counterparts can be found in Refs.~\onlinecite{wangc14, FreedarXiv2016}. Our results are in agreement with the time-reversal classifications,  under the crystalline equivalence principle\cite{WittenRMP2016, ThorngrenPRX2018}.

The scenario of gapped symmetric topologically ordered surface states for 3D SPT phases was first proposed in Ref.~\onlinecite{vishwanath13} for bosonic topological insulators. Later, intensive effort was made on surface SETs for 3D SPT phases, based on either field theoretical analysis, Walker-Wang models, or other methods\cite{MetlitskiPRB2015,Bonderson13d,wangc13b,chen14a,wangc14,  vishwanath13,wangc13,burnell14,chen14,WangJPRX2018,fidkowski13,metlitski14,you14, WangPRX2016,QiPRL2015,LakePRB2016, ChengPRL2018,ChengPRX2016_LSM, ZouPRB2018,WangPRL2017,TachikawaPRL2017, BarkeshliCMP2019,folding, Mao2020,LapaPRB2019,BulmashPRR2020,barkeshli2019relative,FSET1,FSET2, Kobayaski2021arxiv}. While some works concern a specific surface SET state, some others deal with  bulk-boundary relations and 't Hooft anomalies of general surface SETs. Those that are most closely related to this work are several works on anomaly indicators. Time-reversal anomaly indicators for bosonic and fermionic systems were proposed and proved in Refs.~\onlinecite{WangPRL2017, TachikawaPRL2017, BarkeshliCMP2019}. These indicators are the time-reversal counterparts of $\eta_1, \eta_2$ in \eqref{eq:eta_b} and $\eta_{1f}$ in \eqref{eta_f}. The proofs given in Refs.~\onlinecite{TachikawaPRL2017, BarkeshliCMP2019} make use of the path integral on un-oriented manifolds in the limit of topological field theories (i.e., the energy gapped is pushed to infinity). References \onlinecite{folding, Mao2020} developed the folding approach and derived similar expressions of anomaly indicators for the mirror  symmetry. More recently, Ref.~\onlinecite{LapaPRB2019} derived a series of anomaly indicators for bosonic and fermionic time-reversal topological insulators with $U(1)$ symmetries. Also, Ref.~\onlinecite{BulmashPRR2020} developed a general algorithm to compute the $\mathcal{H}^4(G,U(1))$ topological invariants, which can be viewed as anomaly indicators, for bosonic SET phases. The idea is to use SET data to construct a 3D exactly solvable model and extract topological invariants from the model. Unfortunately, deriving explicit expressions of anomaly indicators seem not easy. 

As mentioned above, this work is a direct generalization of the works in Refs.~\onlinecite{folding} and \onlinecite{Mao2020}. We do not have much technical advance compared to these works. Also, we remark that many parts of our results were obtained previously in the context of time-reversal SPT phases in Ref.~\onlinecite{LapaPRB2019}. The connection between this work and Ref.~\onlinecite{LapaPRB2019} can be established by the crystalline equivalence principle\cite{WittenRMP2016,ThorngrenPRX2018}. Nevertheless, our study provides physically clear definitions of the anomaly indicators and the establishment of bulk-boundary correspondence is very direct and systematic. Furthermore, we also deal with the systems with both mirror and $SO(3)$ or $SU(2)$ symmetries, which are not discussed in Ref.~\onlinecite{LapaPRB2019}.

\subsection{Organization of the paper}

The rest of the paper is organized as follows. In Sec.~\ref{sec:general}, we discuss several general aspects of this work, including the dimension reduction and folding approaches, symmetry groups, fundamentals of topological orders, the method of gauging symmetries. In Sec.~\ref{sec:define-eta-u1}, we define a complete set of anomaly indicators for both bosonic and fermionic systems with $U(1)$ and $\mathcal{M}$ symmetries, using the dimensional reduction approach. Next, we discuss properties of surface SETs, derive bulk-boundary relations and the expressions of anomaly indicators for bosonic and fermionic systems with $G=U(1)$, in Secs.~\ref{sec:boson-U(1)} and \ref{sec:fermion-U(1)}, respectively. In Sec.~\ref{sec:SU2-SO3-b}, we define anomaly indicators and derive the bulk-boundary relations for bosonic systems with $G=SU(2)$ or $SO(3)$. In Sec.~\ref{sec:SU2-SO3-f}, we study fermionic systems with $\mathcal{M}$ and $SU_f(2)$ symmetries. We give a brief discussion in Sec.~\ref{sec:discussion}.

The appendices contain some technical discussions. In Appendix \ref{appd:semidirect}, we prove the constraints that certain vortex braiding statistics must obey in the case that $\mathcal{M}$ and $U(1)$ symmetries do not commute.  In Appendix \ref{appd:adjoin-iqh}, we discuss the consequence of adjoining integer quantum Hall states in the dimensional reduction approach in a few cases. In Appendices \ref{sec:app_anyoncond}, \ref{appd:anycond-boson-teta3}, and \ref{appd:anycond-boson-teta45}, we give alternative derivations of $\teta_3$, $\teta_4$ and $\eta_5$ through anyon condensation theory. While these derivations are lengthy and more technical than those in the main text, they do provide a better understanding of the surface SETs and the bulk-boundary correspondence.

\section{Generalities}
\label{sec:general}

In this section, we discuss the background and methods to prepare for the studies in the next several sections.

\begin{figure*}
\centering
\includegraphics[scale=1]{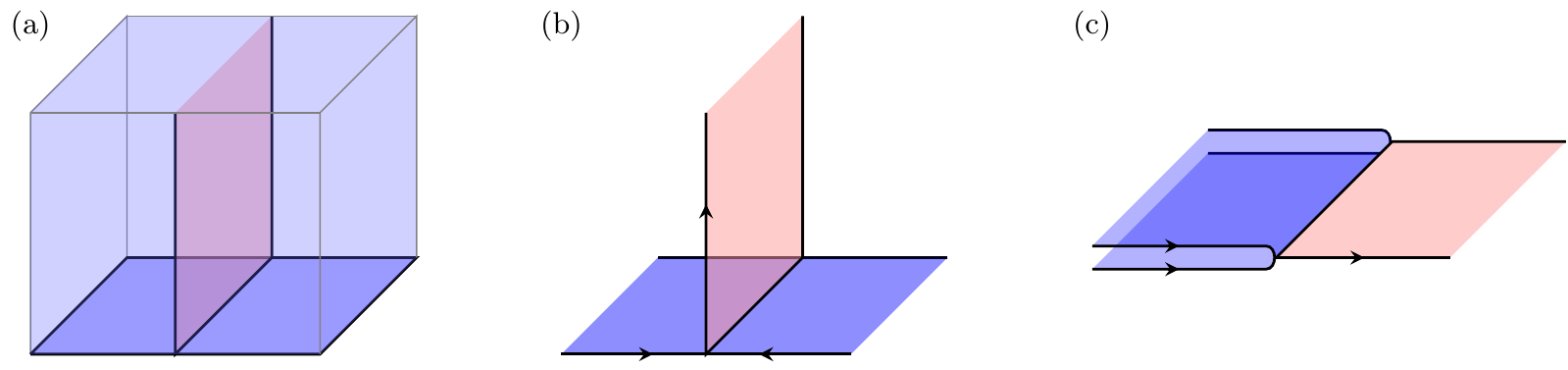}
\caption{Dimensional reduction and folding. The pink plane represents the mirror reflection plane, and the dark blue plane represents the physical surface. Applying symmetric local unitary transformations turns the system in (a) into the inverted T-junction in (b), and further folding the surface turns the T-junction into the 2D system in (c). Arrows represent chirality of the topological states. This procedure reduces the problem of 3D bulk-boundary correspondence to a problem of 2D gapped domain wall. }
\label{fig:folding}
\end{figure*}

\subsection{Dimensional reduction}
\label{sec:dimredc}

One of the approaches to analyze 3D mirror-symmetric SPT phases is the dimensional reduction approach introduced in Ref.~\onlinecite{SongPRX2017}. We  will adopt this approach in this work. Below we briefly review the general idea in the current context.

First of all, we notice that there is no 3D SPT phases protected solely by $G=U(1)$, $SU(2)$ or $SO(3)$ for either bosonic or fermionic systems\cite{chen2013, freed2014, Kapustin2015b, FreedarXiv2016, LanPRB2019_decorationConst}. Then, there exists a $G$-preserving local unitary transformation that can remove all short-range entanglement in the ground state\cite{ChenPRB2010}. To further make $\mathcal{M}$ preserved, we first consider a local unitary transformation $\mathcal{U}_l$, which acts on the left side of the mirror plane in Fig.~\ref{fig:folding}(a) and turns the left part of the state into a trivial product. It respects $G$, i.e., $\mathcal{U}_l g = g\mathcal{U}_l$ for every $g\in {G}$. In addition, we also apply the local unitary transformation $\mathcal{U}_r = \mathcal{M}^{-1} \mathcal{U}_l\mathcal{M}$, which acts only on the right side of the mirror plane in Fig.~\ref{fig:folding}(a). Note that $\mathcal{U}_r$ also preserves $G$, as long as $G$ is a normal subgroup of the overall symmetry group $\hat{G}$, which is always true for $G$ consisting of only internal symmetries. Then, it is not hard to show that the combined unitary transformation $\mathcal{U}_l\mathcal{U}_r$ respects $\mathcal{M}$: $\mathcal{M}^{-1} (\mathcal{U}_l\mathcal{U}_r)\mathcal{M} = \mathcal{U}_l\mathcal{U}_r \nonumber$. Accordingly, $\mathcal{U}_l\mathcal{U}_r$ respects the total symmetry group $\hat{G}$.

The consequence of applying $\mathcal{U}_l\mathcal{U}_r$ onto the ground state is that it turns the state into the trivial product state everywhere except for the degrees of freedom near the mirror plane [Fig.~\ref{fig:folding}(b)]. Near the mirror plane, the supports  of $\mathcal{U}_l$ and $\mathcal{U}_r$ overlap, so entanglement cannot be fully removed. Nevertheless, the degrees of freedom near the mirror plane decouple from elsewhere. Accordingly, we obtain an effective 2D short-range entangled (SRE) state on the mirror plane. In general, a 2D SRE state\footnote{In this paper, we use the convention that SRE states include both SPT states and invertible topological orders.} can be either an SPT or an \emph{invertible topological order}\cite{freed2014}. 2D invertible topological orders are generated under stacking by the $E_8$ state for bosonic systems\cite{e8}, and are generated by integer quantum Hall (IQH) states for fermionic systems. The 2D SRE state in the mirror plane contains all topological properties of the original 3D SPT state. Two nice things of this dimensional reduction are that: (1) since the effective SRE system is 2D, its topological properties are easier to analyze than the original 3D systems; (2) in the mirror plane, $\mathcal{M}$ becomes an internal symmetry so it is easier to deal with too. The latter allows us to gauge $\Z_2^\mathcal{M}$ and extract topological properties by studying gauge fluxes (see Sec.~\ref{sec:gauging}). 

\begin{figure}
\centering
\includegraphics[scale=1]{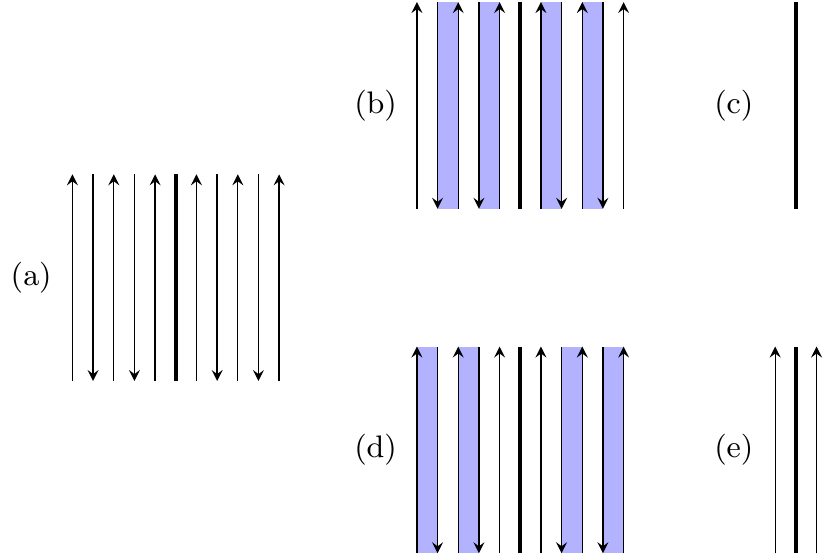}
\centering
\caption{Adjoining operation. The middle line is the mirror plane in the front view.  Other lines are symmetrically located on the two sides  and represent adjoined 2D SRE states. Two lines connected by a blue stripe are removed by local unitary transformations. The existence of two paths, (a)-(b)-(c) and (a)-(d)-(f), demonstrates that (c) and (e) are topologically equivalent from a 3D viewpoint.}
\label{fig:adjoining}
\end{figure}

Classification of 3D SPT states can then be obtained by classifying 2D SRE states with a symmetry group $\hat{G}$, where $\mathcal{M}$ is viewed as internal. However, the latter classification is generally larger than that of the original 3D SPT states. To obtain the correct 3D classification, one needs to consider a reduction by the so-called \emph{adjoining operations}\cite{SongPRX2017}: One adjoins two $G$-symmetric SRE states on the two sides of the mirror plane, where the two states are images of each other under $\mathcal{M}$. The adjoined states can be removed by 3D local unitary transformations(Fig.~\ref{fig:adjoining}), so two 2D SRE states that can be related by adjoining operations are equivalent from a 3D point of view. With adjoining operations, the 3D classification can be obtained from 2D SRE states. Properties of 2D SRE states in the mirror plane will be discussed in Secs.~\ref{sec:define-eta-u1}, \ref{sec:define-eta-su2-b}, and \ref{sec:define-eta-SU2-f} for different symmetries.

\subsection{Folding approach}
\label{sec:folding}

The above analysis can be applied equally well in the presence of a surface (Fig.~\ref{fig:folding}). We will assume that the surface carries a topological order and respects the symmetry group $\hat{G}$, i.e., it is a symmetry-enriched topologically ordered state (SET).\footnote{This assumption may not always be valid. There exist 3D SPT states whose surface is enforced to be gapless\cite{wangc14,CordovaArXiv2019}, if the symmetry is not broken. We will discuss this situation later in the case $G=SU(2)$.} Different from the bulk,  the surface cannot be turned into a trivial product state by local unitary transformations due to the presence of topological order. However, by a similar $\hat{G}$-symmetry local unitary transformation $\mathcal{U}_l\mathcal{U}_r$ as above, all information of symmetry-protected entanglement on the surface can be moved to the intersection line between the surface and the mirror plane. This leaves an inverted T-like junction, which is decoupled from other bulk degrees of freedom and contains all information of symmetry-protected entanglement\cite{folding}.

Let us be more specific on the T-like junction in Fig.~\ref{fig:folding}(b). Let $H_L$ ($H_R$) be the Hamiltonian of the left (right) wing of the junction, $H_{\rm mp}$ be the Hamiltonian of the mirror plane, and $H_{\rm dw}$ be the Hamiltonian of the degrees of freedoms near the intersection line of the surface and mirror plane. The total Hamiltonian is
\begin{equation}
H = H_L + H_R + H_{\rm mp} + H_{\rm dw}.
\end{equation}
Let $U(g)$ be the unitary symmetry operator for $g\in G$. Since $H$ respects $G$ and $\mathcal{M}$, it is required that
\begin{align}
\mathcal{M} H_L \mathcal{M}^{-1} & = H_R, \nonumber\\
\quad \mathcal{M} H_{\rm mp} \mathcal{M}^{-1} & = H_{\rm mp} \nonumber\\
\quad \mathcal{M} H_{\rm dw} \mathcal{M}^{-1} & = H_{\rm dw} \nonumber\\
U(g) H_\alpha U(g)^{-1} & = H_\alpha
\end{align}
where $\alpha = L$, $R$, ${\rm mp}$, or ${\rm dw}$ in the last line. We see that the left and right wings are mirror images of each other. In particular, they have opposite chiral properties such as chiral central charge and Hall conductance.

With the above understanding, Ref.~\onlinecite{folding} proposes to fold the two wings and form the geometry in Fig.~\ref{fig:folding}(c). The system is the same as before, only except that the orientation associated with $H_R$ is reversed. Figure \ref{fig:folding}(c) is a 2D system, consisting of the double-layer system on the left,  the mirror plane on the right and a domain wall between them. It is gapped everywhere and symmetric under the full symmetry $\hat{G}$. It is worth emphasizing that in the double-layer system, $\mathcal{M}$ becomes an internal layer-exchange symmetry too, like in the mirror plane.  Accordingly, the mirror symmetry group $\Z_2^\mathcal{M}$ can be gauged in the double-layer system (see Sec.~\ref{sec:gauging}) and  SET properties can be extracted by studying gauge fluxes.

Therefore, the final setup contains: (a) the double-layer system described by $H_L + H_R$, which represents the original surface; (b) the mirror plane described by $H_{\rm mp}$, which represents the original 3D bulk; and (c) the gapped domain wall $H_\mathrm{dw}$ which describes the boundary condition between (a) and (b). The bulk-boundary correspondence in the original 3D system can then be established by studying the connection between (a) and (b) through the boundary condition (c).  The latter is the problem of gapped domain walls and has been widely studied\cite{Beigi2010,LevinPRB2012, bravyi1998, levin2013, kitaev2012, barkeshli2013defect, barkeshli2013defect2, fuchs2012, lan2015, HungJHEP2015, WanJHEP2017}, e.g.,  by the so-called anyon condensation theory\cite{kitaev2012,  BaisPRB2009, eliens2013,kong2014, NeupertPRB2016}. In the main text, anyon condensation theory will be not be extensively used for our purpose. However, in Appendix \ref{appd:anycond-boson-teta3} and \ref{appd:anycond-boson-teta45}, we will provide alternative derivations for anomaly indicators $\teta_3$, $\teta_4$ and $\teta_5$ using anyon condensation theory. Accordingly, a brief review on this theory is given in Appendix \ref{sec:app_anyoncond}.

\subsection{Symmetry groups}
\label{sec:symmetry}

In this work, we study systems that respect the mirror symmetry $\mathcal{M}$ and an internal unitary symmetry group $G$, with $G = U(1), SU(2)$ or $SO(3)$. All symmetries together form the group $\hat{G}$. Here, we would like to make a few comments on symmetry groups in different scenarios (see Tables \ref{tab:boson} and \ref{tab:fermion}).

First, in fermionic systems, there is a special symmetry, the fermion parity $P_f$, which must be preserved. Moreover, it commutes with all symmetries in $\hat{G}$, i.e., it sits inside the center of $\hat{G}$. Let $\Z_2^f=\{\mathbbm{1}, P_f\}$, where $\mathbbm{1}$ is the identity operator. Then, there is the question of how $\Z_2^f$ is embedded into ${G}$. We will always consider the case that $\Z_2^f$ is a subgroup of $U(1)$ and $SU(2)$, and thereby denote them as $U_f(1)$ and $SU_f(2)$. Note that $P_f$ cannot be an element of $SO(3)$, as it is centerless. More specifically, let $U_\varphi$ be an element in $U_f(1)$ with  $\varphi\in [0, 2\pi)$. Then, $P_f = U_{\pi}$. For $SU_f(2)$, group elements can be represented as $U_{\varphi} = \exp(i\sum_{i=1}^3\varphi_i\sigma_i)$, where $\varphi_i\in[0,2\pi)$ and $\sigma_i$ are Pauli matrices. Then, $P_f = \exp(i\pi\sigma_z) $. In addition, there is also a question of whether $\mathcal{M}^2 = \mathbbm{1}$ or $P_f$. We will consider both cases. We will also call them  ``$\mathcal{M}^2=1$'' and ``$\mathcal{M}^2=-1$'' respectively, in analogy to the time-reversal symmetry.

Second, the total symmetry group $\hat{G}$ is a group extension of $\Z_2^\mathcal{M}$ by the internal symmetry group $G$. Mathematically, group extension is determined by the short exact sequence
\begin{equation}
0 \rightarrow G \rightarrow \hat{G} \rightarrow \Z_2^\mathcal{M} \rightarrow 0
\end{equation}
There may exist several different extensions $\hat{G}$, which we discuss below separately for each $G$. 

(i) For $G= U(1)$, there are three possible extensions, $\hat G = U(1)\times\Z_2^\mathcal{M}$, $U(1)\rtimes\Z_2^\mathcal{M}$, or $[U(1)\rtimes\Z_4^\mathcal{M}]/\Z_2$. The direct product ``$\times$'' corresponds to  $U_\varphi \mathcal{M} = \mathcal{M}U_\varphi$, and the semi-direct product ``$\rtimes$'' corresponds to $U_\varphi \mathcal{M} = \mathcal{M}U_{-\varphi}$.  In the last case, $\Z_4^\mathcal{M}$ means $\mathcal{M}^2 = U_\pi$ and thereby we need to take a quotient over $\Z_2= \{\mathbbm{1},U_\pi\}$. In the case of direct product, there is no actual distinction between $\mathcal{M}^2=\mathbbm{1}$ and $U_\pi$. For bosonic system, we consider $\mathcal{M}^2=\mathbbm{1}$ only. For fermionic systems, we consider all three possible extensions. (In fermionic systems, $\Z_4^\calM$ is denoted as $Z_4^{f\calM}$, since $\mathcal{M}^2=P_f$.)  The three symmetry groups correspond to the AIII, AII, AI Altland-Zirnbauer symmetry classes\cite{AZclass,ff1} of non-interacting fermions\footnote{Altland-Zirnbauer symmetry classes originally concern the time-reversal symmetry $\mathcal{T}$. We adopt the same notation by replacing $\mathcal{T}$ with $\mathcal{M}$. However, $\mathcal{M}^2=\pm 1$ is mapped to $\mathcal{T}^2=\mp 1$, which is necessary to have the correct correspondence between the classifications of mirror and time-reversal topological phases.\cite{WittenRMP2016,ThorngrenPRX2018}}. 
  
(ii) For $G = SU(2)$, there are two possible extensions, $\hat G = SU(2)\times \Z_2^\mathcal{M}$ and $[SU(2)\times \Z_4^\mathcal{M}]/\Z_2$.  The latter case again corresponds to $\mathcal{M}^2 = U_\pi$. For bosonic cases, we consider $\hat G = SU(2)\times \Z_2^\mathcal{M}$ only. For fermionic systems, we consider both extensions. The two symmetry groups correspond to the CI and CII Altland-Zirnbauer symmetry classes. Note that there are different ways to interpret the Altland-Zirnbauer symmetry classes in the context of interacting systems, e.g., see Refs.~\onlinecite{wangc14, MorimotoPRB2015, FreedarXiv2016}. In this paper, we follow the convention of Ref.~\onlinecite{FreedarXiv2016} (see Table \ref{tab:fermion}).

(iii) For $G=SO(3)$, there is only a trivial extension, namely $SO(3)\times\Z_2^\calM$. We will only consider it for bosonic systems. However, understanding properties of $SO(3)$-symmetric bosonic systems will be helpful for the study of fermionic systems with $SU_f(2)$ symmetry.

Third, the double-layer system of the left half in Fig.~\ref{fig:folding}(c) has enhanced symmetries. It is described by the Hamiltonian $H_L + H_R$. Since the two layers are decoupled, each has an internal symmetry $G$, giving rise to a $G\times G$ symmetry. The total symmetry group of the double-layer system is an extension of $\Z_2^\mathcal{M}$ by $G\times G$. This is relevant because in the derivation of bulk-boundary correspondences, in particular with the method of anyon condensation given in Appendices \ref{appd:anycond-boson-teta3} and \ref{appd:anycond-boson-teta45}, we find it convenient to first gauge $G$ (or a subgroup of $G$) in each wing of the T junction and then do the folding. This is equivalent to gauge $G\times G$ in the double-layer system. Nevertheless, the right side of Fig.~\ref{fig:folding}(c) respects a single $G$ only. To match the two sides, it should be understood that $G$ on the right side of Fig.~\ref{fig:folding}(c) is the diagonal subgroup $G$ of $G\times G$, i.e., the subgroup of symmetries with a simultaneous action on the two layers.

Finally, we make a comment on the notation that we will use for $\Z_2^\mathcal{M}$. In the 2D system of Fig.~\ref{fig:folding}(c), $\mathcal{M}$ is an internal symmetry. So, we will drop the superscript $\mathcal{M}$ and simply denote it as $\Z_2$ (e.g., in Table \ref{tab:2d-plane}). Moreover, we will rename $\mathcal{M}$ as $\bfx$ when it is referred to as an internal symmetry in the following discussions and the symmetry group is often referred to as $\Z_2^x$.

\subsection{Topological orders}
\label{sec:topo}

In this work, we assume that surface state is an SET state. Properties of SET states include their topological and symmetry properties\cite{BarkeshliPRB2019, TarantinoNJP2016, TeoAP2015}. Here, we briefly review the topological properties, i.e., physical quantities that characterize a topological order. More detailed review can be found in Ref.~\onlinecite{Kitaev06}. The symmetry properties will be discussed later when we study specific symmetry groups.

A topological order is characterized by (1) a set of anyon labels $1, a, b,\dots$, where ${1}$ or $\mathbbm{1}$ represents the trivial/vacuum anyon, (2) their fusion properties and (3) their braiding properties. We denote the topological order as $\mathcal{C}$. Mathematically speaking, $\mathcal{C}$ is described by a unitary braided fusion category\cite{Kitaev06}. Fusion of anyons are described by the fusion rules 
\begin{equation}
a\times b = \sum N_{ab}^c c
\end{equation}
where $N_{ab}^c$ is a non-negative integer. If $N_{ab}^1=1$, anyon $b$ is the anti-particle of $a$, denoted as $b=\bar{a}$, and vice versa. Two important quantities associated with each anyon $a$ are the quantum dimension $d_a$ and the topological spin $\theta_a$. Quantum dimensions satisfy that $d_ad_b = \sum_c N_{ab}^c d_c$ and $d_a\ge 1$. Braiding properties include the so-called $S$ and $T$ matrices, which are defined as follows
\begin{align}
T_{a,b} & = \theta_a \delta_{a,b} \nonumber\\
S_{a,b} & = \frac{1}{D}\sum_c N_{a\bar{b}}^c\frac{\theta_c}{\theta_a\theta_b} d_c 
\end{align}
where $D=\sqrt{\sum_a d_a^2}$ is called the total quantum dimension of $\mathcal{C}$. If $\mathcal{C}$ is Abelian, i.e., $d_a=1$ for all $a$'s, $T_{a,a}$ is the exchange statistics between to identical anyons $a$ and $S_{a,b}$ is proportional to the complex conjugate of mutual statistics. More generally, if either one among $a,b$ is Abelian, they have an Abelian mutual statistics, given by
\begin{equation}
M_{a,b} = \frac{S_{a,b}^* S_{1,1}}{S_{1,a}S_{1,b}}
\label{eq:M-S-relation}
\end{equation}
Another important relation for bosonic topological order is
\begin{equation}
e^{i2\pi c/8} = \frac{1}{D} \sum_a d_a^2 \theta_a
\label{eq:centralcharge}
\end{equation}
where $c$ is the chiral central charge associated with the edge states of the topological order.

In this work, we will also study fermionic topological orders. Mathematically, fermionic topological orders are described by unitary pre-modular tensor categories\cite{Drinfeld2010, Bruillard2017,LanPRB2016}. One of the key differences to bosonic topological orders is that there exists a special fermion $f$ in fermionic topological orders, such that
\begin{equation}
M_{f,a}=1
\end{equation}
for all $a$'s. That is, $f$ is ``transparent'' to other anyons in terms of mutual statistics. In bosonic topological orders, only the trivial anyon $1$ is transparent. Also, anyons always come in pairs, $a$ and $af$, such that $a\times f = af$ and $\theta_{af} = -\theta_a$. That is $\mathcal{C}=\{1,f,a,af,b,bf,\dots\}$. For later convenience, we will denote the pair $\{a,af\}$ as $[a]$. Another remark is that due to the existence of $f$,  $S$ matrix is degenerate. In contrast, $S$ is non-degenerate and unitary for bosonic topological orders. Fermionic topological orders can be better characterized by gauging the fermion parity. One may consult Ref.~\onlinecite{Mao2020}  for some properties after gauging the fermion parity. Finally, we comment that the relation \eqref{eq:centralcharge} does not hold in fermionic topological orders.

\subsection{Gauging symmetries}
\label{sec:gauging}

We will extensively use another method to study the SRE and SET states, namely the method of gauging symmetries, first introduced in Ref.~\onlinecite{levin2012}. For a quantum many-body system with internal symmetries of a \emph{finite} group $\mathcal{G}$, it is always possible to couple it to a $\mathcal G$ gauge field such that the resulting gauged system remains energetically gapped. In our study, $\mathcal{G}$ will be a finite subgroup of the total symmetry group $\hat{G}$. For an SRE state, the gauged theory becomes a topological order; for an SET state, the original topological order will be enlarged after gauging. For details of the gauging procedure at a microscopic level, we refer the readers to Refs.~\onlinecite{levin2012, wangcj15}. Here, we briefly review the topological excitations in the gauged theory.

For a SRE state with symmetries, there is no anyon before gauging. After gauging,  it becomes a $\mathcal{G}$ gauge theory coupled to matter. The gauged system is topologically ordered. It contains two kinds of anyons: gauge charges (or simply charges) and vortices.  Charge excitations have a one-to-one correspondence to the irreducible representations of the group $\mathcal{G}$. Vortices carry gauge flux. Gauge fluxes have a one-to-one correspondence to the conjugacy classes of $\mathcal{G}$.  For a fixed gauge flux, there are distinct vortices that differ by attaching charges. Let us take the example $\mathcal{G}=\Z_N$ for illustration. In this case, charges are labelled by an integer $q=0,1,\dots,(N-1)$, corresponding to the irreducible representations of $\Z_N$. Fluxes are labeled by $2\pi k$, where $k=0,1,\dots, (N-1)$ corresponding to the conjugacy classes (the same as group elements in this case). The $2\pi$ is incorporated such that the Aharonov-Bohm phase between a charge and a vortex is given by $q\phi = 2\pi qk$. A general vortex is then labeled by the combination $(q,\phi)$, which is frequently named as \emph{dyon}. According to Refs.~\onlinecite{levin2012, wangcj15}, different SPT states are characterized and distinguished by different braiding statistics of vortices.  Topological invariants that uniquely identity the SPT order can be defined through the topological spins and mutual statistics of vortices.\cite{wangcj15} In Secs.~\ref{sec:define-eta-u1}, \ref{sec:define-eta-su2-b} and \ref{sec:define-eta-SU2-f}, we will make use of these topological invariants to define anomaly indicators.

The spectrum of topological excitations in a gauged SET state is more complicated. It contains charges, vortices, and  anyons originating from those before gauging. Charges are the same as in SRE state, being labeled by irreducible representations of $\mathcal{G}$. Anyons in the original theory will be carried over to the gauged theory, however, they may be split and/or identified. The detailed splitting and identification depends on the detailed SET, see e.g., Appendices \ref{appd:anycond-boson-teta3} and \ref{appd:anycond-boson-teta45}. Vortices again carry gauge flux. For a fixed gauge flux, there exist distinct vortices that differ by attaching charges as well as those anyons that originate from the original SET. We refer the refer to Refs.~\onlinecite{BarkeshliPRB2019,TeoAP2015} for detailed discussions.

\section{Defining anomaly indicators for $G=U(1)$}
\label{sec:define-eta-u1}

\begin{table*}
\caption{Classification of 2D SRE phases, including both SPT and invertible topological orders, with a continuous symmetry $G$ and a unitary $\Z_2$ symmetry for bosonic and fermionic systems, where $G=U(1), SU(2)$ or $SO(3)$. The $\Z^2$ component in all cases is associated with the chiral $E_8$ state and IQH state. In the fourth column, we list the smallest possible Hall conductance $\sigma_H^0$ and the Hall conductance $\sigma_H^{\rm root}$ of the root state for the IQH $\Z$ classification, in units of the conductance quantum\footnote{By ``the smallest possible Hall conductance'', we mean when only $U(1)$ or the $z$-component rotation of $SU(2)$ and $SO(3)$ is present.  For $U(1)$ and $SO(3)$, the unit charge or spin is 1, the flux quantum is $2\pi$, and accordingly the conductance quantum is $1/2\pi$. For $SU(2)$, the unit spin is $1/2$, the flux quantum is $4\pi$, such that the conductance quantum is $1/4\pi$.}. The ratio $\sigma_H^{\rm root}/\sigma_H^0$ is the filling factor of the root IQH state. In the last column, we list the reduced classification of SRE states under adjoining operations, which corresponds to the classification of 3D SPT phases in Table \ref{tab:boson} and \ref{tab:fermion}.}
\label{tab:2d-plane}
\begin{tabular}{ccccc}
\hline\hline 
 $\quad$Boson/Fermion $\quad$ & $\quad$ Symmetry $\quad$ & $\quad$ SRE classification  $\quad$&  $\quad$ $(\sigma_H^0,\ \sigma_H^{\rm root})$ $\quad$ & $\quad$  Reduction $\quad$\\
\hline 
Boson & $U(1)\times \Z_2$ & $\Z^2 \times \Z_2^2$ &  $(2, \ 2)$ & $\Z_2^4$\\
Boson & $U(1)\rtimes \Z_2$ & $\Z^2\times \Z_2$ & $(2, \ 2)$  & $\Z_2^3$\\
Boson & $SU(2)\times \Z_2$ & $\Z^2\times \Z_2$ & $(1, \ 1)$ & $\Z_2^3$\\
Boson & $SO(3)\times \Z_2$ & $\Z^2\times \Z_2^2$ & $(2, \ 2)$ & $\Z_2^4$ \\
Fermion(AIII) & $U_f(1)\times \Z_2$ & $\Z^2 \times \Z_4$ & $(1,\ 1)$ & $\Z_2\times \Z_8$\\
Fermion(AII)& $U_f(1)\rtimes \Z_2$ &  $\Z^2 \times \Z_2$ & $(1, \ 1)$ & $\Z_2^3$\\
Fermion(AI) & $[U_f(1)\rtimes \Z_4^f]/\Z_2$ & $\Z^2 $  & $\quad$ $(1, \ 2)$ $\quad$ & $\Z_2$\\
Fermion(CI) & $SU_f(2)\times \Z_2$ & $\Z^2 \times \Z_2$  & $(1/2, \ 1)$ & $\quad$ $\Z_2\times \Z_4$ $\quad$\\
Fermion(CII) & $[SU_f(2)\times \Z_4^f]/\Z_2$ & $\Z^2 \times \Z_2$  & $(1/2, \ 1) $ & $\Z_2^3$\\
\hline
\end{tabular}
\end{table*}

In the following three sections, Secs.~\ref{sec:define-eta-u1}, \ref{sec:boson-U(1)} and \ref{sec:fermion-U(1)}, we study bulk-boundary correspondences and anomaly indicators for $G=U(1)$, i.e., bosonic and fermionic topological crystalline insulators (TCIs). The main purpose of this section is to define a set of topological invariants to characterize the SRE state in the mirror plane in Fig.~\ref{fig:folding}(c) for different total symmetry groups $\mathcal{G}$, along with which we reproduce the classification of 3D SPT phases. These invariants will serve as anomaly indicators.

\subsection{Bosonic systems}
\label{sec:define-eta-u1-b}

We start with 3D bosonic TCIs of symmetries $U(1)\times \Z_2^\calM$ and $U(1)\rtimes \Z_2^\calM$. In the mirror plane, $\mathcal{M}$ becomes an internal symmetry, so we rename it as $\mathbf{x}$ and denote the symmetry groups as $U(1)\times\Z_2$ and $U(1)\rtimes \Z_2$, respectively. In places that clarification is needed, we will also use $\Z_2^x$ to denote the group associated with $\bfx$. For both symmetry groups, we discuss the following three aspects: (1) classification and characterization of strictly 2D SRE states, (2) how the SRE classification reduces to that of the original 3D system under adjoining operations, and (3) definitions of topological invariants, i.e., anomaly indicators. The classifications are summarized in Table \ref{tab:2d-plane}.

\subsubsection{$U(1)\times \Z_2$}
\label{sec:U(1)xZ2-b}

According to Sec.~\ref{sec:dimredc}, 3D TCIs with $U(1)\times\Z_2^\mathcal{M}$ can be reduced to 2D SRE states in the mirror plane with internal symmetry $U(1)\times \Z_2$ using finite-depth local unitary transformations. For strictly 2D SRE states with this symmetry group, the classification is $\Z^2\times \Z_2^2$.\cite{chen2013} The four root states and their basic properties are as follows: 

(i) The root state of the first $\Z$ classification is the so-called  $E_8$ state\cite{e8}. It is an invertible topological order.  It hosts gapless modes on the edge with a chiral central charge $c=8$ (or equivalently, a thermal Hall conductance $\kappa=8$ in units of $\pi^2 T/3h$). The full $U(1)\times \Z_2$ symmetry acts trivially on this state.\footnote{By a trivial symmetry action, we mean at the level of topological properties. For a specific state,  symmetries may have non-trivial actions (i.e., not identity operators), but these actions are local and do not give rise to any constraints on topological properties.} Stacking multiple copies of the root state and its time reversal gives rise to the $\Z$ classification. 

(ii) The root state of the first $\Z_2$ classification is a non-chiral SPT state protected by the symmetry $\mathbf{x}$ alone. The $U(1)$ symmetry acts trivially on this state. Stacking two copies of the root state gives rise to a trivial state, and thereby the classification is $\Z_2$. According to Ref.~\onlinecite{levin2012}, $\Z_2$ SPT phases can be characterized by the topological spin of $\Z_2$ vortices after the symmetry is gauged. Let $x$ be a $\Z_2$ vortex, which we will simply call an ``$x$-vortex''. Then, the non-trivial phase is associated with $\theta_x = \pm i$, while the trivial phase is associated with $\theta_x=\pm 1$. The ``$\pm$'' ambiguity results from the fact that there exist two kinds of vortices, which differ by a $\Z_2$ charge.  

(iii) The root state of the second $\Z$ classification is the bosonic integer quantum Hall (IQH) state with Hall conductance $\sigma_H = 2e^2/h$. Throughout this paper, we use $e$ to denote the unit charge of $U(1)$, regardless if the system is bosonic or fermionic. Bosonic IQH states are protected by $U(1)$ symmetry alone, and $\bfx$ acts trivially. Note that the smallest Hall conductance in bosonic IQH states is $2e^2/h$ and these states are non-chiral\cite{lu2012,LevinPRB2012}. (In contrast, fermionic IQH states have the smallest Hall conductance $e^2/h$ and  they are chiral.)  Stacking multiple copies of the root state gives rise to the $\Z$ classification.

(iv) The root state of the second $\Z_2$ classification is a non-chiral SPT state protected jointly by $U(1)$ and $\Z_2$ symmetries. There are several equivalent ways to characterize this state: (1) If we gauge $\Z_2$ and consider $x$-vortices, they carry  fractional  charge $e/2$ of the $U(1)$ symmetry; (2) if we also gauge the subgroup $\Z_2\subset U(1)$, then the associated $\Z_2$ vortices --- which we will call $w$-vortices --- will have $\pm i$ mutual statistics with respect to $x$-vortices, versus $\pm 1$ in the trivial state; (3) the topological spin of the composite vortices --- referred to as $y$-vortices --- will be $\pm i$. All ``$\pm $'' ambiguities result from the existence of multiple vortices that differ by charge attachments.\cite{levin2012, wangcj15} Since we will make use of the $\Z_2$ subgroup of $U(1)$ below, we will name it $\Z_2^w$ to distinguish it from the $\Z_2$ of the $\mathbf{x}$ symmetry, and also refer the latter as $\Z_2^x$ occasionally.

\begin{subequations}
\label{eq:u1xz2-b-observables}

A general SRE state can be indexed by an integer vector $\mu = (\mu_1, \mu_2, \mu_3, \mu_4)$, with $\mu_1,\mu_3\in \Z$ and $\mu_2,\mu_4=0, 1$ modulo $2$.  It consists of $\mu_i$ copies of the $i$-th root state. Its chiral central charge and Hall conductance are 
\begin{align}
c^{\rm mp} &  =8\mu_1 \\
\sigma_H^{\rm mp} &  = 2\mu_3
\end{align}
where the superscript ``mp'' stands for ``mirror plane'', and we have set the conductance quantum $e^2/h=1$. If we gauge the $\Z_2^w\times \Z_2^x$ subgroup, the $x$-, $w$- and $y$-vortices have the following properties
\begin{align}
\theta_x^2 &  =  (-1)^{\mu_2}\\
\theta_w^2 & =(-1)^{\mu_3} \\
M_{w,x}^2 &  =  (-1)^{\mu_4} \\
\theta_y^2 &  = (-1)^{\mu_2+\mu_3+\mu_4}
\end{align}
\end{subequations}
where $M_{w,x}$ denotes the mutual statistics between $w$- and $x$-vortices. All vortices are Abelian anyons. To get rid of the ``$\pm$'' ambiguity from charge attachment, we have squared the topological spins and mutual statistics. More details on braiding statistics in $\Z_2^w\times \Z_2^x$ gauge theories can be found in  Ref.~\onlinecite{wangcj15}. Note that there is a close relation between the topological spin of $w$-vortices and the Hall conductance, 
\begin{equation}
\theta_w^2 = e^{i\pi \sigma_H^{\rm mp}/2}.
\end{equation}
This is a well known relation in  the ordinary electronic quantum Hall effects\cite{wen-book}.

Next, we consider adjoining operations.  Both $\Z$'s in the above classification will reduce to $\Z_2$. For a state consisting of $\mu_1$ copies of the root state (i), with $\mu_1$ being even, we can adjoin $\mu_1/2$ copies of $E_8$ states on each side of the mirror place to trivialize it. Similarly, we can trivialize a state consisting of even copies of the root state (iii) by adjoining IQH states. Accordingly, with adjoining, non-trivial states are labelled only by $\mu_1$ and $\mu_3$ modulo $2$. That is, $c^{\rm mp}$ is meaningful only  modulo $16$, and $\sigma_H^{\rm mp}$ is meaningful only modulo $4$.  Therefore, the classification reduces to $\Z_2^4$, i.e., 3D bosonic TCIs with $U(1)\times \Z_2^\calM$ symmetry are classified by $\Z_2^4$, in agreement with Ref.~\onlinecite{NLSM}.

It is worth pointing out that since the $\Z$ classification associated with $\sigma_H$ is reduced to $\Z_2$,  $\theta_w$ now contains the same amount of information as $\sigma_H$. We observe that $\theta_x^2, \theta_w^2$ and $M_{w,x}^2$ are enough to distinguish the states built from root states (ii), (iii) and (iv) up to adjoining operations. That is, $w$-, $x$- and $y$-vortices contain enough topological information to characterize root states (ii), (iii) and (iv).

\begin{subequations}
\label{eq:tilde_eta}

We are now ready to define a set of topological invariants, which can uniquely specify a SRE state. We define
\begin{align}
\tilde{\eta}_1 &  = e^{i\pi c^{\rm mp}/8} \label{eq:teta-b1}\\
\tilde{\eta}_2 &  = \theta_x^2 \label{eq:teta-b2}\\
\tilde{\eta}_3 &  = e^{i\pi \sigma_H^{\rm mp}/2} \label{eq:teta-b3}\\
\tilde{\eta}_4 &  = M_{w,x}^2\label{eq:teta-b4}
\end{align}
These quantities are invariant under adjoining operations. They are independent, and all are valued at $\pm 1$. Note that ``$\tilde{\ }$'' is put on all the quantities because ``$\eta$'' is reserved for the anomaly indicators in Eqs.~\eqref{eq:eta_b}. The two sets of indicators, $\{\eta_i\}$ and $\{\teta_i\}$,  are equivalent. Nevertheless,  $\{\tilde{\eta}_i\}$ have better physical meanings as seen above, while $\{\eta_i\}$ have simpler expressions in terms of surface SET quantities which will be discussed in the next section. Note that $\tilde\eta_3$ can alternatively be expressed as
\begin{equation}
\tilde{\eta}_3 = \theta_w^2
\label{eq:teta-b3-alt}
\end{equation}
This alternative definition applies more generally, as it requires only the subgroup $\Z_2^w$ instead of the full $U(1)$ group. In addition, we define the fifth topological invariant
\begin{equation}
\tilde{\eta}_5 = \theta_{y}^2
\label{eq:teta-b5}
\end{equation}
It is not an independent invariant. From Eqs.~\eqref{eq:u1xz2-b-observables}, one can see that $\teta_5 = \teta_2\teta_3\teta_4$. All these topological invariants will serve as anomaly indicators. We will express them in terms of surface quantities in Sec.~\ref{sec:boson-U(1)}.

\end{subequations}

\subsubsection{$U(1)\rtimes \Z_2$}
\label{sec:U(1)rxZ2-b}

3D TCIs with $U(1)\rtimes \Z_2^\calM$ symmetry can be reduced to 2D SRE states with an internal $U(1)\rtimes \Z_2$ symmetry through the dimensional reduction procedure. In this case,  strictly 2D SRE states are classified by $\Z^2 \times \Z_2$.\cite{chen2013,WenPRB2015_SOInf} The three root states are similar to the state (i), (ii) and (iii) in Sec.~\ref{sec:U(1)xZ2-b}. However, symmetry actions on root state (iii) are different: $\bfx$ must have a non-trivial action on it due to the  group structure of  $U(1)\rtimes \Z_2$ (see below). In other states, the symmetry actions remain the same. In addition, there is no SPT state protected jointly by $U(1)$ and $\Z_2$ [i.e., a state similar to root (iv) in Sec.~\ref{sec:U(1)xZ2-b}]. One can easily check that adjoining operations do the same job as above, reducing each $\Z$ in the classification to $\Z_2$. Therefore, the final classification becomes $\Z_2^3$, i.e., 3D bosonic TCIs with $U(1)\rtimes \Z_2^\mathcal{M}$ are classified by $\Z_2^3$, in agreement with Ref.~\onlinecite{NLSM}.

Let $\mu=(\mu_1,\mu_2,\mu_3)$ be an integer vector labelling a general SRE state in the classification. It consists of $\mu_i$ copies of the $i$-th root state. Similarly to $U(1)\times \Z_2$, gauging the subgroup $\Z_2^w\times \Z_2^x \subset U(1)\rtimes \Z_2^x$ is useful for characterizing the state. The chiral central charge, Hall conductance and  braiding statistics of $w$-, $x$-, and $y$-vortices are given as follows:
\begin{subequations}
\begin{align}
c^{\rm mp} & = 8\mu_1 \\
\sigma_H^{\rm mp} & = 2\mu_3 \\
\theta_x^2  & = (-1)^{\mu_2}\\
\theta_w^2 & = (-1)^{\mu_3}\\
M_{w,x}^2  & = (-1)^{\mu_3}\\
\theta_y^2 & = (-1)^{\mu_2}
\end{align}
\end{subequations}
We pay special attention to the mutual statistics $M_{w,x}^2$.  In Appendix \ref{appd:semidirect}, we prove that the following constraint must hold:
\begin{align}
M_{w,x}^2\theta_w^2=1.
\label{eq:mconstraint-b}
\end{align}
It is a consequence of the fact that $\Z_2^w\times\Z_2^x$ has to be lifted to the full $U(1)\rtimes\Z_2^x$ group. In the root state (iii), $\theta_w^2$ is non-trivial, so $M_{w,x}^2$ must also be non-trivial. This implies that $\bfx$ must act nontrivially in the root state (iii), as claimed above.

Lastly, we use the same topological invariants $\teta_1$, $\teta_2$, $\teta_3$, $\teta_4$ and $\teta_5$, given in Eqs.~\eqref{eq:tilde_eta}, for $U(1)\rtimes \Z_2$ symmetric SRE states. They are again invariant under adjoining operations.  While the definitions remain the same, the values that $\teta_i$ takes may be different. In particular, $\teta_4$ and $\teta_5$ are not independent. We have $\teta_4 = \teta_3$ and $\teta_5 = \teta_2$ for $U(1)\rtimes\Z_2$.

\subsection{Fermionic systems}
\label{sec:define-eta-u1-f}

For fermionic systems, we consider 3D TCIs in AIII, AII, and AI classes. After dimensional reduction, the corresponding 2D SRE states in the mirror plane in Fig.~\ref{fig:folding}(c) have a symmetry group $U_f(1)\times \Z_2$, $U_f(1)\rtimes \Z_2$, and $[U_f(1)\rtimes \Z_4^f]/\Z_2$ respectively. Again, since $\mathcal{M}$
becomes internal in the mirror plane, we rename it as $\mathbf{x}$ and refer the corresponding group as $\Z_2^x$ when distinction is needed. In the following discussions, we will frequently use the results from Refs.~\onlinecite{GuLevin2014,WangPRB2016,WangPRB2017}. The anomaly indicators defined in this subsection are summarized in Table \ref{tab:fermion}.

\subsubsection{$U_f(1)\times \Z_2$}

\label{sec:u1xz2-f}
By dimension reduction, 3D fermionic TCIs in AIII class reduce to 2D fermionic SRE states with internal $U_f(1)\times\Z_2$ in the mirror plane. Strictly 2D SRE states with this symmetry group are classified by $\Z^2 \times \Z_4$. As shown in Ref.~\onlinecite{SongPRX2017}, the classification reduces to $\Z_2\times \Z_8$ after taking adjoining operations into accounts. Below we review this classification. In addition, we describe properties of these SRE states, from which we define a set of topological invariants that will serve as our anomaly indicators.

First of all, the three root states in the $\Z^2\times \Z_4$ classification of strictly 2D fermionic SRE states with $U_f(1)\times \Z_2$ symmetry are as follows:

(i) The root state of the first $\Z$ classification is  the $E_8$ state. It is the same $E_8$ state as in bosonic systems. In fermionic systems, one may use two-fermion bound states as bosons to construct this state. Both $U_f(1)$ and $\bfx$ act trivially on this state. It is characterized by a chiral central charge $c=8$ and Hall conductance $\sigma_H = 0$. 

(ii) The root state of the second $\Z$ classification is the famous electronic IQH state at filling factor 1. It is characterized by  the Hall conductance $\sigma_H=1$ (in units of $e^2/h$) and the chiral central charge $c = 1$. While non-trivial topological properties are manifested by $U_f(1)$, this state does not need protection from $U_f(1)$. It is a chiral invertible topological order, similarly to the $E_8$ state. The $\bfx$ symmetry acts trivially on this state.

(iii) The root state of the $\Z_4$ classification is an non-chiral SPT state protected by $\Z_2^{x}$ only (the full symmetry is $\Z_2^f\times \Z_2^x$). According to Ref.~\onlinecite{GuLevin2014}, fermionic SPT states with internal $\Z_2^f\times \Z_2^x$ symmetry are classified by $\Z_8$. For convenience, let us use $\nu=0,1,\dots,7$ to denote the eight $\Z_2^f\times \Z_2^x$ SPT states. Each state consists of $\nu$ pairs of $p_x+ip_y$ and $p_x-ip_y$ superconductors. The $\bfx$ symmetry behaves as the fermion parity of the $p_x+ip_y$ superconductors. According to Ref.~\onlinecite{WangPRB2016}, the odd-$\nu$ states are incompatible with $U_f(1)$ symmetry. Therefore, only the even-$\nu$ states can have an enlarged $U_f(1)\times \Z_2$ symmetry. Equivalently, they consist of $\nu/2$ pairs of $\sigma_H=1$ and $\sigma_H=-1$ IQH states. The total chiral central charge $c=0$ and total Hall conductance $\sigma_H=0$. This leads to a $\Z_4$ classification. 

Two remarks are in order. First, by stacking an $E_8$ state and eight copies of the $\sigma_H=-1$ IQH state, one obtains a state with $c=0$ and $\sigma_H=-8$. It is an SPT state protected by $U_f(1)$ alone, i.e., it becomes  trivial  in the absence of $U_f(1)$. One may use this state and the IQH state to generate the $\Z^2$ classification instead. Second, in contrast to bosonic systems with $U(1)\times\Z_2$, there is no SPT state protected jointly by $U_f(1)$ and $\Z_2$ in fermionic systems.

\begin{subequations}
\label{eq:prop-f1}
A general SRE state can be indexed by an integer vector $\mu = (\mu_1, \mu_2, \mu_3)$, with $\mu_3$ defined only modulo $4$. It consists of $\mu_i$ copies of the $i$th root state. The chiral central charge and Hall conductance are
\begin{align}
c^{\rm mp} & = 8\mu_1 + \mu_2 \\
\sigma^{\rm mp}_H & = \mu_2
\end{align}
Similarly to bosonic systems, many topological properties are captured by gauging the $\Z_2^f\times \Z_2^x$ subgroup. Let us again use $w$-, $x$-, and $y$-vortices to denote the $\Z_2^f$, $\Z_2^x$ vortices and their composite, respectively.  According to Refs.~\onlinecite{WangPRB2017}, all vortices are Abelian anyons, and they satisfy the following properties
\begin{align}
\theta_x^2 & = e^{i\pi \mu_3/2} \label{eq:prop-f1c}\\
\theta_w & = e^{i\pi\mu_2/4 } \\
M_{w,x}^2 & = (-1)^{\mu_3}\\
\theta_y^2 & = e^{i\pi(\mu_2-\mu_3)/2}  \label{eq:prop-f1f}
\end{align}
Several remarks are as follows. First, $c^{\rm mp}$, $\sigma_H^{\rm mp}$ and $\theta_x^2$ uniquely specify $\mu$. Second, different from the bosonic case, there is no $\pm$ sign ambiguity for $\theta_w$, and thereby we do not need to square it to get a topological invariant\cite{WangPRB2017}. Third, different from the bosonic case with $U(1)\times\Z_2$, the mutual statistics $M_{w,x}^2$ is not independent but determined by $\theta_x^2$. It is a manifestation of the fact that there are no SPT states protected jointly by $U_f(1)$ and $\Z_2^x$.
\end{subequations}

Next, we consider adjoining operations, under which the classification reduces to $\Z_2\times \Z_8$\cite{SongPRX2017}. First of all, like in bosonic systems, the $\Z$ classification associated with the $E_8$ state is reduced to $\Z_2$ by adjoining. This makes $\mu_1$ is unambiguous only modulo 2 under adjoining. Second, the $\Z$ associated with IQH states is also reduced to $\Z_2$. However, the reduced $\Z_2$ will extend the $\Z_4$ classification associated with root state (iii), such that they together form a $\Z_8$ classification. This group extension was discussed in Ref.~\onlinecite{SongPRX2017} and reviewed in Appendix \ref{appd:adjoin-iqh}. More specifically, stacking two $\mu=(0,1,0)$ states turns into a state with $\mu = (0,0,-1)$. That is, $\mu=(0,2,0)$ and $\mu=(0,0,-1)$ are equivalent under adjoining operations, making the overall stacking group being $\Z_8$. Accordingly, 3D fermionic TCIs in AIII class are classified by $\Z_2\times\Z_8$.

\begin{subequations}
\label{eq:u1teta-f}
We are now ready to define a set of topological invariants that are invariant under adjoining operations. We define
\begin{align}
\eta_{1f} & = e^{i\pi\sigma_H^{\rm mp}/4}(\theta_x^*)^2 
\label{eq:u1teta-f1}\\
\eta_{2f}  & = e^{i\pi(c^{\rm mp}-\sigma_H^{\rm mp})/8} \label{eq:u1teta-f2} 
\end{align}
In terms of the integer vector $\mu$, we have
\begin{align*}
\eta_{1f} & = e^{i\pi(\mu_2-2\mu_3)/4}\\
\eta_{2f} & = e^{i\pi\mu_1}
\end{align*}
We see  that $\eta_{1f}$ takes a value in $1, e^{i\pi/4}, \dots, e^{i7\pi/4}$ and $\eta_{2f}$ takes a value in $
\pm 1$. The invariant $\eta_{1f}$ is defined to comply the fact that $\mu=(0,2,0)$ and $(0,0,-1)$ are equivalent.   Instead of relying on the Hall conductance $\sigma_H^{\rm mp}$, one may also use $\theta_w$ to define $\eta_{1f}$. Since $\theta_w = e^{i\pi \sigma_H^{\rm mp}/4}$, we have the alternative definition
\begin{equation}
\eta_{1f} =  \theta_w(\theta_x^*)^2
\label{eq:u1teta-f1-alt}
\end{equation}
\end{subequations}
The invariants $\eta_{1f}$ and $\eta_{2f}$ uniquely specify a SRE state in the $\Z_2\times\Z_8$ classification. They will serve as our  anomaly indicators and will be expressed in terms of quantities of the surface topological order in Sec.~\ref{sec:fermion-U(1)}.

We comment that the alternative definition \eqref{eq:u1teta-f1-alt} of $\eta_{1f}$ is exactly the one defined in Ref.~\onlinecite{Mao2020} for fermionic systems with $\mathcal{M}$ symmetry only. Indeed, 3D fermionic TCIs in the $\Z_8$ classification do not need protection from $U_f(1)$. On the other hand, the $\Z_2$ classification does rely on $U_f(1)$. Indeed, in the absence of $U_f(1)$, root state (i) is equivalent to eight copies of root state (ii), which can be trivialized by adjoining.

\subsubsection{$U_f(1)\rtimes \Z_2$}
\label{sec:u1rxz2-f}

By dimensional reduction, 3D fermionic TCIs in AII class reduces to 2D SRE states with internal symmetry $U_f(1)\rtimes \Z_2$ in the mirror plane. To our knowledge, the classification of these 2D SRE states has not been discussed before. We show that classification of strictly 2D SRE states with $U_f(1)\rtimes \Z_2$ symmetry is $\Z^2\times \Z_2$.  The root states are as follows:

(i) The first root state is the $E_8$ state. Both $U_f(1)$ and $\Z_2$ symmetries act trivially on this state. It is characterized by $c=8$ and $\sigma_H=0$. Stacking multiple copies of this root state gives rise to a $\Z$ classification. 

(ii) The second root state is the IQH state at filling factor 1. It is characterized by $c =1$ and $\sigma_H=1$. Similarly to the bosonic case with $U(1)\rtimes\Z_2$, the symmetry $\mathbf{x}$ now must act non-trivially due to its non-commutativity with $U_f(1)$. This can be seen below from vortex braiding statistics. Stacking multiple copies of this root state gives rise the second $\Z$ classification.

(iii) The third root state is a non-chiral SPT state protected by $\Z_2^x$ alone (the full symmetry is $\Z_2^f\times\Z_2^x$). According to Ref.~\onlinecite{GuLevin2014}, fermionic SPT phases protected by $\Z_2^f\times\Z_2^x$ have a $\Z_8$ classification. We use $\nu=0,1,\dots, 7$ to index these states. With the full $U_f(1)\rtimes \Z_2^x$ symmetry, however, we show below that only states with $\nu=0 \modulo{4}$ are allowed and other non-trivial SPTs are incompatible. Accordingly, the root state is the $\nu=4$ state and it leads to a $\Z_2$ classification. In comparison, the corresponding root state for $U_f(1)\times \Z_2$ has $\nu=2$, as discussed in Sec.~\ref{sec:u1xz2-f}.

We remark that there is no non-trivial SPT state protected jointly by $U_f(1)$ and $\Z_2^x$, i.e., all $U_f(1)\rtimes\Z_2^x$ SPT states  are protected solely by $U_f(1)$ or $\Z_2^x$.  We expect that jointly-protected SPT states should be detected by independent mutual statistics between $U_f(1)$ vortices and $x$-vortices. By ``independent'', we mean that the mutual statistics is not fully determined by individual properties of  $U_f(1)$ vortices and $x$-vortices. However, for $U_f(1)\rtimes\Z_2^x$, one can argue that the mutual statistics is not independent. To see that, one may first gauge $\Z_2^f$ and turn the SPT state into an SET state with a remaining $U'(1)\rtimes\Z_2^x$ symmetry, where $U'(1)\equiv U_f(1)/\Z_2^f$. Then, information of vortex mutual statistics is included in (a) those between $\Z_2^f$ vortices and $x$-vortices and (b) those between $U'(1)$ vortices and $x$-vortices, after we further gauge $U'(1)\rtimes\Z_2^x$. According to Refs.~\onlinecite{GuLevin2014,WangPRB2017}, the mutual statistics in (a) is determined by the topological spins of $x$-vortices. The mutual statistics in (b) is not independent either, due to the fact that there is no jointly-protected bosonic SPT phases for $U'(1)\rtimes\Z_2^x$, as discussed in Sec.~\ref{sec:U(1)rxZ2-b}. Closely related and more detailed discussions along this line can be found in Appendix \ref{appd:semidirect}.

\begin{subequations}
\label{eq:u1rtimesz2-prop}
A general SRE state is then indexed by an integer vector $\mu = (\mu_1, \mu_2, \mu_3)$, with $\mu_3$ defined  modulo $2$. It consists of $\mu_i$ copies of the $i$th root state. The chiral central charge and Hall conductance are given by
\begin{align}
c^{\rm mp} & = 8\mu_1 + \mu_2\\
\sigma_H^{\rm mp} & = \mu_2
\end{align}
More properties can be probed by gauging the $\Z_2^f\times \Z_2^x$ subgroup and studying braiding statistics between $w$-, $x$- and $y$-vortices. One complication is that these vortices may be non-Abelian. 
Nevertheless, regardless of being Abelian or non-Abelian, we find that the vortices always satisfy
\begin{align}
\theta_x^2 & = e^{i\pi\mu_3 + i\pi \mu_2/4}\label{eq:18c}\\
\theta_w & = e^{i\pi\mu_2/4} \\
 M_{w,x}^2 & =e^{-i\pi \mu_2/2} \\
\theta_y^2 & =  e^{i\pi\mu_3 + i\pi \mu_2/4}\label{eq:18f}
\end{align}
\end{subequations}
More details on vortex braiding statistics can be found in Appendix \ref{appd:semidirect}. A few remarks are in order. First, $c^{\rm mp}, \sigma_H^{\rm mp}$ and $\theta_x$ uniquely specify a state in the $\Z^2\times \Z_2$ classification. Other quantities are not independent. Second, both $\theta_x^2$ and $M_{w,x}^2$ depend on $\mu_2$. It means that $\mathbf{x}$ must have a non-trivial action on root state (ii), in contrast to the $U_f(1)\times \Z_2$ case. Third, in Appendix \ref{appd:semidirect}, we show the following constraint must hold
\begin{align}
M_{w,x}^2\theta_w^2 =1
\end{align}
which is the same as Eq.~\eqref{eq:mconstraint-b} for bosonic systems with $U(1)\rtimes\Z_2$. This constraint is important to derive the expressions in \eqref{eq:18c}-\eqref{eq:18f}. In root state (iii) where $\theta_w=1$, we immediately have $M_{w,x}^2=1$. For a general $\Z_2^f\times\Z_2^x$ SPT state with an index $\nu$, the mutual statistics $M_{w,x}^2=e^{-i\nu\pi/2}$\cite{GuLevin2014}. Accordingly, we see that $\nu$ must be a multiple of $4$, as already claimed above.

Next, we consider reduction of the classification under adjoining operations. This is very much similar to in the case of $U_f(1)\times \Z_2$. Both $\Z$'s will reduce to $\Z_2$ under adjoining. However, for $U_f(1)\rtimes\Z_2$,  the $\Z_2$ classification associated with root state (ii) will not extend the $\Z_2$ classification associated with root state (iii). That is, stacking two copies of root state (ii) turns into a trivial state. Derivation of this result is given in Appendix \ref{appd:adjoin-iqh}. Accordingly, under adjoining operations, the overall classification reduces to $\Z_2^3$, i.e., 3D TCIs in AII class is classified by $\Z_2^3$. We remark that  mirror TCIs in this class correspond to the famous 3D time-reversal topological insulators with $\mathcal{T}^2=-1$, according to the topological equivalence principle. Classification of the latter in the presence of interaction is indeed $\Z_2^3$\cite{WangScience2014}.

Finally, we define topological invariants.  The two invariants $\eta_{1f}$ and $\eta_{2f}$, defined in Eqs.~\eqref{eq:u1teta-f1} and \eqref{eq:u1teta-f2}, apply to $U_f(1)\rtimes \Z_2$ too. With these definitions and Eqs.~\eqref{eq:u1rtimesz2-prop}, we have
\begin{equation}
\eta_{1f} = e^{i\pi \mu_3}, \quad \eta_{2f} = e^{i\pi\mu_1}\nonumber
\end{equation}
In addition, we define the third invariant $\eta_{3f}$ 
\begin{align}
\eta_{3f} &  = e^{i\pi \sigma_H^{\rm mp}} = \theta_w^4
\label{eq:u1teta-f3}
\end{align}
That is, $\eta_{3f}= e^{i\pi\mu_2}$.  All these quantities are invariant under adjoining operations and take values of $\pm 1$. Evaluating $\eta_{1f}, \eta_{2f}$ and $\eta_{3f}$ uniquely specifies the SRE state of the mirror plane. We remark that in the language of Ref.~\onlinecite{wangc14} for time-reversal symmetric topological insulators, the indicators $\eta_{1f}, \eta_{2f}$ and $\eta_{3f}$ detect the $eTmT$, three-fermion and non-interacting electronic topological insulators respectively.

Similarly the bosonic case, one may define additional topological invariants using $M_{w,x}^2$ and $\theta_y^2$. However, they are not independent. For both $U_f(1)\times\Z_2$ and $U_f(1)\rtimes \Z_2$, one can check that $M_{w,x}^2 = (\theta_x^*)^4$. In addition, one may define
\begin{equation}
\eta_{4f} = \theta_w(\theta_y^*)^2
\label{eq:u1teta-f4}
\end{equation}
However, $\eta_{4f} = \eta_{1f}^*$ holds for both $U_f(1)\times\Z_2$ and $U_f(1)\rtimes \Z_2$.

\subsubsection{$[U_f(1)\rtimes \Z_4^f]/\Z_2$}

\label{sec:u1rxz4-f}

Now we discuss 3D TCIs  in AI class, whose symmetry group is $[U_f(1)\rtimes\Z_4^{f\calM}]/\Z_2$. They reduce to 2D SRE states in the mirror plane with internal symmetry $[U_f(1)\rtimes \Z_4^f]/\Z_2$. 

We show that the strictly 2D SRE states are classified by $\Z^2$. First, according to Ref.~\onlinecite{WangPRB2016}, there is no non-trivial SPT state protected by $\Z_4^f$ alone. Second, there is also no non-trivial SPT state protected jointly by $\mathbf{x}$ and $U_f(1)$. It can be argued in a similar way as for $U_f(1)\rtimes\Z_2$(see Sec.~\ref{sec:u1rxz2-f}). Therefore, the only possible SRE states are stacks of the $E_8$ and IQH states. This leads to a $\Z^2$ classification. Nevertheless, the root state associated with the second $\Z$ is not associated with the $\sigma_H=1$ IQH state, but the $\sigma_H = 2$ state. In other words, the $\sigma_H=1$ state is incompatible with $[U_f(1)\rtimes \Z_4^f]/\Z_2$ symmetry. The reason behind this is the so-called $H^3(G,\Z_2)$ obstruction\cite{GuPRB2014,WangPRX2018}. We do not give a detailed reasoning here, but instead refer the readers to Ref.~\onlinecite{esb}, which shows that the quaternion group, i.e., $(\Z_4^f\rtimes\Z_4^f)/\Z_2$, is incompatible with the $\sigma_H=1$ IQH state due to $H^3(G,\Z_2)$ obstruction. Since $(\Z_4^f\rtimes\Z_4^f)/\Z_2$ is a subgroup of $[U_f(1)\rtimes \Z_4^f]/\Z_2$, the latter is incompatible to the $\sigma_H=1$ IQH state either.

A general state can be indexed by an integer vector $\mu=(\mu_1, \mu_2)$. It consists of $\mu_i$ copies of the $i$th root state. The chiral central charge and Hall conductance are 
\begin{align}
c^{\rm mp } & = 8\mu_1 + 2 \mu_2\nonumber\\
\sigma_H^{\rm mp}& = 2\mu_2
\end{align}
Under adjoining operations, the $\Z$ classification associated with $E_8$ state again reduces to $\Z_2$. On the other hand, the $\Z$ classification associated with IQH states all become trivial. This is not hard to understand: as the root state has  $\sigma_H=2$, adjoining a $\sigma_H =1$ state on each side of the mirror plane can trivialize the root state.  Therefore, the overall classification becomes $\Z_2$. 

We observe that topological invariant $\eta_{2f}$ in \eqref{eq:u1teta-f2} still applies, and it distinguishes the states in the $\Z_2$ classification. Therefore, $\eta_{2f}$ will serve as the anomaly indicator for 3D TCIs with in AI class. At the same time, $\eta_{1f}$ is not applicable as there is no $\Z_2^f\times\Z_2^x$ subgroup, and $\eta_{3f}$ is always equal to 1.

\section{Anomaly indicators for bosonic systems with $G=U(1)$}
\label{sec:boson-U(1)}

In this section, we use the folding approach outlined in Sec.~\ref{sec:folding} to derive expressions for the anomaly indicators $\teta_{1}, \teta_2, \teta_3, \teta_4$ and $\teta_5$, defined in Eqs.~\eqref{eq:tilde_eta}, in terms of SET quantities for bosonic TCIs. During the derivation, we will define a set of equivalent anomaly indicators $\eta_1, \eta_2,\eta_3$ and $\eta_4$, which are the ones listed in Sec.~\ref{sec:main-results}. Relations between the two sets of indicators are listed in Table \ref{tab:eta-relation}. Both $U(1)\times \Z_2^\mathcal{M}$ and $U(1)\rtimes \Z_2^\mathcal{M}$ symmetry groups are considered.

\subsection{Surface SETs}
\label{sec:STO-b}

Let us first define a few quantities to describe surface topological orders in the presence of $U(1)$ and $\mathcal{M}$ symmetry.  The topological properties are reviewed in Sec.~\ref{sec:topo}, so here we only discuss symmetry properties. General theories on SET phases can be found in Refs.~\cite{BarkeshliPRB2019, TeoAP2015, TarantinoNJP2016}.

Consider a general bosonic topological order $\mathcal{C}=\{1, a, b, \dots\}$. Symmetry properties of $\mathcal{C}$ consists of two pieces:\footnote{Strictly speaking, one also need to consider stacking SPT phases. However, it does not affect our discussions of anomaly.} (i) how a symmetry permutes anyon types and (ii) what fractional quantum number is carried by certain anyons. For $U(1)$ group, there is no non-trivial permutation, as all group members are continuously connected to the identity. Then, symmetry properties of $\mathcal{C}$ are all encoded in the fractional quantum number carried by each anyon, which is the fractional charge $q_a$ (defined modulo 1) for every $a$. We will take the convention $0\le q_a<1$. The fractional charges $\{q_a\}$ should satisfy the following property
\begin{equation}
q_a + q_b = q_{c} \modulo{1}
\label{eq:q-cond}
\end{equation}
for all $a,b,c$ satisfying $N_{ab}^c \neq 0$. In particular, since $N_{a\bar{a}}^1=1$,  we have $q_{\bar{a}} + q_a = 0 \modulo{1}$, where we understand that the vacuum anyon can never carry fractional charge.

For the mirror symmetry $\mathcal{M}$, a non-trivial permutation on anyons is allowed. We denote it as $\rho_m^{\mathcal{C}}$, which is an invertible map from $\mathcal{C}$ to itself. It is actually an anti-autoequivalence of the UMTC $\mathcal{C}$, considering that $\mathcal{M}$ reverses the orientation. Fusion and braiding properties should satisfy the following relations under the action of $\rho_m^{\mathcal{C}}$:
\begin{align}
N_{\rho_m^{\mathcal{C}}(a),\rho_m^{\mathcal{C}}(b)}^{\rho_m^{\mathcal{C}}(c)}& = N_{ab}^c\nonumber\\
\theta_{\rho_m^{\mathcal{C}}(a)} & = \theta_a^* \nonumber\\
\rho_m^{\mathcal{C}}(\bar{a}) & = \overline{\rho_m^\mathcal{C}(a)}  
\label{eq:permute-prop}
\end{align}
where the complex conjugation is due to the fact that $\mathcal{M}$ is an orientation-reversing symmetry. Since $\mathcal{M}^2 =1$, we also require that $\rho_m^\mathcal{C}(\rho_m^\mathcal{C}(a)) =a$.

Anyons may also carry fractional mirror quantum number. To define it,  consider a two-anyon wave function $|a,\bar a\rangle$, where $a$ and $\bar{a}$ are located symmetric on the two sides of the mirror axis, and $\rho_m^{\mathcal{C}}(a) = \bar{a}$. This state is symmetric under $\mathcal{M}$\footnote{Note that not every state $|a,\bar{a}\rangle$ with $\rho_m^{\mathcal{C}}=\bar{a}$ is an eigenstate of $\mathcal{M}$. However, one can always modify $|a,\bar{a}\rangle$ \emph{locally} around $a$ and $\bar{a}$ such that it becomes a mirror eigenstate.}, so we have a well defined mirror eigenvalue 
\begin{equation}
\mathcal{M}|a, \bar{a} \rangle=\mu_a|a, \bar a\rangle
\end{equation}
where $\mu_a=\pm 1$. The quantity $\mu_a$ is the ``fractional mirror quantum number'' that describes the mirror SET. If $\rho_m^\mathcal{C}(a) \neq \bar{a}$, there is no physical way to define $\mu_a$. However, for later convenience, we define
\begin{equation}
\mu_a =0, \quad \text{if } \rho_m^\mathcal{C}(a) \neq \bar{a}.
\label{eq:mu=0}
\end{equation}
The quantity $\mu_a$ should satisfy the following property
\begin{equation}
\mu_a\mu_b = \mu_c
\label{eq:mu-prop}
\end{equation}
if all three anyons have a well defined mirror eigenvalue and $N_{ab}^c \neq 0$. Another constraint is that if an Abelian $a = b\times \bar{\rho}_m^\mathcal{C}(b)$, then we must have  $\mu_a = \theta_a$. These constraints are believed to be complete for Abelian topological order, but incomplete for non-Abelian topological orders.

Different choices of $(q_a, \rho_m^\mathcal{C}, \mu_a)$ describe different SETs with $U(1)$ and $\mathcal{M}$ symmetries. Difference between $U(1)\times \Z_2^\mathcal{M}$ and $U(1)\rtimes\Z_2^\mathcal{M}$ lies in the constraints on these quantities. We understand that $U(1)$ charges reverse sign under $\mathcal{M}$ for $U(1)\rtimes \Z_2^{\mathcal{M}}$, but not for $U(1)\times \Z_2^\mathcal{M}$. Accordingly, we have 
\begin{align}
q_{\rho_m^\mathcal{C}(a)} = \zeta q_a \ \modulo{1}
\label{eq:q-mirror-1}
\end{align}
where $\zeta=1$ for $U(1)\times \Z_2^\mathcal{M}$ and $\zeta=-1$ for $U(1)\rtimes \Z_2^\mathcal{M}$. With this constraint, we see that for $U(1)\times \Z_2$, if $\rho_m^\mathcal{C}(a) = \bar{a}$, then $q_a=q_{\bar{a}}$, leading to $q_a = 0$ or $1/2$. On the other hand, no such constraint exists for $U(1)\rtimes\Z_2$. One may also consider how $U(1)$ flux transforms under $\mathcal{M}$. One can check $U(1)$ flux flips the sign under $\mathcal{M}$ for $U(1)\times\Z_2^\mathcal{M}$, and does not change for $U(1)\rtimes\Z_2^\mathcal{M}$. This is due to the orientation-reversing nature of $\mathcal{M}$.

Finally, it is worth considering another mirror symmetry
\begin{equation}
\mathcal{M}' = \mathcal{M}\mathcal{U}_{\pi}
\label{eq:mprime}
\end{equation}
Like $\mathcal{M}$, we can specify the associated permutation $\rho_{m'}^\mathcal{C}$ and $\mu_{a}'$. Since $U(1)$ symmetries does not permute anyons, we have $\rho_{m'}^\mathcal{C} = \rho_m^\mathcal{C}$. The mirror eigenvalue $\mu_a'$ can be defined similarly as above. The key problem is how to relate $\mu_a'$ to $\mu_a$. We note that the two anyon state $|a,\bar{a}\rangle$ should be symmetric under both $\mathcal{M}$ and $U(1)$. Let the \emph{absolute} $U(1)$ charge around $a$ and $\bar{a}$ in this state be $Q_a$ and $Q_{\bar{a}}$, respectively. Under $\mathcal{M}$, the charge $Q_a$ is mapped to $\zeta Q_a$. To respect $\mathcal{M}$, it is then required that $Q_{\bar{a}}=\zeta Q_a$. The action of $\mathcal{U}_\pi$ on $|a,\bar{a}\rangle$ is determined by the total charge of $a$ and $\bar{a}$:
\begin{equation}
U_\pi |a,\bar{a}\rangle = e^{i\pi (Q_a+Q_{\bar{a}})}|a,\bar{a}\rangle = e^{i\pi (1+\zeta) Q_a}|a,\bar{a}\rangle
\end{equation}
Accordingly, we have
\begin{equation}
\mu_a' = \left\{
\begin{array}{lc}
\mu_a e^{i2\pi q_a},   & \quad U(1)\times \Z_2^\mathcal{M} \\
\mu_a,  & \quad U(1)\rtimes \Z_2^\mathcal{M}
\end{array} 
\right.
\label{eq:muprime}
\end{equation}
where we have used the relation $e^{i2\pi Q_a} = e^{i2\pi q_a}$.

\subsection{Review on $\teta_1$ and $\teta_2$}
\label{sec:teta12}

Our goal is to express the indicators $\teta_1,\teta_2,\teta_3,\teta_4$ and $\teta_5$ in terms of SET quantities $(q_a,\rho_m^\mathcal{C},\mu_a)$ and $(d_a,\theta_a, N_{ab}^c, S, T)$. The expressions for $\teta_1$ and $\teta_2$ [equivalently $\eta_1$ and $\eta_2$ in Eqs.~\eqref{eq:eta_b1} and \eqref{eq:eta_b2}] were previously discussed in Refs.~\cite{WangPRL2017,BarkeshliCMP2019, folding}. Below we review some basic facts regarding $\teta_1$, $\teta_2$ and their derivations. 

After dimensional reduction and folding discussed in Sec.~\ref{sec:dimredc} and \ref{sec:folding}, the main setup of our systems is shown in Fig.~\ref{fig:folding}(c). The indicators $\teta_1$ and $\teta_2$ are defined in Eqs.~\eqref{eq:teta-b1} and \eqref{eq:teta-b2} through quantities of the mirror plane. The left half of Fig.~\ref{fig:folding}(c) is a double-layer topological order $\mathcal{C}\boxtimes \mathcal{C}$.  Since the two halves of Fig.~\ref{fig:folding}(c) are connected by a gapped domain wall, we shall have
\begin{equation}
c^{\rm mp} = 2 c
\end{equation}
where $c$ is the chiral central charge associated with the topological order $\mathcal{C}$ and $c^{\rm mp}$ is the chiral central charge of the mirror plane. Note that this is already a ``bulk-boundary relation'': $c^{\rm mp}$ is a quantity of the bulk and $c$ is a quantity of SET.  Since $c^{\rm mp}$ must be a multiple of $8$, we have that $c$ must be a multiple of 4. Using the relation \eqref{eq:centralcharge} and the definition \eqref{eq:teta-b1}, one obtains the following expression
\begin{equation}
\eta_1\equiv \teta_1= \frac{1}{D}\sum_a d_a^2 \theta_a
\end{equation}
One can easily check that due to the existence of $\mathcal{M}$ symmetry, the right-hand side can only take values $\pm 1$, in agreement with that of $\teta_1$.  

The second indicator $\teta_2$ was exhaustively studied in Ref.~\onlinecite{folding}. It makes use of anyon condensation theory. The key point in the derivation is that the $x$-vortex in the mirror plane can be lifted to some $\Z_2$ vortex $X$ in the left half of Fig.~\ref{fig:folding}(c). The two vortices must have the same topological spin $\theta_x = \theta_X$. Through anyon condensation theory, one is able to identify $X$ and compute its topological spin. The final result, obtained in Ref.~\onlinecite{folding}, is that the indicator $\teta_2$ in  \eqref{eq:teta-b2} can be expressed as
\begin{equation}
\teta_2 = (\theta_{X})^2 \equiv \eta_2
\end{equation}
where 
\begin{equation}
\eta_2 = \frac{1}{D} \sum_a d_a\theta_a \mu_a
\end{equation}
where $\eta_2$ can only be $\pm1$ too. If one is interested in more detail, we refer the reader to Ref.~\onlinecite{folding}. We remark that Ref.~\onlinecite{folding} assumes $c=0$ in the derivation, but it is easy to generalize to the case that $c\neq 0$ (see a closely  related discussion in the derivation of $\teta_5$ in Appendix \ref{appd:anycond-boson-teta45}).

\subsection{$\teta_3$}
\label{sec:teta3-derive}

Now we move on to the indicator $\teta_3$, which is defined in Eq.~\eqref{eq:teta-b3}. We show that $\teta_3 = \eta_1\eta_3$, and $\eta_3$ is given in Eq.~\eqref{eq:eta_b3}. For convenience, we repeat the expression here:
\begin{equation}
\eta_3 = \frac{1}{D}\sum_{a\in\mathcal{C}} d_a^2 \theta_a e^{i2\pi q_a}
\label{eq:eta3-repreat}
\end{equation}
Below we derive this result by considering properties of Hall conductance. In Appendix \ref{appd:anycond-boson-teta3}, we make use of the alternative definition \eqref{eq:teta-b3-alt} of $\teta_3$ and derive the same result from anyon condensation theory. 

The right half of Fig.~\ref{fig:folding}(c) is characterized by the Hall conductance $\sigma_{H}^{\rm mp}$. The double-layer topological order $\mathcal{C}\boxtimes\mathcal{C}$ on the left is characterized by a Hall conductance $2\sigma_H$, where $\sigma_H$ is the Hall conductance of a single $\mathcal{C}$. Since the two halves are connected by a gapped domain, we have
\begin{equation}
\sigma_H^{\rm mp} = 2\sigma_H
\label{eq:sigmaH-relation}
\end{equation}
With the definition \eqref{eq:teta-b3}, we obtain the following relation
\begin{equation}
\teta_3 = e^{i\pi \sigma_H}
\label{eq:sigmaH-relation2}
\end{equation}
The is an equation that connects the ``bulk'' quantity $\teta_3$ to the surface quantity $\sigma_H$. Since $\sigma_H^{\rm mp}$ must be even, Eq.~\eqref{eq:sigmaH-relation} implies that $\sigma_H$ must be an integer. Accordingly, the right-hand side of \eqref{eq:sigmaH-relation2} can only take values $\pm 1$, in agreement of the values that $\teta_3$ can take.

To proceed, we need to express $\sigma_H$ in terms of $q_a, d_a, \theta_a$, etc. This problem was studied in Ref.~\onlinecite{LapaPRB2019} in the context of time-reversal topological insulators. We repeat their argument here for the paper to be more self-contained. To proceed, we make use of a result from fractional quantum Hall (FQH) states\cite{wen-book}: in FQH systems, adiabatically inserting a $2\pi$ flux will create an excitation $m$, which is an Abelian anyon in $\mathcal{C}$. This anyon satisfies 
\begin{equation}
\theta_m = e^{i\pi\sigma_H}
\label{eq:theta-sigma}
\end{equation}
In general, inserting a flux $\phi$ accumulates a charge $q=\sigma_H\phi/2\pi$ and the topological spin of this flux is $ \frac{1}{2}q\phi =\sigma_H\phi^2/4\pi$.  The mutual statistics between $m$ and any other anyon $a$ is given by 
\begin{equation}
M_{m,a} = e^{i 2\pi q_a}
\label{eq:mma}
\end{equation}
which is the usual Aharonov-Bohm phase.  Equation \eqref{eq:mma} can also be written in terms of the $S$ matrix, through the general relation \eqref{eq:M-S-relation}:
\begin{equation}
S_{m,a} =\frac{d_a}{D} e^{-i 2\pi q_a}
\end{equation}
where the fact that $m$ is Abelian is used.

With these results, one first notices that
\begin{equation}
\delta_{b,m} = \sum_{a}S_{b,a}^\dag S_{a,m} = \frac{1}{D}\sum_a S_{\bar b,a} d_a e^{-i2\pi q_a}
\label{eq:delta-f}
\end{equation}
Then, the indicator $\teta_3$ follows
\begin{align}
\teta_3 = \theta_m & = \sum_{b} \theta_b d_b\delta_{b,m} \nonumber\\
& = \frac{1}{D}\sum_{a,b} \theta_b d_b S_{\bar b,a}d_a e^{-i2\pi q_a}  \nonumber\\
& = \frac{1}{D^2}\sum_{a,b} \theta_b d_b  d_ae^{-i2\pi q_a}\sum_c N_{ab}^c\frac{\theta_c}{\theta_a\theta_b} d_c  \nonumber\\
& =\frac{1}{D^2}\sum_{a,c} \theta_c\theta_a^* d_cd_a e^{-i2\pi q_a} \sum_b N_{\bar a c}^b d_b \nonumber\\
& = \left(\frac{1}{D}\sum_{c} d_c^2 \theta_c\right)\left(\frac{1}{D}\sum_{a} d_a^2 \theta_a^* e^{-i2\pi q_a}\right) \nonumber\\
& = \eta_1 \eta_3^* 
\label{eq:teta3-derive}
\end{align}
where in the second line, we have inserted \eqref{eq:delta-f}; in the third line, we have used the definition of $S$ matrix; in the four line, we used the property that $d_ad_c = \sum_{b}N_{\bar{a}c}^b d_b$; finally, the expressions of $\eta_1$ and $\eta_3$ are used. Since $\eta_1$ and $\teta_3$ only take values $\pm 1$, so is $\eta_3$. Accordingly, the complex conjugation on $\eta_3$ in \eqref{eq:teta3-derive} does not matter.

A few comments are in order. First, the above derivation holds for both $U(1)\times \Z_2^\mathcal{M}$ and $U(1)\rtimes \Z_2^\mathcal{M}$. Second, the fact that $\eta_3 =\pm 1$ can also be seen by SET properties. To see that, for $U(1)\times\Z_2^\mathcal{M}$, we can replace the summation in \eqref{eq:eta3-repreat} with a summation over $\rho_m^\mathcal{C}(\bar a)$, and then
\begin{align}
\eta_3 & = \frac{1}{D}\sum_{\rho_m^\mathcal{C}(\bar a)} d_{\rho_m^\mathcal{C}(\bar a)}^2 \theta_{\rho_m^{\mathcal{C}}(\bar a)} e^{i2\pi q_{\rho_m^\mathcal{C}(\bar a)}} \nonumber\\
& = \frac{1}{D} \sum_{a} d_a^2 \theta_{a}^* e^{-i2\pi q_a}\nonumber\\
& = \eta_3^*
\end{align}
Accordingly, $\eta_3$ must be real. Considering $\eta_3 = \eta_1\teta_3^*$ must be a phase, we obtain $\eta_3=\pm 1$. For $U(1)\rtimes\Z_2^\mathcal{M}$, it can be argued similarly by replacing the summation in \eqref{eq:eta3-repreat} with a summation over  $\rho_m^\mathcal{C}(a)$ instead. Third, the Abelian anyon $m$ is the footprint left on the surface by a $U(1)$ monopole, when it travels from the vacuum into the 3D bulk. We will discuss more about properties of $m$ in Sec.~\ref{sec:monopole-b}

\subsection{$\teta_4$ and $\teta_5$}

In this subsection, we derive expressions for $\teta_4$ and $\teta_5$. For $U(1)\rtimes \Z_2^\calM$, the two indicators are not independent: $\teta_4=\teta_3$ and $\teta_5=\teta_2$ (Sec.~\ref{sec:U(1)rxZ2-b}). There is no need to derive their expressions. Therefore, we focus on $U(1)\times \Z_2^\calM$ in this section. 

Among $\teta_4$ and $\teta_5$, only one is independent once other indicators are given. They are related by $\teta_5 = \teta_2\teta_3\teta_4$.  Here, we claim that $\teta_5 = \eta_4$, and the expression of $\eta_4$ in terms of SET quantities is given in Eq.~\eqref{eq:eta_b4}. For convenience, we repeat the expression here:
\begin{equation}
\eta_4 = \frac{1}{D}\sum_{a\in\mathcal{C}} d_a \theta_a \mu_a  e^{i2\pi q_a}
\label{eq:eta4-repeat} 
\end{equation}
This claim will be proved shortly. The indicator $\eta_4$ take values of $\pm 1$. At the same time, we obtain $\teta_4 = \teta_2\teta_3\teta_5 =  \eta_1\eta_2\eta_3\eta_4$. In Appendix \ref{appd:anycond-boson-teta45}, we will derive the expressions of $\teta_4$ and $\teta_5$ from anyon condensation theory, which provides an alternative understanding.

The proof of the above claim is straightforward. We will make use of the $y$-vortices, which correspond to the mirror symmetry $\mathcal{M}'$ defined in \eqref{eq:mprime}. Since $\teta_5 = \theta_y^2$, we can directly use the formula of $\teta_2$ by replacing the quantities $(\rho_m^\mathcal{C},\mu_a)$ with $(\rho_{m'}^\mathcal{C}, \mu_a')$. Given $\rho_{m'}^\mathcal{C} = \rho_{m}^\mathcal{C}$ and $\mu_a' = \mu_a e^{i2\pi q_a}$ [see Eq.~\eqref{eq:muprime}], we immediately obtain the result $\teta_5 =  \eta_4$, where $\eta_4$ follows from $\eta_2$ by replacing $\mu_a$ with $\mu_a'$. We remark that if we apply this proof to $U(1)\rtimes\Z_2$, we will see that $\teta_5$ and $\teta_2$ have the same expression. This verifies the relation $\teta_5 = \teta_2$ from a surface viewpoint.

\subsection{Properties of $m$}
\label{sec:monopole-b}

As mentioned above, the anyon $m$ in \eqref{eq:mma} is a surface avatar of the bulk $U(1)$ monopoles. Properties of monopoles in $U(1)$ gauge theory are widely studied for the purpose of detecting topological phases (see e.g. Refs.~\cite{MetlitskiPRB2013, wangc14, ZouPRB2018, WangCPRX201_U1SpinLiquid}). The connection between $m$ and monopoles  can be established by this thought experiment\cite{MetlitskiPRB2013, wangc13}: imagine that a monopole adiabatically moves from the outside to the inside of a 3D SPT system and leaves a footprint on the surface. Since this is a process equivalent to adiabatically inserting a $2\pi$ flux, the footprint on the surface is the anyon $m$. Due to this connection, below we discuss properties of $m$ and express its topological spin $\theta_m$, $U(1)$ charge $Q_m$, and mirror fractionalization $\mu_m$ in terms of the anomaly indicators.

As discussed above in Sec.~\ref{sec:teta3-derive}, the topological spin is 
\begin{align}
\theta_m=\eta_1\eta_3
\end{align}
Next, following Laughlin's flux insertion argument, the charge accumulated by adiabatically inserting a $2\pi$ flux is $\sigma_H$. Accordingly, the $U(1)$ charge $Q_m=\sigma_H$. Note that $Q_m$ is tied to the specific state with the $2\pi$ flux being adiabatically inserted.  However, the surface Hall conductance $\sigma_H$ can be modified by attaching 2D bosonic IQH states, which does not affect any SET properties. Since $\sigma_H$ of bosonic IQH states is an even integer, $Q_m$ is well defined only modulo $2$. So, it is more convenient to consider $e^{i\pi Q_m}$. Due to the relation \eqref{eq:theta-sigma}, we have
\begin{align}
e^{i\pi Q_m} = e^{i\pi \sigma_H} = \theta_m = \eta_1\eta_3.
\end{align}
To see how $m$ is permuted by mirror symmetry $\mathcal{M}$, we consider the mutual statistics
\begin{align}
M_{\rho_m^\mathcal{C}(m), a} & = M_{m, \rho_m^\mathcal{C}(a)}^* = e^{-i2\pi q_{ \rho_m^\mathcal{C}(a)}} = e^{- i 2\pi \zeta q_a}
\label{eq:m-permute}
\end{align}
where the first equality is due to the fact $\rho_m^\mathcal{C}$ is an anti-autoequivalence, the second equality is due to \eqref{eq:mma} and the last equality is due to \eqref{eq:q-mirror-1}. Due to braiding non-degeneracy of $\mathcal{C}$, it is not hard to see
\begin{align}
\rho_m^\mathcal{C}(m)= \left\{
\begin{array}{ll}
\bar m, & \zeta=1\\
m, & \zeta =-1
\end{array}
\right.
\end{align}
This agrees with the intuition that $U(1)$ fluxes are flipped by $\mathcal{M}$ for $U(1)\times\Z_2^\mathcal{M}$, but are unchanged under $\mathcal{M}$ for $U(1)\rtimes\Z_2^\mathcal{M}$. Accordingly, for $\zeta=1$, i.e., $U(1)\times \Z_2^\calM$, we can further define the mirror eigenvalue $\mu_m$. We show in Appendix \ref{appd:anycond-boson-teta45} that $\mu_m=\tilde \eta_4$ [see discussions around Eq.~\eqref{eq:monople-teta4}]. Accordingly, we have
\begin{align}
\mu_m =\teta_4= \eta_1\eta_2\eta_3\eta_4.
\end{align}
We remark that all above properties hold  for the time-reversal counterparts. In particular, $\mu_m$ corresponds to the Kramers degeneracy $\mathcal{T}^2_m$.

\begin{table}
	\caption{Different SETs of the toric code topological order with $U(1)\times \mathbb{Z}_2^{\mathcal{M}}$ or $U(1)\rtimes\mathbb{Z}_2^{\mathcal{M}}$. The quantity $\sigma_{e}=e^{i\pi q_e}$ and  $\sigma_{m}=e^{i\pi q_m}$. In the SET names, we have followed the notation of Ref.~\onlinecite{wangc13}: ``C'' represents that the anyon has a fractional charge $1/2$ and ``M" represents that the anyon carries a mirror eigenvalue $-1$. We note that $\eta_4$ only applies to $U(1)\times \mathbb{Z}_2^{\mathcal{M}}$ but not $U(1)\rtimes \mathbb{Z}_2^{\mathcal{M}}$; otherwise, all quantities are the same for the two symmetry groups. The last row is the ``eFmF'' example from Ref.~\onlinecite{vishwanath13}. }
	\label{tab:toric-code}
	\begin{centering}
		\begin{tabular}{c|cccccccc}
			\hline 
			\hline 
			Phase & $\sigma_{e}$ & $\sigma_{m}$ & $\mu_{e}$ & $\mu_{m}$ & $\eta_{1}$ & $\eta_{2}$ & $\eta_{3}$ & $\eta_{4}$\tabularnewline
			\hline 
			e0m0 & 1 & 1 & 1 & 1 & 1 & 1 & 1 & 1\tabularnewline
			\hline 
			eM & 1 & 1 & $-1$ & 1 & 1 & 1 & 1 & 1\tabularnewline
			\hline 
			eC & $-1$ & 1 & 1 & 1 & 1 & 1 & 1 & 1\tabularnewline
			\hline 
			eCM & $-1$ & 1 & $-1$ & 1 & 1 & 1 & 1 & 1\tabularnewline
			\hline 
			eCmM & $-1$ & 1 & 1 & $-1$ & 1 & 1 & 1 & $-1$\tabularnewline
			\hline 
			eMmM & 1 & 1 & $-1$ & $-1$ & 1 & $-1$ & 1 & $-1$\tabularnewline
			\hline 
			eCMmM & $-1$ & 1 &$-1$ & $-1$ & 1 & $-1$ & 1 & 1\tabularnewline
			\hline 
			eCMmCM & $-1$ & $-1$ & $-1$ & $-1$ & 1 & $-1$ & $-1$ & 1\tabularnewline
			\hline 
			eCMmC & $-1$ & $-1$ & $-1$ & 1 & 1 & 1 & $-1$ & 1\tabularnewline
			\hline 
			eCmC & $-1$ & $-1$ & 1 & 1 & 1 & 1 & $-1$ & $-1$\tabularnewline
			\hline 
			eFmF & 1 & 1 & 1 & 1 & $-1$ & $-1$ & $-1$ & $-1$\tabularnewline
			\hline 
		\end{tabular}
		\par\end{centering}
\end{table}

\subsection{Examples}
In this section, we explore a few examples for the anomaly indicators. We consider the SET examples in Ref.~\cite{wangc13} for the toric code topological order,
\begin{equation}
\mathcal{C}=\{1,e,m,\epsilon\},
\end{equation}
where $e$ and $m$ are bosons, and $\epsilon=e\times m$ is a fermion. All anyons are Abelian with $d_a=1$, and the total quantum dimension is $D=2$. We consider the symmetry group $U(1)\times\Z_2^\mathcal{M}$ or  $U(1)\rtimes\Z_2^\mathcal{M}$. Following the notation of Ref.~\onlinecite{wangc13}, we list various SET states for $\mathcal{C}$ in Table \ref{tab:toric-code}, according to different values of $q_e$,  $q_m$, $\mu_e$ and $\mu_m$.  The SET data of $\epsilon$ are determined by using Eqs.~\eqref{eq:q-cond} and \eqref{eq:mu-prop}. We can easily obtain all the indicators of different SETs by using Eq.~\eqref{eq:eta_b} and the results are shown in Table \ref{tab:toric-code}. We note that, in Table \ref{tab:toric-code} , $\eta_4$ only applies to $U(1)\times \mathbb{Z}_2^{\mathcal{M}}$ but not $U(1)\rtimes \mathbb{Z}_2^{\mathcal{M}}$. One can see that ``eCmM" is anomaly-free with $U(1)\rtimes\Z_2^\mathcal{M}$, but anomalous with $U(1)\times\Z_2^\mathcal{M}$. For other SETs, they show the same anomaly characteristic with either $U(1)\times\Z_2^\mathcal{M}$ or $U(1)\rtimes\Z_2^\mathcal{M}$ group.

In addition, we list the example of ``eFmF'' SET in Table \ref{tab:toric-code}\cite{vishwanath13}. It contains the same four Abelian anyons as the toric code topological order. However, $e$ and $m$ are fermions instead. We have considered the case that all $q_e$,  $q_m$, $\mu_e$ and $\mu_m$ are trivial, and obtained the anomaly indicators $\eta_i$. With the relations between $\eta_i$ and $\teta_i$ in Table \ref{tab:eta-relation}, we have
\begin{align}
\teta_1  =\teta_2=\teta_5=-1,\quad \teta_3 =\teta_4 =1
\end{align}
Then, $\theta_x^2=\teta_2=-1$. Accordingly, the mirror symmetry, that corresponds to the $x$-vortices, must have a non-trivial action on the eFmF state. On the contrary, both mirror permutation and mirror fractionalization are ``trivial'' on every anyon. This example demonstrates that there is no well-defined concept of ``a trivial SET state''.

\section{Anomaly indicators for fermionic systems with $G=U_f(1)$}
\label{sec:fermion-U(1)}

In this section, we use the folding approach to derive the expressions for the indicators $\eta_{1f}, \eta_{2f}$ and $\eta_{3f}$ in fermionic systems. We  mainly discuss the symmetry groups $U_f(1)\times\Z_2^\calM$ and $U_f(1)\rtimes\Z_2^\calM$. The case of $[U_f(1)\rtimes\Z_4^{f\calM}]/\Z_2$ is slightly different and will be discussed separately in Sec.~\ref{sec:AI-indicators}. We remark that many technical parts of our derivations are simply repetitions of those in Ref.~\onlinecite{LapaPRB2019}, which derived expressions of anomaly indicators for time-reversal topological insulators. Nevertheless, it is still worth studying them in the context of topological crystalline  insulators and providing an alternative viewpoint.

\subsection{Surface SETs}
\label{sec:u1f-set}
Consider a fermionic topological order $\mathcal{C} = \{1, f, a, af, \dots\}$. We denote the pair $\{a,af\}$ as $[a]$. Similarly to the bosonic case, symmetry properties of $\mathcal{C}$ contain two pieces of data: (i) permutation of anyons by the symmetries and (ii) fractional quantum numbers carried by anyons. Again, $U_f(1)$ does not permute anyons. Fractionalization of $U_f(1)$ is described by the fractional charge $q_a$. Compared to the bosonic case, the charge $q_a$ can be defined in the range $0\le q_{a} < 2$. This is because the unit charge is the fermion $f$, which is viewed as an anyon in our notation. The fractional charges satisfy 
\begin{equation}
q_a+q_b = q_c \modulo{2}
\end{equation}
whenever $N_{ab}^c\neq 0$. In particular, $q_{af} = q_a+ 1\modulo{2}$ and $q_{\bar{a}} +q_a = 0 \modulo{2}$. 

The mirror symmetry $\mathcal{M}$ can permute anyons in a non-trivial way. Like in the bosonic case, we denote the permutation as $\rho_m^\mathcal{C}$. It is an invertible map that maps $\mathcal{C}$ to itself. It satisfies $\rho_m^\mathcal{C}\circ\rho_m^\mathcal{C} = \mathbbm{1}$, $\rho_m^\mathcal{C}(1) = 1$ and $\rho_m^\mathcal{C}(f) = f$. The fusion and braiding properties satisfy the same relations in \eqref{eq:permute-prop}. Mirror symmetry fractionalization is defined in the same way as in the bosonic case, using a two-anyon state. Differently from the bosonic case, we now have two kinds of two-anyon states to consider: $|a,\bar{a}\rangle$ and $|a,\bar{a}f\rangle$. The latter is well-defined state because $f$ is a local excitation, and it is a state of odd fermion parity. To respect the mirror symmetry, we must have $\rho_m^\mathcal{C}(a) =\bar{a}$ and $\rho_m^\mathcal{C}(a) = \bar af$, respectively. Note that for a given $a$, at most one of the two conditions can be satisfied. So, let us define
\begin{align}
\xi_a = \left\{\begin{array}{ll}
1, & \text{if } \rho_m^\mathcal{C}(a) =\bar{a}\\
-1, & \text{if } \rho_m^\mathcal{C}(a) =\bar{a}f\\
0, & \text{otherwise}
\end{array}
\right.
\end{align}
Then, as long as $\xi_a\neq 0$, we define the mirror fractionalization as follows:
\begin{align}
\mathcal{M}|a, \rho_m^\mathcal{C}(a)\rangle = \mu_a|a, \rho_m^\mathcal{C}(a)\rangle
\end{align}
where $\mu_a=\pm 1$. For convenience, we also define $\mu_a=0$ if $\xi_a=0$. The property \eqref{eq:mu-prop} is still satisfied in the fermionic case. Note that the above discussions on $\mathcal{M}$ do not apply to the AI class, where $\mathcal{M}^2 = P_f$. We will discuss it separately in Sec.~\ref{sec:AI-indicators}.

Different choices of $(q_a, \rho_m^\mathcal{C}, \mu_a)$ describe different SET states with $U_f(1)$ and $\mathcal{M}$ symmetries. Like in the bosonic case, the difference between $U_f(1)\times\Z_2^\calM$ and $U_f(1)\rtimes\Z_2^\calM$ is the following condition on fractional charge
\begin{equation}
q_{\rho_m^\mathcal{C}(a)} = \zeta q_a \modulo{2}
\label{eq:qf-prop}
\end{equation}
where $\zeta=1$ for $U_f(1)\times \Z_2^\calM$ and $\zeta=-1$ for $U_f(1)\rtimes\Z_2^\calM$. This condition puts a very strong constraint on the possible permutations in the case of $U_f(1)\rtimes\Z_2^\calM$: it forbids those with $\xi_a=-1$. If $\xi_a=-1$, we have $q_{\rho_m^\mathcal{C}(a)} = q_{\bar{a}f} = -q_a+1$. Meanwhile, Eq.~\eqref{eq:qf-prop} leads to $q_{\rho_m^\mathcal{C}(a)} =-q_a$. The two equations can never be satisfied simultaneously. Accordingly, $\xi_a=-1$ is forbidden for $U_f(1)\rtimes\Z_2^\mathcal{M}$.

\subsection{$\eta_{1f}$}
\label{sec:eta-1f}

The first indicator $\eta_{1f}$ was defined and studied in Ref.~\onlinecite{Mao2020} by two of us, with the expression given in \eqref{eq:eta-f1}. The time-reversal counterparts were previously studied in Refs.~\cite{WangPRL2017,TachikawaPRL2017}. For convenience, we repeat the expression  here:
\begin{align}
\eta_{1f}= \frac{1}{\sqrt{2}D} \sum_{a\in\mathcal{C}} d_a\theta_a\mu_a
\label{eq:eta1f-repeat}
\end{align}
It was derived through the folding approach and anyon condensation theory with the definition \eqref{eq:u1teta-f1-alt}, for systems with only $\mathcal{M}$ symmetry. In that case, the indicator can take $16$ distinct values $1, e^{i\pi/8}, \dots, e^{i15\pi/8}$, both by its definition and by evaluating the right-hand side of \eqref{eq:eta1f-repeat}. As the derivation is technically complicated, we do not repeat it here and refer readers to Ref.~\cite{Mao2020}.

In the presence of $U_f(1)$ symmetry, possible values that $\eta_{1f}$ can take will be reduced. According to Sec.~\ref{sec:define-eta-u1-f}, $\eta_{1f}$ takes values in $1, e^{i\pi/4}, \dots, e^{i7\pi/4}$ for $U_f(1)\times\Z_2^\mathcal{M}$, and takes values $\pm 1$ for $U_f(1)\rtimes \Z_2^\mathcal{M}$. Here, we show that the same results can be obtained from \eqref{eq:eta1f-repeat} by using constraints on the SET quantities $(q_a,\rho_m^\mathcal{C}, \mu_a)$. 

First, we show that the presence of $U_f(1)$ forbids $\eta_{1f}$ to take the values $e^{i\nu\pi/8}$ with add $\nu$. To do that, we cite a result from Ref.~\onlinecite{Mao2020}: to take $e^{i\nu\pi/8}$ with an odd $\nu$, the topological order $\mathcal{C}$ must allow Majorana-type vortices after gauging $\Z_2^f$. A Majorana-type vortex $v$ carries fermion-parity flux  and satisfies a fusion rule of the following form:
\begin{align}
v\times \bar{v} = 1 + f + \dots
\end{align}
where ``$\dots$'' represents other anyons in $\mathcal{C}$. After gauging $\Z_2^f$, $\mathcal{C}$ is enlarged to a bosonic topological order $\mathcal{B}$ and there remains a global symmetry $U'(1)\equiv U_f(1)/\Z_2^f$. Let $q_\alpha'$ be the fractional charge of $\alpha\in\mathcal{B}$ associated with the $U'(1)$ symmetry. The charge $q_\alpha'$ is measured in units of $2e$, which is the elementary charge of $U'(1)$. In particular, we have $q'_1=0$ and $q_{f}'=1/2$. If there exists a Majorana-type vortex $v$, then $q_v' + q_{\bar{v}}'=q_1'$ \emph{and} $q_v' + q_{\bar{v}}' = q_f'$, according to the constraint \eqref{eq:q-cond}. However, $q_1'\neq q_f'$, so Majorana-type vortices cannot exist. Therefore, $e^{i\nu\pi/8}$ with an odd $\nu$ can not be taken by $\eta_{1f}$.

Second, for $U(1)\rtimes\Z_2^\mathcal{M}$, we further show that $\eta_{1f}$ can only take $\pm1$. According to the discussions in Sec.~\ref{sec:u1f-set}, $\xi_a=-1$ is not allowed for $U(1)\rtimes\Z_2^\mathcal{M}$. Accordingly, to have nonzero $\mu_a$, only $\xi_a=1$ is allowed, i.e., $\rho_m^\mathcal{C}(a) = \bar{a}$. In this case, we must have $\theta_a = \pm 1$, following from $\theta_{\rho_m^\mathcal{C}(a)} = \theta_a^* =\theta_a$. With this, the right-hand side of \eqref{eq:eta1f-repeat} must be real. Then, we conclude that $\eta_{1f}$ can only be $\pm 1$ for $U(1)\rtimes\Z_2^\calM$.

\subsection{$\eta_{2f}$}
Next we derive the bulk-boundary relation associated with $\eta_{2f}$ using the folding approach. First, we recall that $\eta_{2f}$ is defined in \eqref{eq:u1teta-f2} through the chiral central charge $c^{\rm mp}$ and Hall conductance $\sigma_H^{\rm mp}$ of the mirror plane. We need to express them in terms of SET quantities. Using the fact that the left and right parts of Fig.~\ref{fig:folding}(c) are connected by a gapped domain wall, we immediately have
\begin{align}
c^{\rm mp}  & = 2 c \nonumber \\
\sigma_H^{\rm mp} & = 2 \sigma_H
\end{align}
where $c$ and $\sigma_H$ are the chiral central charge and Hall conductance of a single layer of topological order $\mathcal{C}$. Then, the definition \eqref{eq:u1teta-f2} of $\eta_{2f}$ gives rise to
\begin{equation}
\eta_{2f} = e^{i\pi(c-\sigma_H)/4}
\label{eq:eta2f-cHall}
\end{equation}
This expression establishes a bulk-boundary relation: $\eta_{2f}$ is a bulk quantity, and $c,\sigma_H$ are quantities of the surface SET. Since both $c^{\rm mp}$ and $\sigma_H^{\rm mp}$ must be integers, the surface $c$ and $\sigma_H$ are multiples of $1/2$.

The above relation is identical to the counterpart obtained in Ref.~\onlinecite{LapaPRB2019} for time-reversal topological insulators. In addition, Ref.~\onlinecite{LapaPRB2019} shows that the right-hand side of \eqref{eq:eta2f-cHall} can be further expressed in terms of $d_a,\theta_a$ and $q_a$ in \eqref{eq:eta-f2}. We repeat expression here:
\begin{align}
\eta_{2f} = \frac{1}{\sqrt{2}D} \sum_a d_a^2 \theta_a e^{i\pi q_a}
\label{eq:eta2f-repeat}
\end{align}
To be self-contained, we briefly sketch the derivation of $\eta_{2f}$. The basic ideal is to gauge $\Z_2^f$ and extend the fermionic topological order $\mathcal{C}$ into a bosonic topological order $\mathcal{B}$. Let us denote $\mathcal{B} = \mathcal{C}\oplus\mathcal{C}_1$, where $\mathcal{C}_1$ contains fermion-parity vortices. Due to the presence of $U_f(1)$, there exists a special Abelian vortex $V\in\mathcal{C}_1$ --- the one obtained by adiabatically inserting a $\pi$ flux of $U_f(1)$. Following the usual quantum Hall physics\cite{wen-book}, $V$ satisfies the following properties
\begin{equation}
\theta_V = e^{i \pi \sigma_H/4}, \ M_{V,a} = e^{i\pi q_a}
\label{eq:V-prop}
\end{equation}
where the latter is simply the Aharonov-Bohm phase. All other vortices can be obtained by fusing $V$ with $a\in\mathcal{C}$. We denote them as $V_a$ such that $V\times a = V_a$. The total quantum dimension of $\mathcal{B}$ is $D_\mathcal{B}=\sqrt{2}D$. Since $\mathcal{B}$ is a bosonic topological order, we employ the formula \eqref{eq:centralcharge} for the chiral central charge:
\begin{align}
e^{i\pi c/4} & = \frac{1}{\sqrt{2}D} \left(\sum_{a\in \mathcal{C}} d_{a}^2 \theta_a + \sum_{V_a\in\mathcal{C}_1} d_{V_a}^2 \theta_{V_a} \right) \nonumber\\
& = \frac{1}{\sqrt{2}D}  \sum_{a\in\mathcal{C}} d_a^2 \theta_{a}\theta_VM_{V,a}\nonumber\\
& = e^{i\pi \sigma_H/4}  \frac{1}{\sqrt{2}D}  \sum_{a\in\mathcal{C}} d_a^2 \theta_{a}e^{i\pi q_a}
\label{eq:ceq}
\end{align}
where we have used the fact that the first term on the right-hand side of the first line vanishes and have inserted \eqref{eq:V-prop} in the third line. Then, the expression \eqref{eq:eta2f-repeat} immediately follows.

Two remarks are in order. First, the above derivation applies for both $U_f(1)\times\Z_2^\mathcal{M}$ and $U_f(1)\rtimes\Z_2^\mathcal{M}$. In fact, it works for AI class too (see Sec.~\ref{sec:AI-indicators}). Second, the fact that the right-hand side of \eqref{eq:eta2f-repeat} can only take values $\pm1$ can be proven using constraints of SET quantities. The argument is very similar to that for $\eta_3$ in the bosonic case; see the last paragraph in Sec.~\ref{sec:teta3-derive}.

\subsection{$\eta_{3f}$}
\label{sec:eta3f-u1f}

The indicator $\eta_{3f}$ can be obtained similarly to $\eta_{2f}$. Following the relation $\sigma_H^{\rm mp} = 2\sigma_H$ and the definition \eqref{eq:u1teta-f3} of $\eta_{3f}$, we immediately have
\begin{equation}
\eta_{3f} = e^{i2\pi\sigma_H}
\label{eq:eta3f-Hall}
\end{equation}
As above, this equation is a bulk-boundary relation. Because of the properties \eqref{eq:V-prop} of the special vortex $V$, we may rewrite \eqref{eq:eta3f-Hall} as $\eta_{3f} = (\theta_V)^8$.  Alternatively, we may consider adiabatically inserting a $2\pi$ flux into the fermionic topological order $\mathcal{C}$, which creates an anyon $m$. We should have $V\times V =m$. Accordingly, 
\begin{align}
\eta_{3f} = \theta_m^2.
\label{eq:eta3f-m}
\end{align}
where  $\theta_m = \theta_V^4$ is used.

The next task is to express \eqref{eq:eta3f-Hall} or \eqref{eq:eta3f-m} in terms of SET quantities. This is again done in Ref.~\onlinecite{LapaPRB2019}. The expression is given in \eqref{eq:eta-f3} and we repeat here:
\begin{align}
\eta_{3f} = \frac{1}{2D}\sum_{a,b\in\mathcal{C}}d_ad_be^{i2\pi q_a} e^{i2\pi q_b} S_{ab}
\label{eq:eta3f-repeat}
\end{align}
Let us briefly review the derivation to be self-contained. The starting point is the following relation
\begin{align}
M_{m,a} = e^{i2\pi q_a} 
\label{eq:mma-f}
\end{align}
which is the Aharonov-Bohm phase between anyon $a$ and the $2\pi$ vortex $m$. It is the same as the bosonic counterpart \eqref{eq:mma}. 
A special case is that $M_{m,m} = \theta_m^2  = e^{i2\pi q_m}$. Accordingly, \eqref{eq:eta3f-m} can be rewritten as $\eta_{3f} = e^{i2\pi q_m}$. To proceed, we make use of the following property of fermionic topological order: $\hat S_{[a],[b]} \equiv \sqrt{2} S_{a,b}$ is a unitary matrix, where $[a]$ denotes the pair $(a,af)$. With the property that $q_{af}=q_a+1$ and the relation \eqref{eq:M-S-relation}, Eq.~\eqref{eq:mma-f} can be rewritten as
\begin{align}
\hat{S}_{[m],[a]} = \frac{\sqrt{2}d_a}{D} e^{i2\pi q_{[a]}}
\end{align}
where we have denoted $q_{[a]} = q_a \modulo{1}$. The unitarity of $\hat{S}$ leads to
\begin{align}
\delta_{[m],[b]} &  = \sum_{[a]} \hat S_{[m],[a]}^* \hat S_{[a],[b]}\nonumber\\
 & = \sum_{[a]} \frac{\sqrt{2}d_a}{D} \hat S_{[a],[b]}e^{i2\pi q_{[a]}}
\end{align} 
Then, the indicator $\eta_{3f}$ follows as
\begin{align}
\eta_{3f} & = \theta_m^2 = e^{i2\pi q_m} \nonumber\\
& = \sum_{[b]} \delta_{[m],[b]}d_b e^{i2\pi q_{[b]}} \nonumber\\
& = \frac{\sqrt{2}}{D} \sum_{[a],[b]} d_ad_b e^{i2\pi q_{[a]}}e^{i2\pi q_{[b]}} \hat S_{[a],[b]}\nonumber\\
& = \frac{1}{2D} \sum_{a,b} d_ad_b e^{i2\pi q_{a}}e^{i2\pi q_{b}}  S_{a,b}
\end{align}
This proves the expression of $\eta_{3f}$.

We remark that the above expression applies to all topological crystalline insulators, as the mirror symmetry does not appear explicitly. Nevertheless, $\mathcal{M}$ guarantees that $\eta_{3f}$ can at most take two values, $+1$ or $-1$. This can be shown through the expression \eqref{eq:eta3f-repeat} by replacing the summation as an alternative summation over $\rho_{m}^{\mathcal{C}}(a)$ or $\rho_{m}^{\mathcal{C}}(\bar a)$, similarly to the argument  for $\eta_3$ and $\eta_{2f}$.  However, as discussed in Sec.~\ref{sec:define-eta-u1-f},  the following relations are imposed: $\eta_{3f}=\eta_{1f}^4$ in AIII class and $\eta_{3f}=1$ in AI class. There should exists certain way to show these relations from constraints on surface SET quantities, which we have not figured out yet unfortunately. Similarly to the bosonic case, one can alternatively derive $\eta_{3f}$ using anyon condensation theory. The derivation is similar to that in Appendix \ref{appd:anycond-boson-teta3} and Ref.~\onlinecite{Mao2020}, and we will not describe the details.

\subsection{Indicators for AI class}
\label{sec:AI-indicators}

The AI class deserves a special attention. In this case, the mirror symmetry satisfies $\mathcal{M}^2 =P_f$. Accordingly, the description of mirror symmetry properties of the SETs are slightly different from that in Sec.~\ref{sec:u1f-set}. The permutation $\rho_m^\mathcal{C}$ satisfies the same properties as in the case $\mathcal{M}^2=1$. For example, $\rho_m^\mathcal{C}\circ\rho_m^\mathcal{C}=1$ still holds, as fermion parity cannot permute anyons. However, the mirror symmetry fractionalization will be different. For the two-anyon state $|a,\bar{a}f\rangle$ with $\rho_m^\mathcal{C}(a) = \bar{a}f$, we should have $\mathcal{M}^2 = P_f = -1$. Accordingly, $\mu_a$ shall take values $\pm i$. Accordingly, $\mu_a$ is generally given by
\begin{align}
\mu_a = \left\{\begin{array}{ll}
\pm1, & \text{if } \xi_a=1\\
\pm i, & \text{if } \xi_a=-1\\
0, & \text{otherwise}
\end{array}
\right.
\label{eq:mua-AI}
\end{align}
Since the expression \eqref{eq:eta1f-repeat} of the indicator $\eta_{1f}$ explicitly involves $\mu_a$, it is not applicable any more in the AI class. In fact, this inapplicability has already been clear in its bulk definition, as discussed in Sec.~\ref{sec:u1rxz4-f}. 

Alternatively, the expression \eqref{eq:eta2f-repeat} of $\eta_{2f}$ and the expression \eqref{eq:eta3f-repeat} of $\eta_{3f}$ only involve the fractional charge $q_a$, so they are still applicable. This agrees with their definitions from the viewpoint of the mirror plane, discussed in Sec.~\ref{sec:u1rxz4-f}. The indicator $\eta_{2f}$ takes values $+1$ and $-1$, which distinguishes the $\Z_2$ classification of topological crystalline insulators in AI class. On the other hand, $\eta_{3f}$ must be equal to 1, according to the result in Sec.~\ref{sec:u1rxz4-f}. There should exists certain way to prove this from constraints on surface SET quantities, which we have not figured out yet unfortunately.

\subsection{Properties of $m$}
Because of the connection between monopole physics and its surface avatar $m$, it is again interesting to discuss the properties of $m$ and relate them to the indicators. First of all, we see that Eq.~\eqref{eq:mma-f} can only determine $[m]$ but not $m$ itself. Equivalently, it means that the SET quantities $\theta_a, \mu_a, q_a$, etc, cannot determine $m$, so are the indicators. This ambiguity can also be seen from the fact that attaching a 2D IQH state with unit Hall conductance, which does not change SET quantities, changes $m$ to $mf$.  Accordingly, only $\theta_m^2$ and $e^{i2\pi q_m}$ can be determined by the anomaly indicators.    With (\ref{eq:eta3f-m}), \eqref{eq:mma-f} and $M_{m,m}=(\theta_m)^2$, we immediately have
\begin{align}
\theta_m^2=e^{i2\pi q_m} = \eta_{3f}.
\end{align}
Nevertheless, we always have 
\begin{align}
\theta_m=e^{i\pi q_m}
\end{align}
which is not affected by the ambiguity. It follows from \eqref{eq:V-prop} that $\theta_m = e^{i\pi \sigma_H} = e^{i\pi q_m}$. 


Next, we consider mirror properties of $m$. Following the same line of calculation as in \eqref{eq:m-permute}, one can show that $[m]$ is invariant under $\rho_m^\mathcal{C}$ in AI and AII classes, and $[m]$ is mapped to $[\bar{m}]$ in AIII class. Moreover, for AIII class, further considering \eqref{eq:qf-prop}, we obtain
\begin{align}
\rho_m^\mathcal{C}(m) = \left\{
\begin{array}{ll}
\bar{m}, & \text{if $\sigma_H$ is integer }\\
\bar{m}f, & \text{if $\sigma_H$ is half-integer }
\end{array}
\right.
\end{align} 
For either cases, we can further define $\mu_m$. However, similarly to $\theta_m$ and $q_m$, $\mu_m$ is ambiguous due to possible stacking of 2D IQH states on the surface, which effectively changes it to $mf$  such that $\mu_m$ changes the sign. Nevertheless, one can show that the product of $\mu_m$ and $\theta_m$ is determined by the anomaly indicator $\eta_{1f}$:
\begin{align}
\mu_m \theta_m = \eta_{1f}^2. 
\label{eq:mu-eta}
\end{align}
To see that, we first notice from Eqs.~\eqref{eq:prop-f1} that $\eta_{1f}^2 = e^{i\pi \sigma_H^{\rm mp}/2}M_{w,x}^2$. With the relation that $\sigma_H^{\rm mp} = 2 \sigma_H$ and $\theta_m = e^{i\pi \sigma_H}$, we have $\eta_{1f}^2 = \theta_m M_{w,x}^2$. Furthermore, one can show that $M_{w,x}^2=\mu_m$. The proof, which we do not discuss in detail here, is similar to the discussion in the bosonic case in Appendix \ref{eta4_indc2} and one needs to use the results from anyon condensation theory in Ref.~\onlinecite{Mao2020}. Combining all, we derive \eqref{eq:mu-eta}.

\begin{table*}
	\caption{Topological and symmetry data of T-Pfaffian$_{\pm}$. The empty entries of $\mu_a$ correspond to the fact that these anyons are not invariant under $\bar\rho_m^\mathcal{C}$. } \label{tab:Pfaffian}
	\begin{centering}
	\begin{tabular}{c|cccccccccccc}
		\hline 
		\hline 
		& $I_{0}$ & $I_{2}$ & $I_{4}$ & $I_{6}$ & $\psi_{0}$ & $\psi_{2}$ & $\psi_{4}$ & $\psi_{6}$ & $\sigma_{1}$ & $\sigma_{3}$ & $\sigma_{5}$ & $\sigma_{7}$\tabularnewline
		\hline 
		$\theta_a$ & $1$ & $i$ & $1$ & $i$ & $-1$ & $-i$ & $-1$ & $-i$ & $1$ & $-1$ & $-1$ & $1$\tabularnewline
		\hline 
		$q_a$ & $0$ & $1/2$ & $1$ & $3/2$ & $0$ & $1/2$ & $1$ & $3/2$ & $1/4$ & $3/4$ & $5/4$ & $7/4$\tabularnewline
		\hline 
		$\mu_a\left(\text{T-Pfaffian}_{+}\right)$ & $1$ &  & $-1$ &  & $1$ &   & $-1$ &   & $1$ & $-1$ & $-1$ & $1$\tabularnewline
		\hline 
		$\mu_a\left(\text{T-Pfaffian}_{-}\right)$ & $1$ &  & $-1$ &  & $1$ &  & $-1$ &  & $-1$ & $1$ & $1$ & $-1$\tabularnewline
		\hline 
		\end{tabular}
\par\end{centering}	
\end{table*}

\subsection{Examples}

We consider the mirror version of the T-Pfaffian$_\pm$ topological order as examples, which are first proposed in Refs.~\onlinecite{chen14a,Bonderson13d}.  These examples are already illustrated in Ref.~\cite{LapaPRB2019}.   The symmetry group is $U(1)\rtimes \mathbb{Z}_2^\mathcal{M}$. 
 The topological and symmetry data of the ``M-Pfaffian$_\pm$'' states are as shown in Tab.~\ref{tab:Pfaffian}. The $S$ matrix of the T-Pfaffian topological order can be found in Ref.~\cite{Bonderson13d}. We calculate the indicators by using the expressions in Eq.~\eqref{eta_f} and obtain the following: 
\begin{center}
	\begin{tabular}{c|ccc}
		\hline 
		& \ $\eta_{1f}$ \ & \ $\eta_{2f}$  \ & \ $\eta_{3f}$\ \tabularnewline
		\hline 
		\ T-Pfaffian$_{+}$ \ & 1 & 1 & $-1$\tabularnewline
	  \ T-Pfaffian$_{-}$\  & $-1$ & 1 &$ -1$\tabularnewline
		\hline 
	\end{tabular}
\par\end{center}
For the same symmetry group, we also consider fermionic SETs, ``eMmM+f", ``eCmC+f", ``eFmF+f",  which are obtained by appending the trivial electron ``f" to the ``eMmM", ``eCmC" and ``eFmF" bosonic SETs in Table \ref{tab:toric-code}. Using expressions in \eqref{eta_f}, we easily obtain
\begin{center}
	\begin{tabular}{c|ccc} 
		\hline 
		&\  $\eta_{1f}$ \  & \ $\eta_{2f}$\  &  \ $\eta_{3f}$ \ \tabularnewline
		\hline 
		\ eCmC+f\  & 1 & $-1$ & $1$\tabularnewline
		\ eMmM+f\  & $-1$ & 1 & $1$\tabularnewline
		\ eFmF+f\  & $-1$ & $-1$ & 1\tabularnewline
		\hline
	\end{tabular}
\par\end{center}
More examples can be found in Refs.~\onlinecite{MetlitskiPRB2015,Bonderson13d,wangc13b,chen14a,wangc14} and \onlinecite{LapaPRB2019} and we do not bother to list more.

\section{Bosonic systems with $G=SU(2)$ or $SO(3)$}
\label{sec:SU2-SO3-b}
We now move on to $SU(2)$ and $SO(3)$. We study bosonic systems in this section, and study fermionic systems in the next section. For bosonic systems, we will consider two cases, $SU(2)\times \Z_2^\mathcal{M}$ and $SO(3)\times \Z_2^\mathcal{M}$. We begin with definitions of anomaly indicators in the two cases, and then derive bulk-boundary relations and explicit expressions of anomaly indicators in Secs.~\ref{sec:eta-SU2-b} and \ref{sec:eta-SO3-b}, respectively.

\subsection{Defining indicators}
\label{sec:define-eta-su2-b}

We first discuss the physics on the mirror plane. The classification is summarized in Table \ref{tab:2d-plane}.

\subsubsection{$SU(2)\times \Z_2$}
\label{sec:SU2-def}

Let us first study the classification of 2D SRE states on the mirror plane. Our strategy is to find invertible topological orders, SPT states protected by $\mathbf{x}$ alone, SPT states protected by $SU(2)$ alone, and SPT states protected jointly by $SU(2)$ and $\bfx$.  We show that the SRE states with $SU(2)\times\Z_2$ internal symmetry are classified by $\Z^2 \times \Z_2$.  The three root states that are similar to those in Sec.~\ref{sec:define-eta-u1-b}:

(i) The root state associated with the first $\Z$ in the classification is the $E_8$ state. The full symmetry group $SU(2)\times\Z_2$ acts trivially on the state.

(ii) The root state associated with the $\Z_2$ classification is the non-trivial SPT state protected by $\bfx$ only. Again, we denote $\mathcal{M}$ as $\bfx$ on the mirror plane. The $SU(2)$ symmetry acts trivially on this state.

(iii) The root state associated with the second $\Z$ is the ``IQH state'' protected by $SU(2)$. The mirror symmetry $\mathcal{M}$ acts trivially on this state. These IQH states were discussed in Ref.~\onlinecite{LiuPRL2013_SU2SO3SPT}. One may understand it by considering the $U(1)$ subgroup
\begin{equation}
U_{SU(2)}(1) = \left\{e^{i\alpha \hat S^z}| \alpha\in [0,4\pi)\right\}
\end{equation}
where $\hat S^z = \sigma^z/2$ is the $z$-component of spin. Since the angle $\alpha$ has a periodicity $4\pi$, the unit charge associated with $U_{SU(2)}(1)$ is ``$1/2$''. At the same time, the Hall conductance $\sigma_H$ is an integer, instead of even integer like in the regular $U(1)$ case where the angle periodicity is $2\pi$. It can be seen from the fact that $\sigma_H$ is equal to the amount of charge accumulated after inserting a flux quantum. In bosonic IQH states, the accumulated charge is twice of the unit charge, which is an integer in the current convention.

We remark that there is no SPT state protected jointly by $SU(2)$ and $\bfx$. It can be understood as follows. We first gauge  $\Z_2$ and obtain an $SU(2)$-enriched $\Z_2$ gauge theory. There are four anyons in $\Z_2$ gauge theories $1, e, x, xe$, where $e$ denotes the $\Z_2$ charge and $x$ is a $\Z_2$ vortex. If there is any nontrivial SPT phase protected mutually by $SU(2)$ and $\Z_2$, the vortex $x$ should carry a fractional quantum number of $SU(2)$. However, as we will discuss below in Sec.~\ref{sec:eta-SU2-b}, no nontrivial symmetry fractionalization exists for $SU(2)$ symmetry. Accordingly, there is no mutual SPT state.

Let $\mu=(\mu_1, \mu_2, \mu_3)$ be an integer vector indexing a general state in the $\Z^2\times\Z_2$ classification. It consists of $\mu_i$ copies of the $i$-th root state. Similarly to the discussions in Sec.~\ref{sec:define-eta-u1}, this state can be characterized by the chiral central charge $c^{\rm mp}$, Hall conductance $\sigma_H^{\rm mp}$ and braiding statistics of $w$-, $x$- and $y$-vortices of the  $\Z_2^w\times \Z_2^x$, where $\Z_2^{w} = \{1, e^{i2\pi S^z}\}$ is a subgroup of $SU(2)$. These quantities are given by
\begin{subequations}
\begin{align}
c^{\rm mp} & = 8\mu_1  \\
\sigma_H^{\rm mp} & = \mu_3 \\
\theta_{x}^2 & = (-1)^{\mu_2} \\
\theta_{w}^2 & = (-1)^{\mu_3} \\
M_{w,x}^2 & = 1 \label{eq:su2-mu5}\\
\theta_{y}^2 & = (-1)^{\mu_2+\mu_3}
\end{align}
\end{subequations}
where \eqref{eq:su2-mu5} follows from the fact that there is no nontrivial mutual SPT state. All the quantities are very similar to those in Sec.~\ref{sec:define-eta-u1}, except $\sigma_H^{\rm mp}$ due to a different convention of the angle periodicity. 

Under adjoining operations, the $\Z^2$ classification associated with $E_8$ and IQH states reduces to $\Z_2^2$, similarly as before. Therefore, the final classification of 3D systems with $SU(2)\times\Z_2^\calM$ is $\Z_2^3$. Similarly to the indicators in \eqref{eq:tilde_eta}, we define
\begin{subequations}
\label{eq:teta-su2}
\begin{align}
\teta_1 & = e^{i\pi c^{\rm mp}/8} \label{eq:su2-teta1}\\
\teta_2 & = \theta_x^2 \label{eq:su2-teta2}\\
\teta_3 & = e^{i\pi \sigma_H^{\rm mp}}\label{eq:su2-teta3}\\
\teta_4 & = M_{w,x}^2 \\
\teta_5 & = \theta_y^2 
\end{align}
\end{subequations}
All the quantities are invariant under adjoining operations. We remark that $\teta_4=1$ due to $\eqref{eq:su2-mu5}$, and $\teta_5=\teta_2\teta_3$. Also, we note that the definition \eqref{eq:su2-teta3} differs from that in \eqref{eq:teta-b3} by a factor of 2 in the exponent. Regardless of this difference, we use the same symbol ``$\teta_3$'' to denote the indicator.

\subsubsection{$SO(3)\times \Z_2$}
\label{sec:SO3-def}

Under dimensional reduction, 3D crystalline insulators with $SO(3)\times \Z_2^\calM$ symmetry reduce to 2D SRE states with $SO(3)\times\Z_2$ internal symmetry. We show that the latter are classified by $\Z^2\times \Z_2^2$. Our classification strategy is the same as above. The four root states are:

(i) The root state associated with the first $\Z$ in the classification is the $E_8$ state. The full symmetry group $SO(3)\times\Z_2$ acts trivially on the state.

(ii) The root state associated with the $\Z_2$ classification is the non-trivial SPT state protected by $\bfx$ only. The $SO(3)$ symmetry acts trivially on this state.

(iii) The root state associated with the second $\Z$ classification is the $\nu=1$ bosonic IQHE protected by $SO(3)$ symmetry. The symmetry $\bfx$ acts trivially on this state. Similarly to the $SU(2)$ case, this state can be understood by considering the $U(1)$ subgroup
\begin{equation}
U_{SO(3)}(1) = \left\{e^{i\alpha \hat S^z}| \alpha\in [0,2\pi)\right\}
\end{equation}
where $\hat S^z$ is the $z$-component spin operator associated with \emph{integer} spins. We emphasize that angle periodicity is $2\pi$. Accordingly, the unit charge is ``$1$'' and the conductance quantum is $\sigma_H=2$.

(iv) The root state associated with the second $\Z_2$ classification is an SPT state protected jointly by $SO(3)$ and $\bfx$. One way to understand the state is to gauge $\Z_2$ and obtain a $SO(3)$-enriched $\Z_2$ gauge theory. The root SPT state is characterized by the fact that the $x$-vortex carries a half-integer spin --- a fractional quantum number of $SO(3)$ symmetry (see Sec.~\ref{sec:eta-SO3-b} below). Alternatively, we may gauge $\Z_2^2\times\Z_2^x$, where $\Z_2^w = \{1, e^{i \pi S^z}\}$. The root state is characterized by $M_{w,x}^2=-1$. 

These SRE states have a one-to-one correspondence to those in the $U(1)\times\Z_2$ case in Sec.~\ref{sec:U(1)xZ2-b}. One just simply identify $U(1)$ in Sec.~\ref{sec:U(1)xZ2-b} with  in $U_{SO(3)}(1)$ in this section. Then, all properties follow from our discussions in Sec.~\ref{sec:U(1)xZ2-b}, including: (a) the chiral central charge $c^{\rm mp}$, Hall conductance $\sigma_H^{\rm mp}$, and properties of $w$-, $x$-, and $y$-vortices are given by Eq.~\eqref{eq:u1xz2-b-observables}; (b) under adjoining operations, the classification reduces to $\Z_2^4$; and (c) the anomaly indicators $\teta_1, \teta_2,\teta_3,\teta_4$  and $ \teta_5$ are defined by Eq.~\eqref{eq:tilde_eta}.

\subsection{Indicators for $SU(2)\times \Z_2^\mathcal{M}$}
\label{sec:eta-SU2-b}

We now turn to expressing the anomaly indicators in \eqref{eq:teta-su2} through surface SET quantities. We need to introduce quantities that characterize SET phases with $SU(2)\times \Z_2^\calM$. Mirror symmetry properties are characterized by the permutation $\rho_m^\mathcal{C}$ and mirror fractionalization $\mu_a$, as introduced in Sec.~\ref{sec:STO-b}. 

Accordingly, we only need to study $SU(2)$ symmetry properties. As all group elements are connected to the identity in $SU(2)$, no permutation on anyons is allowed. Moreover, the second cohomology group $H^2(SU(2),A)$, which characterizes $SU(2)$ fractionalization on anyons, is trivial for any finite Abelian group $A$\cite{EssinPRB2013,BarkeshliPRB2019}.  Accordingly, there is no symmetry fractionalization data either. That is, given a topological order, there is only one kind of symmetry enrichment by $SU(2)$.  While anyons may carry integer or half-integer spins, both are considered as ``integer quantum numbers'' of $SU(2)$. In addition, there is no distinct phase enriched mutually by $\mathcal{M}$ and $SU(2)$. A consequence is that no surface SET supports a 3D bulk SPT that needs protection from $SU(2)$ symmetry. In the current context,  the root state (iii) in Sec.~\ref{sec:SU2-def} cannot support surface SETs. Accordingly, the only option is gapless surface states.  This point has already been emphasized previously in Refs.~\cite{wangc14, CordovaArXiv2019}. 

 In terms of anomaly indicators, the above discussion means that $\teta_3 = 1$ for any $SU(2)\times\Z_2^\calM$ surface SETs. Accordingly, while $\teta_3=-1$ is valid in the bulk, it can never be achieved by evaluating $\teta_3 = \eta_1\eta_3$ though the SET expressions in Eqs.~\eqref{eq:eta_b}. Indeed, with all $q_a$'s associated with $U_{SU(2)}(1)$ being 0, we always have $\eta_3=\eta_1$. Similarly, we have $\eta_4=\eta_2$. The remaining indicators $\teta_1$ and $\teta_2$ only involve the mirror symmetry, so are the same as before, which are discussed in Sec.~\ref{sec:teta12}.

\subsection{Indicators for $SO(3)\times \Z_2^\mathcal{M}$}
\label{sec:eta-SO3-b}

Next, we consider anomaly indicators for $SO(3)\times\Z_2^\calM$ SET states. Again, we only need to study symmetry enrichment by $SO(3)$. Like $U(1)$ and $SU(2)$, anyon permutation is not allowed by $SO(3)$. Symmetry fractionalization is characterized by a ``fractional spin'' $s_a$ of anyon $a\in\mathcal{C}$ under $SO(3)$ action. We understand that $SO(3)$ supports either integer or half-integer spins. The latter are ``fractional spins'', as they are projective representations of $SO(3)$. We take the convention that $s_a=0$ or $1/2$.  Like the fractional charge $q_a$, the spins $\{s_a\}$ satisfy
\begin{align}
s_a+s_b = s_{c} \modulo{1}
\end{align}  
when $N_{ab}^c \neq 0$. The fractional spin is actually fully characterized by its $z$ component $s_a^z$, which is understood as the fractional charge under $U_{SO(3)}(1)$. Since we have the relation
\begin{equation}
s_a^z = s_a \modulo{1},
\label{eq:ssz-relation}
\end{equation} 
the two quantities are equivalent in terms of characterizing fractionalization in the SET phase.

With the above understanding, we now express the indicators in terms of SET quantities. The indicators $\teta_1$ and $\teta_2$ only involve mirror symmetry properties and so are the same as those in Sec.~\ref{sec:teta12}. By interpreting $U(1)$ in Sec.~\ref{sec:boson-U(1)} as $U_{SO(3)}(1)$, making the substitution $q_a\rightarrow s_a^z$ in \eqref{eq:eta3-repreat} and \eqref{eq:eta4-repeat}, and using the relation \eqref{eq:ssz-relation}, we immediately obtain the following expressions
\begin{align}
\eta_3 &= \frac{1}{D} \sum_{a\in\mathcal{C}} d^2_a \theta_a e^{i2\pi s_a} \\
\eta_4 & =\frac{1}{D} \sum_{a\in\mathcal{C}} d_a \theta_a \mu_a e^{i2\pi s_a}
\label{eq:eta4-SO3-b}
\end{align}
Also, we have $\teta_3 = \eta_1\eta_3$, $\teta_4 = \eta_1\eta_2\eta_3\eta_4$ and $\teta_5 = \eta_4$, the same as before. Note that $s_a$ is the spin of an anyon under the action of $SO(3)$, which should be distinguished from the topological spin $\theta_a$. The latter is often denoted as $\theta_a = e^{i2\pi s_a}$, so readers should not be confused with the two kinds of spins.

\section{Fermionic systems with $G=SU_f(2)$}
\label{sec:SU2-SO3-f}

In this section, we study 3D fermionic systems in CI and CII classes. In interacting systems, the two classes correspond to symmetry groups $SU_f(2)\times \Z_2^\mathcal{M}$ and $[SU_f(2)\times \Z_4^{f\calM}]/\Z_2$, respectively\cite{FreedarXiv2016}.  We begin with defining the indicators, and then derive the bulk-boundary correspondence and explicit expressions of anomaly indicators in Sec.\ref{sec:eta-CI} and \ref{sec:eta-CII}, respectively.

\subsection{Defining indicators}
\label{sec:define-eta-SU2-f}

We discuss the physics on the mirror plane first. The classification is summarized in Table \ref{tab:2d-plane}.
\subsubsection{$SU_f(2)\times \Z_2$}
\label{sec:su2timesz2m_f}

Let us first study the 2D SRE states on the mirror plane. We claim that the SRE states with $SU_f(2)\times \Z_2$ internal symmetry are classified by $\Z^2\times \Z_2$. The three root states are similar to those in Sec.~\ref{sec:u1xz2-f}:

(i) The  root state of the first $\Z$ classification is the $E_8$ state. The full symmetry group $SU_f(2)\times \Z_2$ acts trivially on this state. It is characterized by a chiral central charge $c=8$ and Hall conductance $\sigma_{H}=0$.

(ii) The root state of the second $\Z$ classification is the ``IQH state'' at a filling factor $\nu=2$. One way to understand these IQH states\footnote{Strictly speaking, these states should be called ``spin quantum Hall states''\cite{LiuPRL2013_SU2SO3SPT,Wen2017RMP_Zoo}, because $U_{SU_f(2)}(1)$ corresponds to spin conservation rather than charge conservation. However, in this work, we refer to all the states associated with $U(1)$ symmetry as IQH states, regardless of the microscopic origin of the $U(1)$ symmetry. Also,  the $U(1)$ Chern number is referred to as the ``Hall conductance'', regardless if it is associated with spin or charge.}  is to use the $U(1)$ subgroup
\begin{equation}
U_{SU_f(2)}(1) = \left\{e^{i\alpha \hat S^z}| \alpha\in [0,4\pi)\right\}
\end{equation}
where $\hat S^z$ is the $z$-component of spin. The special element $e^{i 2\pi \hat S^z} = -1$ is the fermion parity $P_f$.  The full $SU_f(2)$ group imposes the condition that the root state must have a filling factor $\nu=2$. Intuitively, the root state can be viewed as an $SU_f(2)$-symmetric spinful IQH state with the spin-up and spin-down fermions individually form a $\nu=1$ IQH state.  The rigorous reason is similar to the AI class. In fact, the symmetry group $[U_f(1)\rtimes \Z_4^f]/\Z_2$ of the AI class (after dimensional reduction) is a subgroup of $SU_f(2)$, with ``$U_f(1)$'' being identified with $U_{SU_f(2)}(1)$  and  ``$\Z_4^f$'' being generated by $e^{i\pi\hat S^x}$. Since AI class forbids  the $\nu=1$ state, so is $SU_f(2)$. The $\Z_2^x$ symmetry acts trivially this root state. Similarly to bosonic $SU(2)\times \Z_2^\mathcal{M}$ case in Sec.\ref{sec:SU2-def}, the unit charge is $1/2$ due to the $4\pi$ angle periodicity. Accordingly, the Hall conductance $\sigma_H=1/2$ for the state at filling factor $\nu=1$. Then, the root state is characterized by $\sigma_{H}=1$. The chiral central charge of this state is $c=2$.

(iii) The root state of the $\Z_2$ classification is an non-chiral SPT state protected by $\Z_2^x$ only (the full symmetry is $\Z_2^f\times \Z_2^x$). Fermionic SPT states with internal $\Z_2^f\times \Z_2^x$ symmetry are classified by $\Z_8$\cite{GuLevin2014}, and we use $\nu=0,1,...,7$ to label these eight states.  We show that, to be compatible with $SU_f(2)$, the root state must be the $\nu=4$ state, which then leads to the $\Z_2$ classification. To see that, we gauge the $\Z_2^x$ symmetry, giving rise to an $SU_f(2)$-symmetric  $\Z_2^x$ gauge theory. The  $x$-vortex may carry spin $0$ or $1/2$ under $SU_f(2)$, both of which correspond to ``integer'' charges of $U_{SU_f(2)}(1)$. Therefore, if we gauge $\Z_2^f\times \Z_2^x$, the mutual statistics $M_{w,x}^2$ between $w$- and $x$-vortices must be 1. According to Ref.~\onlinecite{WangPRB2017},  only the $\nu=0$ and 4 states have $M_{w,x}^2=1$. Therefore, the root state is the $\nu=4$ SPT state. This argument also shows that there is no SPT state protected jointly by $SU_f(2)$ and $\Z_2$ in fermionic system.

\begin{subequations}
\label{eq:su2timesz2-prop}
A general SRE state can be indexed by an integer vector $\mu = (\mu_1, \mu_2, \mu_3)$, with $\mu_3$ defined only modulo $2$. It consists of $\mu_i$ copies of the $i$th root state. The chiral central charge and (spin) Hall conductance are
\begin{align}
c^{\rm mp} & = 8\mu_1 + 2\mu_2 \label{eq:su2timesz2-prop-a}
 \\
\sigma^{\rm mp}_{H} & = \mu_2 \label{eq:su2timesz2-prop-b}
\end{align}
Similarly to Sec.~\ref{sec:u1xz2-f}, we can capture many topological properties by gauging $\Z_2^f\times \Z_2^x$ subgroup.  We still use $w$-, $x$-, and $y$- vortices to denote the $\Z_2^f$, $\Z_2^x$ vortices and their composite, respectively. According to Refs.~\onlinecite{WangPRB2017}, all these vortices are Abelian anyons. Similarly to Eqs.~\eqref{eq:prop-f1c}-\eqref{eq:prop-f1f}, we have 
\begin{align}
\theta_x^2 & = (-1)^{\mu_3} \\
\theta_w & = e^{i\pi\mu_2/2 } \\
M_{w,x}^2 & = 1\\
\theta_y^2 & = (-1)^{\mu_2-\mu_3}  
\end{align}
We remark that all quantities here are squares of those in in Eqs.~\eqref{eq:prop-f1c}-\eqref{eq:prop-f1f}.

\end{subequations}

Next we consider the adjoining operations, under which the classification reduces to $\Z_2\times \Z_4$. First, under adjoining, the $\Z$ classification associated with the $E_8$ state reduces to $\Z_2$. The argument is similar as before. Secondly, the $\Z$ classification associated with the IQH states reduces to $\Z_2$.  However, this reduced $\Z_2$ will extend the $\Z_2$ classification associated with the root state (iii), such that they together form the $\Z_4$ classification. The reason behind this extension is the same as that for $U_f(1)\times \Z_2^\mathcal{M}$ in Sec.~\ref{sec:u1xz2-f}. Therefore, the classification of 3D SPT states with $SU_f(2)\times\Z_2^\mathcal{M}$ is $\Z_2\times\Z_4$, in agreement with Ref.~\onlinecite{FreedarXiv2016} under crystalline equivalence principles.

We now define a set of topological invariants that are invariant under adjoining operations, which will serve as anomaly indicators. First, we define the two invariants $\eta_{1f}$ and $\eta_{2f}$ as follows,
\begin{align}
\eta_{1f} & = e^{i\pi\sigma_{H}^{\rm mp}/2}(\theta_x^*)^2 
\label{eq:suf2teta-f1}\\
\eta_{2f}  & = e^{i\pi(c^{\rm mp}-2\sigma_{H}^{\rm mp})/8}. \label{eq:suf2teta-f2} 
\end{align}
Compared to Eqs.~\eqref{eq:u1teta-f1} and \eqref{eq:u1teta-f2}, the difference is a factor of 2 ahead of $\sigma_H^{\rm mp}$ due to different angle periodicities between $U_{SU_f(2)}(1)$ and $U_f(1)$ in AIII class. However, the underlying physics does not change, so our notations remain the same. The alternative expression (\ref{eq:u1teta-f1-alt}) of $\eta_{1f}$ is still valid, as $\theta_w=e^{i\pi \sigma_H^{\rm mp}/2}$ in the current case. With these definitions and Eqs.~\eqref{eq:su2timesz2-prop}, we have
\begin{equation}
\eta_{1f} = e^{i\pi(\mu_2-2 \mu_3)/2}, \quad \eta_{2f} = e^{i\pi\mu_1}.
\label{eq:eta1f-values-def}
\end{equation}
We see that $\eta_{1f}$ can take $1,e^{i\pi/2}, e^{i\pi}, e^{i3\pi/2}$, indicating the $\Z_4$ classification. Also, $\eta_{2f}=\pm 1$, indicating the $\Z_2$ classification. We can also define the anomaly indicator $\eta_{3f}$, similarly to (\ref{eq:u1teta-f3}). However, it is easy to see that $\eta_{3f}=1$ in the current case.

\subsubsection{$[SU_f(2)\times \Z_4^{f}]/\Z_2$}
\label{sec: fsu2M2Pf_def_indicators}

Now we discuss 3D SPT phases in CII class, whose symmetry group is $[SU_f(2)\times \Z_4^{f\mathcal{M}}]/\Z_2$. They reduce to 2D SRE states in the mirror plane with internal symmetry $[SU_f(2)\times \Z_4^f]/\Z_2$.

We show that  strictly 2D SRE states are classified by $\Z^2\times \Z_2$. The two root states for $\Z^2$ classification are the same as those of $SU_f(2)\times \Z_2$, namely the $E_8$ state and the IQH state with filling factor $\nu=2$. According to Ref.~\onlinecite{WangPRB2016}, there is no SPT state protected by $\Z_4^f$ alone. However, we claim that there exists and only exists a non-trivial SPT state protected jointly by $SU_f(2)$ and $\Z_4^f$. It gives rise to a $\Z_2$ classification. 

To see our claim, we gauge the $\Z_4^f$ group and there remains an $SO(3)$ symmetry.  For the moment, we consider SRE states with $c^{\rm mp}=\sigma_H^{\rm mp}=0$, i.e., states with no components from root states (i) and (ii). Then, the gauged $\Z_4^f$ is unique, with $16$ anyons in total, labelled by $x^if^j$, with $i,j=0,1,2,3$. Here, $f$ is the original fermion, $x$ is a unit $\Z_4^f$ vortex. We take a convention that $x$ is the bosonic vortex, and the other unit vortices have $\theta_{xf}=-i$, $\theta_{xf^2}=-1$ and $\theta_{xf^3}=i$. Then, the question of classifying the original SRE states becomes classifying $SO(3)$-enriched $\Z_4^f$ gauge theories. According to our discussion in Sec.~\ref{sec:eta-SO3-b}, $SO(3)$ SET phases are distinguished by spins $s_f$ and $s_{x}$. Because of the fact that $SO(3)$ is extended to $SU_f(2)$ by the fermion parity $\Z_2^f$, we have $s_f=1/2$. Accordingly, there are only two SET phases, distinguished by $s_{x}=0$ or $1/2$. Back to the language of the original fermionic system, the two phases actually differ by the root state (iv) of $SO(3)\times\Z_2$ symmetry (see Sec.~\ref{sec:SO3-def}). The latter can be thought of as formed by strongly-bound fermion pairs, which are spin-singlet or triplet bosons. In the Hilbert space of the fermion pairs, the group $[SU_f(2)\times \Z_4^f]/\Z_2$ acts like $SO(3)\times \Z_2$. Moreover, the root state (iv) of $SO(3)\times\Z_2$ is characterized by the property that the bosonic $\Z_2$ vortex, corresponding to $x$ here, carries a spin $1/2$.  In light of this idea, one way to construct the third root state is to stack a trivial fermionic insulator the $SO(3)\times \Z_2$ root state (iv) formed by strongly-bound fermion pairs.  Stacking the trivial fermionic insulator does not trivialize the root state (iv) of $SO(3)\times \Z_2$ symmetry, which will be clear below after we define topological invariants.

\begin{subequations}
A generic SRE state can be indexed by an integer vector $\mu=(\mu_1,\mu_2,\mu_3)$, with $\mu_3$ defined only modulo 2. It consists of $\mu_i$ copies of the $i$th  root state. The chiral central charge and Hall conductance are again given by \eqref{eq:su2timesz2-prop-a} and \eqref{eq:su2timesz2-prop-b}. To further characterize the SRE states, we consider gauging the subgroup $(\Z_4^f\times\Z_4^f)/\Z_2 \equiv \Z_4^f\times\Z_2$. Let $p$ be a unit vortex of $\Z_4^f\subset SU_f(2)$, and $x$ be a $\Z_4^f$ unit vortex originating from the mirror symmetry. Then, the mixed vortex $y$ is a $\Z_2$ vortex. Note that two $x$-vortices fuse into the fermion parity vortex $w$, so do two $p$-vortices. Classification and topological invariants of $\Z_4^f\times\Z_2$ SRE states have been studied in Ref.~\onlinecite{WangPRB2017}. It finds that all vortices are Abelian. Translating the results of Ref.~\onlinecite{WangPRB2017} into the current notation,  we have that the SRE states are characterized by
\begin{align}
\theta_x^4 & = e^{i\pi\mu_2/2} \\
\theta_p^4 & = e^{i\pi\mu_2/2}\\
\theta_y^2 & = (-1)^{\mu_3}
\end{align}
\end{subequations}
Since both $x$ and $p$ are now $\Z_4^f$ vortices, we need to take the forth power of their topological spins to make them unambiguous to charge attachments. The topological spin $\theta_w$ of a fermion parity vortex satisfy $\theta_w =\theta_x^4 = \theta_p^4$. The quantity $\theta_y^2$ can be alternatively written as
\begin{align}
\theta_y^2 = \theta_p^2\theta_x^2 M_{p,x}^2 
\end{align}
In this form, we can see that the the root state with $\mu=(0,0,1)$ can be obtained by stacking a trivial fermionic insulator with either the root state (ii), (iii) or (iv) of bosonic $SO(3)\times \Z_2$ SRE states in Sec.~\ref{sec:SO3-def}. The case of the root state (iv) is discussed above.

Next we consider the adjoining operations, under which  the classification reduces to $(\Z_2)^3$. Firstly, similarly to Sec.\ref{sec:su2timesz2m_f}, under adjoining, the $\Z$ classification associated to the $E_8$ state reduces to $\Z_2$, so the $\mu_1$ is only unambiguous modulo 2.  Secondly, the $\Z$ classification associated to the IQH states also reduces to $\Z_2$. However, different from that in Sec.\ref{sec:su2timesz2m_f}, this reduced $\Z_2$ does not extend the  $\Z_2$ classification related to the third root state, which can be easily understood by the fact that there is no nontrivial SPT protected mirror symmetry alone. Therefore, both $\mu_2$ and $\mu_3$ take values modulo 2.

Now we define the topological invariants that are invariant under adjoining operations. First of all, $\eta_{2f}$ defined in (\ref{eq:suf2teta-f2}) still applies here. It identifies the root state associated with the $E_8$ state.  Similarly to (\ref{eq:u1teta-f3}),  we  define  the indicator  
 \begin{align}
 {\eta}_{3f}=e^{i\pi \sigma_{H}^{\text{mp}}}=\theta_w^2 = \theta_x^8
 \end{align}
which identifies the root state associated with the IQH state. To distinguish the third root state, we define the indicator $\eta_{4f}$ 
\begin{align}
\eta_{4f} = \theta_{y}^2 = \theta_p^2\theta_x^2 M_{p,x}^2.
\end{align}
In terms of the integer vector $\mu$, they are given by
\begin{align}
\eta_{2f} & =(-1)^{\mu_1}\nonumber\\
{\eta}_{3f} & =(-1)^{\mu_2}\nonumber\\
\eta_{4f} & =(-1)^{\mu_3}
\end{align}
Two remarks are in order. First, the indicator $\eta_{4f}$ is the ``fermionic version" of the indicator $\teta_5$ for bosonic systems with $SO(3)\times\Z_2$. Indeed, $y$-vortices of $[SU_f(2)\times\Z_4^f]/\Z_2$ becomes those of $SO(3)\times\Z_2$ after quotient out the fermion parity $\Z_2^f$. Second, the definition $\eta_{1f}$ in \eqref{eq:suf2teta-f1} -- which is used to identify SPT phases protected by $\mathcal{M}$ alone --- does not work for the current symmetry group. Instead, one may define $\teta_{1f} = e^{i\pi \sigma_H^{\rm mp}/2} (\theta_{x}^*)^4$. However, we always have $\teta_{1f}=1$ as there is no non-trivial SPT state protected by $\Z_{4}^{f\mathcal{M}}$ alone.

\subsection{Indicators for CI class: $SU_f(2)\times \Z_2^\mathcal{M}$}
\label{sec:eta-CI}

Now we express the anomaly indicators  in terms of surface SET quantities.  We first discuss quantities that characterize SET states with $SU_f(2)\times \Z_2^\mathcal{M}$.  Mirror symmetry properties are characterized by the mirror permutation $\rho_m^{\mathcal{C}}$ and mirror eigenvalues $\mu_a$, as introduced in Sec.~\ref{sec:u1f-set}.

Like in the bosonic case, there is no nontrivial anyon permutation for $SU_f(2)$ and every anyon $a\in\mathcal{C}$ carries a spin $s_a$. The spin $s_a$ is either $0$ or $1/2$, as we will take it modulo $1$. In particular, the local fermion $f$ has $s_f=1/2$ and the vacuum anyon has $s_{\mathbbm{1}}=0$. The $SU_f(2)$ spins respect the fusion structure,  
\begin{align}
s_{ a}+s_{ b}=s_{c} \modulo{1}
\label{eq:spin-relation-total}
\end{align}
for any $a, b, c\in\mathcal{C}$ with $N_{ a  b}^{ c}\neq 0$. Accordingly, we always have $s_{af}=s_a+1/2 \modulo{1}$ and $s_{\bar a}=-s_a\modulo{1}$. The $z$-component spin $s_a^z$ associated with $U_{SU_f(2)}(1)$ is in analogy to $q_a$ of $U_f(1)$ in Sec.~\ref{sec:u1f-set}. However, $q_a$ is define modulo 2, while $s_a^z$ is defined modulo 1 due to the $4\pi$ periodicity of the $U_{SU_f(2)}(1)$ angle.  The $z$-component $s_a^z$ and the total spin $s_a$ always satisfy
\begin{align}
s_a^z=s_a (\text{mod }1).
\label{eq:spin-relation-f}
\end{align} 
Similarly to Eq.~\eqref{eq:qf-prop}, under mirror permutation, the spins should satisfy
\begin{align}
s_{\rho_m^c(a)} = s_a,  \quad s_{\rho_m^c(a)}^z = s_a^z, \modulo{1}
\end{align}
An important consequence is that $\rho_m^\mathcal{C}(a)$ cannot be $\bar{a}f$. Indeed, if $\rho_m^\mathcal{C}(a)=\bar{a}f$, then $s_{a}= s_{\rho_m^\mathcal{C}(a)} = s_{\bar{a}f} =-s_a +1/2 \modulo{1}$,  which is impossible for $s_a=0$ or $1/2$. Remarkably, one can further show\cite{wangc14} that $\mathcal{C}$ can be decomposed into $\{1,f\}\boxtimes \mathcal{C}_b$, i.e., a stack of a bosonic topological order $\mathcal{C}_b$ and a trivial fermionic topological order  $\{1,f\}$. To see that, one pick the integer-spin anyon out of each pair, $a$ and $af$. These integer-spin anyons are closed under fusion, and therefore form a topological order, denoted as $\mathcal{C}_b$. Moreover, since $\rho_m^\mathcal{C}$ preserves $SU_f(2)$ spin modulo 1, $\mathcal{C}_b$ is also closed under the action of $\rho_m^\mathcal{C}$.

With the above understanding, we now express $\eta_{1f}$ and $\eta_{2f}$ in terms of surface SET quantities.  The indicator $\eta_{1f}$ only involves mirror symmetry properties, so the expression \eqref{eq:eta1f-repeat} in Sec.~\ref{sec:eta-1f} remains valid. We reproduce the expression here,
\begin{align}
\eta_{1f}= \frac{1}{\sqrt{2}D} \sum_{a\in\mathcal{C}} d_a\theta_a\mu_a
\label{eq:eta1f-su2f}
\end{align}
However, $\eta_{1f}$ can only take $\pm 1$ in this expression, in contrast to four possible values $\pm 1, \pm i$ through its definition [see Eq.~\eqref{eq:eta1f-values-def}]. The reason is due to the property that $\rho_m^\mathcal{C}(a)$ can only be $\bar{a}$ but not $\bar{a}f$, as discussed above. For $\rho_m^\mathcal{C}(a)=\bar{a}$, it is required that $\theta_a = \theta_{\rho_m^\mathcal{C}(a)}^* = \theta_{\bar{a}}^* = \theta_a^*$. Therefore, $\theta_{a}=\pm 1$. Accordingly, for $\mu_a=\pm 1$, we must have $\rho_m^{\mathcal{C}}(a)=\bar{a}$ and $\theta_a=\pm1$, i.e., all non-vanishing terms in \eqref{eq:eta1f-su2f} are real. This makes $\eta_{1f}$ a real number, and thereby $\pm 1$.  Similarly to the bosonic case in Sec.\ref{sec:eta-SU2-b}, it implies that if $\eta_{1f}=\pm i$, a symmetric surface must be gapless, i.e., the so-called symmetry-enforced gaplessness.\cite{wangc14, CordovaArXiv2019}

The indicator $\eta_{2f}$ can be obtained from \eqref{eq:eta2f-repeat} by interpreting $U_f(1)$ there as $U_{SU_f(2)}(1)$, and substituting $q_a$ with $2s_a^z$.  Further with the relation (\ref{eq:spin-relation-f}), we have
\begin{align}
\eta_{2f}=\frac{1}{\sqrt{2}D} \sum_a d_a^2 \theta_a e^{i2\pi s_a}.
\label{eq:eta2f-su2-f}
\end{align}
We note that the substitution of $q_a$ by $2s_a^z$, instead of $s_a^z$, is due to the fact that the angle period of $U_{SU_f(2)}(1)$ is $4\pi$ instead of $2\pi$. 

\subsection{Indicators for CII class: $[SU_f(2)\times \Z_4^{f\calM}]/\Z_2$}
\label{sec:eta-CII}
We now discuss $SU_f(2)$ and mirror properties of the anyons in CII class. The $SU_f(2)$ properties are the same as in CI class (Sec.~\ref{sec:eta-CI}). In particular, the fermionic topological order $\mathcal{C}$ can be decomposed into $\{\mathbbm{1}, f\}\boxtimes \mathcal{C}_b$, where ${\mathcal{C}}_b=\{\mathbbm{1}, a, b, c,...\}$ forms a modular tensor category, i.e., a bosonic topological order. In addition, all anyons in $\mathcal{C}_b$ carry integer spins.

Mirror properties again include mirror permutation and mirror eigenvalues. In CII class, $\mathcal{M}^2=P_f$, for which we have briefly discussed the mirror permutation $\rho_m^\mathcal{C}$ and mirror eigenvalues $\mu_a$ in Sec.~\ref{sec:AI-indicators} when we study AI class. Generally speaking, $\mu_a$ can be $\pm 1, \pm i$, as shown in Eq.~\eqref{eq:mua-AI}. However, in the presence of $SU_f(2)$ symmetry, the case $\rho_m^\mathcal{C}(a)=\bar a f$ does not exist, making $\mu_a=\pm i$ impossible. This is due to the fact that $\mathcal{M}$ cannot mix anyon in $\mathcal{C}_b$ and $f\mathcal{C}_b$, as discussed in Sec.~\ref{sec:eta-CI}. Different to the CI class,  the local fermion has $\mu_f=1$ in CII class. Accordingly, $\mu_{af} = \mu_{a}$ while $s_{af} = s_{a}+1/2 \modulo{1}$.

Let us now discuss the anomaly indicators. The indicator $\eta_{2f}$ is exactly the same as in CI class, and the explicit expression in terms of SET quantities is given by (\ref{eq:eta2f-su2-f}). For $\eta_{3f}$, we claim that it is enforced to be 1 if the surface is an SET state. This is another example of ``symmetry-enforced gaplessness'' if $\eta_{3f}=-1$ in the bulk. To see the claim, we follow the discussion in Sec.~\ref{sec:eta3f-u1f} and obtain $\sigma_H^{\rm mp}=2\sigma_H$, where $\sigma_H$ is the surface Hall conductance.  Therefore,
\begin{align}
{\eta}_{3f}=e^{i2\pi\sigma_H}
\label{eq:eta3f-su2f-ind}
\end{align}
Due the the decomposition $\mathcal{C}=\{1,f\}\boxtimes \mathcal{C}_b$, we think of the surface as a stack of a fermionic IQH state with $[SU_f(2)\times\Z_4^{fM}]/\Z_2$ symmetry and a bosonic fractional quantum Hall state with $SO(3)\times\Z^\mathcal{M}$ symmetry. The former must have an integer $\sigma_H$ due to $SU_f(2)$ symmetry, as argued in Sec.~\ref{sec:su2timesz2m_f}. The latter must also have an integer $\sigma_H$ due to the mirror symmetry, following the argument in Sec.~\ref{sec:teta3-derive}. Combining the two, we have the total $\sigma_H$ must be integer and so $\eta_{3f}=1$ if evaluated through \eqref{eq:eta3f-su2f-ind}. This fact also implies that the ``monopole'' anyon $m$, generated by adiabatically inserting a flux quantum of the $U_{SU_f(2)}(1)$ group, must carry an integer spin.  Another way to see the symmetry-enforced gaplessness is that the bulk SPT with $\eta_{3f}=-1$ is an intrinsically fermionic state, while the surface SET is simply a bosonic SET stacked with a trivial fermionic insulators. Moreover, the symmetries do not mix $f$ with anyons in $\mathcal{C}_b$ at all. So, the bosonic surface SET enforces $\eta_{3f}=1$.

Finally, we study the indicator $\eta_{4f}$ in terms of surface SET quantities. To proceed, we again make use of the decomposition $\mathcal{C}=\{1,f\}\boxtimes \mathcal{C}_b$. We then evaluate $\eta_{4f}$ for $\mathcal{C}_b$ and the trivial fermionic insulator separately, which we denote as $\eta_{4f}^{\mathcal{C}_b}$ and $\eta_{4f}^{\rm tri}$, respectively. For the bosonic topological order $\mathcal{C}_b$, the symmetry group acts like $SO(3)\times\Z_2^\mathcal{M}$. Then, the definition of $\eta_{4f}$ reduces precisely to $\teta_5$ of $SO(3)\times\Z_2^\mathcal{M}$. With the expression \eqref{eq:eta4-SO3-b}, we have  
\begin{align}
\eta_{4f}^{\mathcal{C}_b} = \frac{1}{D_{\mathcal{C}_b}} \sum_{a\in\mathcal{C}_b} d_a \theta_a \mu_a e^{i2\pi s_a}
\end{align}
Next, $\eta_{4f}^{\rm tri}$ for the trivial fermionic insulator can be directly evaluated. One way is to perform an edge theory analysis, similar to those in Appendix \ref{appd:adjoin-iqh}. One can show that $\eta_{4f}^{\rm tri}=1$. A simpler way is to fold the trivial fermionic insulator as in Fig.~\ref{fig:folding}(c). One can see that the folded trivial fermionic insulators is the same state as what we apply in the adjoining operations. Since the adjoined states must have trivial anomaly indicators, so $\eta_{4f}^{\rm tri}=1$. Taking the product of the two parts, we have
\begin{align}
\eta_{4f} & = \eta_{4f}^{\mathcal{C}_b} \times \eta_{4f}^{\rm tri}  \nonumber\\
& = \frac{1}{D_{\mathcal{C}_b}} \sum_{a\in\mathcal{C}_b} d_a \theta_a \mu_a e^{i2\pi s_a} \nonumber \\
& = \frac{1}{\sqrt{2}D} \sum_{a\in\mathcal{C}} d_a \theta_a \mu_a e^{i2\pi s_a} 
\end{align}  
where in the last line we have replaced the summation over anyons in $\mathcal{C}_b$ with a summation over anyons in $\mathcal{C}$. The replacement is valid because $\theta_{af}=-\theta_{a}$, $\mu_{af}=\mu_a$, $e^{i2\pi s_{af}} = -e^{i2\pi s_a}$, and $D= \sqrt{2}D_{\mathcal{C}_b}$, where $D$ is the total quantum dimension of $\mathcal{C}$.


\section{Discussions}
\label{sec:discussion}

In summary, we have established bulk-boundary correspondences for 3D bosonic and fermionic SPT phases that are protected by the mirror symmetry $\mathcal{M}$ and a Lie group symmetry $G$, where $G=U(1), SU(2)$ or $SO(3)$, through the folding approach. In particular, the mirror cases of Altland-Zirnbauer symmetry classes that support non-trivial 3D fermionic SPT phases are explored in the presence of strong interaction. For each symmetry group, we have defined a set of bulk topological invariants, a.k.a. anomaly indicators, and expressed them in terms of surface SET quantities. These expressions allow us to easily determine the type of anomaly for any SET state. The main results have been summarized in Sec.~\ref{sec:main-results}.

For future studies, there are a few directions to explore. First, it is interesting to explore general Lie group symmetry protected topological states.  In this work, we only explored the most common groups $U(1)$, $SU(2)$ and $SO(3)$.  Recent development on classification fermionic SPT phases has focused on finite groups\cite{KapustinThorngren,WangPRX2018, WangQRPRX2020_GeneralGf}. Classification of SPT phases in the presence of general Lie group symmetries has not be very clear, in particular in fermionic systems.  Second, it is also interesting to extend the folding approach to other spatial symmetries, such as rotation. The bulk state of rotation and other spatial SPT phases have been explored in a similar spirit to this work recently\cite{ChengArxiv2018_rotationSPT, ZhangPRB2020_pointgroup}. Extending these studies in presence of a surface would be interesting.

\emph{Note added.} Just before we are about to submit this work to arXiv, we become aware of the work Ref.~\onlinecite{Tata2021arXiv}, which generalizes Ref.~\onlinecite{BulmashPRR2020} to fermionic SET phases. While the topics overlap between this work and Ref.~\onlinecite{Tata2021arXiv}, the methods are very different.

\begin{acknowledgments}
We are grateful to Z.-C. Gu and Q.-R. Wang for very helpful discussions on the classification of fermionic SPT phases. This work was supported by Research Grant Council of Hong Kong (ECS 21301018) and URC, HKU (Grant No. 201906159002).
\end{acknowledgments}

\appendix
\section{Constraints on vortex braiding statistics}
\label{appd:semidirect}

In Sec.~\ref{sec:define-eta-u1}, we characterize 2D bosonic $U (1)\rtimes\mathbb{Z}_2$ SRE states by gauging its $\Z_2^w \times\Z_2^x$ subgroup and studying the braiding statistics of vortices. Similarly, we characterize 2D fermionic $U_f (1)\rtimes\mathbb{Z}_2$ SRE states by gauging its $\Z_2^f\times\Z_2^x$ subgroup. In this appendix, we show that the full symmetry group, $U (1)\rtimes\mathbb{Z}_2$ or $U_f(1)\rtimes\mathbb{Z}_2$, enforces constraints on the vortex braiding statistics. In particular, in both bosonic and fermionic cases, the following constraint must be satisfied:
\begin{align}
M_{w,x}^2\theta_w^2=1
\label{appA:eq1}
\end{align}
where $w$ is any vortex associated with $\Z_2^w$ (or $\Z_2^f$) and $x$ is any vortex associated with $\Z_2^x$. This constraint is absent for $U (1)\times\mathbb{Z}_2$ and $U_f(1)\times\mathbb{Z}_2$.

\subsection{Bosonic SRE states with $U(1)\rtimes\mathbb{Z}_2$}  
In general, vortices in a 2D theory obtained by gauging $\mathbb{Z}_2^w\times\mathbb{Z}_2^x$ SRE states can have eight different types of braiding statistics, described by the following eight choices:
\begin{equation}
\theta_{w}^2=\pm1,\ \ \theta_{x}^2=\pm1, \ \  M_{w,x}^2=\pm1 \nonumber
\end{equation}
where $w$ is a vortex associated with $\mathbb{Z}_2^w$, $x$ is a vortex associated with $\mathbb{Z}_2^x$, and all quantities are squared to remove ambiguities from charge attachments. All vortices are Abelian anyons. These eight types have a one-to-one correspondence to the eight 2D $\Z_2^w\times\Z_2^x$ SPT phases.\cite{chen2013, wangcj15} However, we show below that an enlarged $U(1)\rtimes \Z_2^x$ symmetry, with $\Z_2^w\subset U(1)$,  enforces the constraint \eqref{appA:eq1}. It means that, among the eight $\mathbb{Z}_2^w\times\mathbb{Z}_2^x$ SPT phases, only four can be lifted to  $U(1)\rtimes \Z_2^x$ SPT phases.

Consider a 2D $U(1)\rtimes \Z_2^x$ symmetric SRE state. To show Eq.~\eqref{appA:eq1}, we first turn this state into a 2D SET state by gauging $\Z_2^w$. Since $\Z_2^w$ is the center of $U(1)\rtimes\Z_2^x$, gauging it does not break any symmetry. The resulting $\Z_2^w$ gauge theory contains four anyons $\{1, e, w, we\}$, where $e$ is the charge and $w, we$ are the vortices. The topological order is either toric-code-like or double-semion-like, depending on if $\theta_w^2=1$ or $-1$. At the same time, we have $\theta_e=1$, $M_{e,w}=-1$, and $M_{w,w}=\theta_w^2$. To be more specific, let $w$ be the vortex obtained by \emph{adiabatically} inserting a $\pi$ flux of $U(1)$. Then, $w$ carries a $U(1)$ charge $Q_w=\sigma_H/2$ and $we$ carries a charge $Q_{we}=1+\sigma_H/2$.  The gauged theory has a remaining global symmetry $[U(1)\rtimes \Z_2^x]/\Z_2^w \equiv U'(1)\rtimes \Z_2^x$. If group elements of $U(1)$ are parametrized by $\mathcal{U}_\alpha$ with $0\le \alpha <2\pi$ and $\mathcal{U}_0=\mathcal{U}_{2\pi}$, then group elements of $U'(1)$ are parametrized by $\mathcal{U}_\alpha$ with $0\le \theta <\pi$ and $\mathcal{U}_0=\mathcal{U}_\pi$. The unit charge associated with $U'(1)$ is twice of that of $U(1)$, i.e., $2e$. Accordingly, anyons carries fractional charge of $U'(1)$: 
\begin{align}
q'_{e} = \frac{1}{2}, \quad q'_w=\frac{\sigma_H}{4},\quad q'_{we} = \frac{\sigma_H}{4} + \frac{1}{2} \label{appA:eq2}
\end{align}
which are measured in units of $2e$ (general SET properties are reviewed in Sec.~\ref{sec:STO-b}). In fact, for the purpose of obtaining Eq.~\eqref{appA:eq1}, it is enough to consider the subgroup $\Z_2^p\times\Z_2^x \subset U'(1)\rtimes \Z_2^x$, where
\begin{align}
\Z_2^p = \{1, \ \mathcal{U}_{\pi/2}\}.
\end{align}
In other words, we will study $\Z_2^p\times\Z_2^x$ enriched $\Z_2^w$ gauge theory. 

Properties of this SET state can be described by further gauging the $\Z_2^p\times\Z_2^x$ group. It is not hard to see that there is no anyon permutation by $\Z_2^p\times \Z_2^x$. Then, according to Ref.~\onlinecite{WangPRX2016}, the SET state is characterized by the following six quantities. The first four quantities are
\begin{align}
M_{e,p}^2 & = -1, \quad M_{e,x}^2=1 \nonumber\\
M_{w,p}^2 & = \theta_w^2, \quad M_{w,x}^2=\pm 1 \label{appA:eq3}
\end{align}
where $M_{\alpha,\beta}^2$ is the statistical phase obtained by braiding $\alpha$ round $\beta$ twice, with $\alpha=e,w$ and $\beta$ being a $p$- or $x$-vortex. Note that after gauging $\Z_2^p\times\Z_2^x$, the resulting topological order may be non-Abelian. Nevertheless, it is shown in Ref.~\onlinecite{WangPRX2016}, the quantity $M_{\alpha,\beta}^2$ is always an Abelian phase and it is independent of charge attachments. (Note that we have used different notations from those in Ref.~\onlinecite{WangPRX2016}.) The quantities $M_{e,p}^2$ and $M_{w,p}^2$ are determined by Aharonov-Bohm phases with the fractional charges in \eqref{appA:eq2} and the relation that $\theta_w^2 = e^{i\pi\sigma_H/2}$.  The anyon $e$ does not carries fractional charge of $\bfx$, so $M_{e,x}^2=1$. At the same time, $M_{w,x}^2 = \pm 1$ is not determined. In this discussion, we have implicitly used the fact that braiding statistics of the vortices are the same when we gauge symmetries in different orders. 

The other two quantities that describe the SET state are given by
\begin{align}
x_e = 0, \quad x_w = 1.
\label{appA:eq4}
\end{align}
The physical meanings of these two quantities are a bit indirect. Consider the anyon $e^{x_e}w^{x_w}$, with $x_e,x_m = 0,1$. If the mutual statistics between $e^{x_e}w^{x_w}$ and $a$ is $-1$, it signals that $a$ carries a projective representation of $\Z_2^p\times\Z_2^x$, which is manifested by a symmetry-protected local two-fold degeneracy on $a$\cite{LevinPRB2012}; otherwise, if the mutual statistics is 1, $a$ does not carry a protected two-fold degeneracy. We claim that in our SET, $e$ carries a protected two-fold degeneracy, while $w$ carries a protected degeneracy if and only if $\sigma_H/2$ is an odd integer. One can easily check that this claim leads to \eqref{appA:eq4}. To see the claim, we note that $\bfx$ behaves as charge conjugation. It maps a state with a $U(1)$ charge $Q$ to a state with charge $-Q$ and vice versa,  making the two states degenerate. When $Q$ is an odd integer, the two states form a projective representation of $\Z_2^p\times\Z_2^x$ after gauging $\Z_2^w$; on the other hand, if $Q$ is an even integer, it is not a projective representation of $\Z_2^p\times\Z_2^x$. The two-fold degeneracy associated with a projective representation cannot be lifted by any local perturbation that preserves $\Z_2^p\times \Z_2^x$.  Accordingly, we see that the unit charge $e$ carries a protected two-fold degeneracy. At the same time, $w$ carries a charge $\sigma_H/2$. So, if $\sigma_H/2$ is odd, $w$ carries a protected two-fold degeneracy; if $\sigma_H/2$ even, $w$ does not carry a protected two-fold degeneracy.

With the above understanding, we now make use of a result from Ref.~\onlinecite{WangPRX2016}. It was shown that if the SET is a valid two-dimensional state, the following two equations must be satisfied
\begin{align}
1 & = \left(M_{e,p}^2\right)^{x_e} \left(M_{w,p}^2\right)^{x_w} \left(\theta_e\right)^{2x_e} \left(\theta_w\right)^{2x_w},\label{appA:eq5}
\\
1 & = \left(M_{e,x}^2\right)^{x_e} \left(M_{w,x}^2\right)^{x_w} \left(\theta_e\right)^{2x_e} \left(\theta_w\right)^{2x_w}.
\label{appA:eq6}
\end{align}
Roughly speaking, the two equations are consequences of two dimensionality of the theory. According to Ref.~\onlinecite{WangPRX2016}, if the right-hand side of either equation is not equal to 1, it means the SET cannot live in strictly 2D but instead only on the surface of a 3D system. Our theory originates from strictly 2D SRE state, so Eqs.~\eqref{appA:eq5} and \eqref{appA:eq6} must be satisfied. Readers may consult Ref.~\onlinecite{WangPRX2016} for details (note that we have adapted the equations into the current notation). Inserting Eqs.~\eqref{appA:eq3} and \eqref{appA:eq4}, we find that Eq.~\eqref{appA:eq5} is automatic and Eq.~\eqref{appA:eq6} leads to the constraint Eq.~\eqref{appA:eq1}. This completes our proof.

\subsection{Fermionic SRE states with $U_f(1)\rtimes\mathbb{Z}_2$}
A general fermionic SRE state with $\mathbb{Z}_2^f\times\mathbb{Z}_2^x$ symmetry can be constructed by stacking $\mu$ copies of the chiral $p_x + ip_y$ superconductor and $\nu$ copies of the non-chiral root SPT state. The latter has a $\Z_8$ classification\cite{GuLevin2014}.  Gauging the symmetry can result in 128 different types of braiding statistics associated with the vortices\cite{Kitaev06,GuLevin2014, WangPRB2017}. They are described by
\begin{equation}
\theta_{w}=e^{i\pi \mu/8},\ \theta_{x}^2=e^{i\pi \nu/4},\  M_{w,x}^2=e^{-i\pi \nu/2}
\label{appA:eq7}
\end{equation}
where $w$ denotes a $\mathbb{Z}_2^f$ vortex and $x$ denotes a $\mathbb{Z}_2^x$ vortex.  Note that the mutual statistics $M_{w,x}^2$ is not independent but always equal to $(\theta_x^*)^4$. We remark that $w$- and $x$-vortices may be non-Abelian depending on the values of $\mu$ and $\nu$. 

Similarly to the bosonic case, we look for constraints enforced on the braiding statistics by an enlarged $U_f(1)\rtimes\Z_2^x$ symmetry group. There are two constraints. First, to lift $\Z_2^f$ to $U_f(1)$, it is required that $\mu$ must be even. This result was shown in Ref.~\onlinecite{WangPRB2016} and extensively used in the main text. When $\mu$ is even, the chiral superconductors are topologically equivalent to IQH states, with the Hall conductance $\sigma_H=\mu/2$. Second, the constraint \eqref{appA:eq1} still holds in the fermionic case. The rest of this section is devoted to showing \eqref{appA:eq1} for $U_f(1)\rtimes\Z_2$.

To show Eq.~\eqref{appA:eq1}, we take the same strategy as in the bosonic case. We consider a SRE state with $U_f(1)\rtimes\mathbb{Z}_2^x$ symmetry, then gauge the center $\Z_2^f$ and turn it into an SET state. The resulting gauge theory contains four Abelian anyons, which we again denote as $\{1, e, w, we\}$, where $e$ is the fermionic charge and $w$ is the fermion parity vortex. We have $\theta_e=-1$ and $\theta_w=\theta_{we}=e^{i\sigma_H\pi/4}$.  To be more specific, let $w$ be the vortex that is obtained by adiabatically inserting a $\pi$ flux of $U_f(1)$. Accordingly, $w$ carries a $U_f(1)$ charge $q_w=\sigma_H/2$.  The fusion rules depend on whether $\sigma_H$ is even or odd. In particular, for even $\sigma_H$, $w\times w =1$; for odd $\sigma_H$, $w\times w = e$. The remaining global symmetry is $[U_f(1)\rtimes\Z_2^x]/\Z_2^f \equiv U'(1)\rtimes\Z_2^x$, similarly to the bosonic case.  So, we arrive at a $U'(1)\rtimes \Z_2^x$ symmetry enriched topological order.  

Properties of $U'(1)\rtimes \Z_2^x$ SET states are generally discussed in Sec.~\ref{sec:STO-b}. First, the symmetry $\bfx$ may do the permutation $w\leftrightarrow we$. Similarly to Eq.~\eqref{eq:q-mirror-1}, this permutation is allowed only if $q_w' = - q_{we}' \modulo{1}$, where $q'_a$ is the fractional charge associated with $U'(1)$. The fractional charges carried by $e,w$ and $we$ are still given by Eq.~\eqref{appA:eq2} (except that $\sigma_H$ is not restricted to be even in fermionic systems). Accordingly,  $\sigma_H$ must be an odd integer to allow the permutation $w\leftrightarrow we$ by $\bfx$. In fact, when $\sigma_H$ is odd, $\bfx$ must permute $w$ and $we$. This can be proven by contradiction.  Assume that $w$ is not permuted by $\bfx$ and $\sigma_H$ is odd. Let us focus on the subgroup $\Z_2^p\times \Z_2^x\subset U'(1)\rtimes \Z_2$. Similarly to the bosonic case, the anyon $e$ carries a projective representation of $\Z_2^p\times\Z_2^x$. On the other hand, for odd $\sigma_H$, we have the fusion rule $w\times w =e$, which means that the representation on $e$ is a tensor product of the representations on two $w$'s. The latter cannot be a projective representation. So, we arrive at a contradiction. We summarize the result: When $\sigma_H$ is even, $\bfx$ does not permute $w$ and $we$; when $\sigma_H$ is odd, $\bfx$ does the permutation.

Below we consider even and odd $\sigma_H$ separately. When $\sigma_H$ is even, the characterization of SET is very much the same as the bosonic case. For the $\Z_2^p\times \Z_2^x$ subgroup, the SET is again described by six quantities
\begin{align}
x_e & =0, \quad M_{e,p}^2 = -1, \quad M_{e,x}^2=1,\nonumber\\
x_w & =1, \quad M_{w,p}^2 = \theta_w^2, \quad M_{w,x}^2= \pm1, \pm i.
\end{align}
All  quantities have the same physical meanings as in the bosonic case. The conditions \eqref{appA:eq5} and \eqref{appA:eq6} still apply. Then, the constraint \eqref{appA:eq1} follows immediately.

For odd $\sigma_H$, let us first consider $\sigma_H=1$. We will see that it is enough to study properties associated with the $\bfx$ symmetry. In the presence of permutation $w\leftrightarrow we$ by $\bfx$, it was shown in Ref.~\onlinecite{BarkeshliPRB2019} that there are four types of $\Z_2^x$ SET phases, characterized by (1) whether the un-permuted anyons $e$ carries a $\Z_2^x$ fractional charge and (2) whether a $\Z_2^x$ bosonic SPT state is glued. However, if $e$ carries a fractional charge of $\mathbb{Z}_2^x$, the original symmetry group should be $[U_f(1)\rtimes \mathbb{Z}^f_4]/\mathbb{Z}_2$. Accordingly, the symmetry group $U_f(1)\rtimes\Z_2^x$ results in only two SETs after gauging $\Z_2^f$, which differs relatively by a $\Z_2^x$ SPT state. The two topological orders obtained by further gauging $\Z_2^x$ have been worked out in Ref.~\onlinecite{dijkgraaf1989} (see also Ref.~\onlinecite{BarkeshliPRB2019}). Both contain $9$ anyons in total: $1^\pm$, $e^\pm$, $[w]$, $x_1^\pm$, $x_2^\pm$, where $1^+$ is the new vacuum, $1^-$ is the $\Z_2^x$ charge. Anyons $a^+$ and $a^-$ differ by the charge $1^-$. The anyon $[w]$ is obtained by identifying $w$ and $we$ after gauging. The anyons $x_1^\pm$ and $x_2^\pm$ are different $\Z_2^x$ vortices. The quantum dimensions and topological spins are as follows:
\begin{align}
d_{1^\pm}  & = d_{e^\pm}  = 1 \nonumber\\
d_{[w]} & =2, \nonumber\\
d_{x_1^\pm} &  =d_{x_2^\pm}  = \sqrt{2} \nonumber\\
\theta_{1^\pm} & =1 \nonumber\\
\theta_{e^\pm} & = -1 \nonumber\\
\theta_{[w]} & = \theta_w =e^{i\pi/4}\nonumber\\
\theta_{x_1^\pm} &  = \pm \kappa e^{i\pi/8}\nonumber\\
\theta_{x_2^\pm} &  = \pm \kappa^* e^{i\pi/8}\nonumber
\end{align}
where $\kappa= 1$ or $i$ for the two gauged SETs respectively. We observe that regardless of the value of $\kappa$,  we always have $\theta_x^4=e^{i\pi/2}$, where $x$ is any of the four $x$-vortices. From Eq.~\eqref{appA:eq7}, we obtain that $M_{w,x}^2=(\theta_x^*)^4 = e^{-i\pi/2}$. For general odd $\sigma_H$, we make use of a result from Ref.~\cite{WangPRB2017} it showed that $\theta_w$, $\theta_x^2$ and $M_{w,x}^2$ are multiplicative under stacking. Accordingly, by stacking $\sigma_H$ copies of the $\sigma_H=1$ state, we obtain 
\begin{equation}
\theta_w=e^{i\sigma_H\pi/4},\ \theta_x^2 = \kappa^2 e^{i\sigma_H\pi/4}, \ M_{w,x}^2=e^{-i\sigma_H\pi/2}.
\end{equation}
Wee see that Eq.~\eqref{appA:eq1} holds for any odd $\sigma_H$.

Combining our analyses of the even and odd $\sigma_H$ cases, we prove that \eqref{appA:eq1} holds generally. 

\section{Adjoining IQH states}
\label{appd:adjoin-iqh}

In this appendix, we explicitly show the classification reduction of the 2D SRE states by adjoining IQH states in fermionic systems. Unlike the $\Z$ classification generated by the $E_8$ state which always reduces to $\Z_2$ under adjoining, the reduction by adjoining IQH states is more interesting and differs for different symmetries.

\subsection{$U_f(1)\times\mathbb{Z}_2^\calM$}
\label{appB:1}

As discussed in Sec.~\ref{sec:u1xz2-f}, 2D fermionic SRE states with internal $U_f(1)\times\mathbb{Z}_2$ symmetry are classified by $\mathbb{Z}^2\times\mathbb{Z}_4$. A general SRE state is indexed by an integer vector $\mu=(\mu_1,\mu_2,\mu_3)$. Under adjoining operations, the $\Z$ classification associated with root state (ii) collapses to $\Z_2$. This $\Z_2$ extends the original $\Z_4$ classification, and together they form a $\Z_8$ classification.  To explicitly show this result, we perform an analysis from edge theory. Our analysis is not new but basically a review of that in Ref.~\onlinecite{SongPRX2017}.

Let us consider adjoining a pair of IQH states on the two sides of the mirror plane (see Fig.~\ref{fig:adjoining}). It adds two chiral fermions $\psi_{R_1}$ and $\psi_{R_2}$ to the edge of the mirror plane, where the subscript ``$R$'' means ``right-moving''. The two edge modes are described by the Hamiltonian
\begin{align}
H=-i\sum_{n=1,2}\psi_{R_n}^\dagger\partial_x\psi_{R_n}.
\label{eq:H1}
\end{align}
The $U_f(1)$ symmetry has the action $\psi_{R_n}\rightarrow e^{i\alpha}\psi_{R_n}$, where $\alpha$ a rotation angle. To have the full symmetry being $U_f(1)\times\Z_2$, the symmetry $\bfx$ should have the action $\psi_{R_1}\leftrightarrow\psi_{R_2}$. 

We need to read out the index $\mu$ for the state with the above edge. To do that, we define two new fermions:
\begin{equation}
\begin{split}
\tilde{\psi}_{R_1}=&\frac{1}{\sqrt{2}}\left(\psi_{R_1}+\psi_{R_2}\right)\\
\tilde{\psi}_{R_2}=&\frac{1}{\sqrt{2}}\left(\psi_{R_1}-\psi_{R_2}\right)
\end{split}
\label{eq:psi_tilde}
\end{equation}
Under $\bfx$ action, the new fermions $\tilde\psi_{R_1}$ and $\tilde\psi_{R_2}$ transforms as
\begin{align}
\tilde{\psi}_{R_1}\rightarrow\tilde{\psi}_{R_2}, \ \tilde{\psi}_{R_2}\rightarrow-\tilde{\psi}_{R_2}
\end{align}
We see that the $\bfx$ symmetry behaves as the fermion parity of the right-moving fermion $\tilde\psi_{R_2}$. Then, the square of the topological spin of $x$-vortices can be easily computed, which is $\theta_{x}^2 = e^{i\pi/2}$\cite{GuLevin2014}. At the same time, we have $c=2$ and $\sigma_H=2$. Therefore, with the data in \eqref{eq:prop-f1}, we find that the adjoined state corresponds to $\mu=(0,2,1)$.

Hence, adjoining IQH states establishes the following equivalence relation for the index $\mu$:
\begin{align}
\mu \sim \mu + (0,2,1).
\end{align}
With this relation, we see that a stack of two copies of root state (ii), i.e. the state indexed by $\mu=(0,2,0)$, is equivalent to the state $(0,0,-1)$ --- the inverse state of the root state (iii). Accordingly, the root state (ii) and (iii) together generate a $\Z_8$ classification, after taking adjoining operations into accounts.

\subsection{$U_f(1)\rtimes\mathbb{Z}_2^\calM$}
Strictly 2D fermionic SRE states with  $U_f(1)\rtimes\mathbb{Z}_2$ symmetry are classified by $\mathbb{Z}^2\times\mathbb{Z}_2$.  A general SRE state can be indexed by an integer vector $\mu=(\mu_1,\mu_2,\mu_3)$. Below we show that adjoining IQH states reduces the $\Z$ classification  associated with root state (ii) reduces to $\Z_2$, by a similar edge theory analysis as above.

By adjoining two IQH states, we again obtain an edge with two chiral fermions $\psi_{R_1}$ and $\psi_{R_2}$, described by the Hamiltonian \eqref{eq:H1}. Nevertheless, the symmetry actions will be different to the case of $U_f(1)\times\Z_2$. To have the correct symmetry group $U_f(1)\rtimes\Z_2$, we shall have the following symmetry action: under $U_f(1)$, 
\begin{equation}
\psi_{R_1}\rightarrow e^{i\alpha}\psi_{R_1},\quad\psi_{R_2}\rightarrow e^{-i\alpha}\psi_{R_2}
\end{equation}
and under $\bf{x}$, 
\begin{equation}
\psi_{R_1}\leftrightarrow\psi_{R_2}.
\end{equation}
We now read off the index $\mu$ associated with this edge. It is obvious that $c=2$. Also, the Hall conductance $\sigma_H=2$. This is somewhat not obvious, as $\psi_{R_2}$ is positively charged under $U_f(1)$. Nevertheless, it is not hard to show that a right-moving positively charged fermion gives $\sigma_H=1$. The quantity $\theta_x^2$ can be obtained exactly the same as in Appendix \ref{appB:1}, which is equal to $e^{i\pi/2}$. Comparing with the data in \eqref{eq:u1rtimesz2-prop}, we obtain that $\mu=(0,2,0)$.

Hence, adjoining IQH states establishes the equivalence relation 
\begin{align}
\mu \sim \mu + (0,2,0)
\end{align}
So, the $\Z$ classification associated with root state (ii) reduces to $\Z_2$.

\section{Review on anyon condensation}
\label{sec:app_anyoncond}

\begin{figure}
\centering
\includegraphics[scale=1]{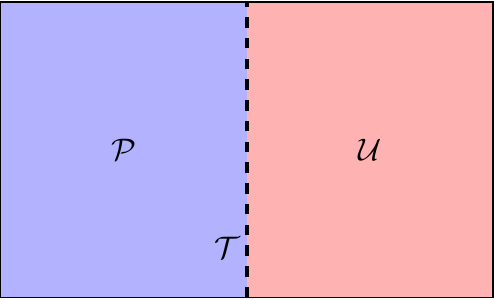}
\caption{Schematics of a gapped domain wall between topological orders $\mathcal{P}$ and $\mathcal{U}$.}
\label{fig:anyoncond}
\end{figure}

In this appendix, we review the basics of anyon condensation theory\cite{BaisPRB2009,kong2014}. This theory will be extensively used in Appendices \ref{appd:anycond-boson-teta3} and \ref{appd:anycond-boson-teta45} to provide alternative derivations of the expressions for anomaly indicators $\teta_{3}, \teta_{4}$, $\teta_{5}$ and $\eta_{3f}$. 

Anyon condensation is an algebraic theory that describes properties of a gapped domain wall that connects two topological orders (see Fig.~\ref{fig:anyoncond}). Let $\mathcal{P}$ be the topological order ``before'' condensation, and $\mathcal{U}$ be the topological order ``after'' condensation.  Intuitively, anyon condensation is such an event: when parameters of the microscopic Hamiltonian vary in an appropriate manner, certain anyons in $\mathcal{P}$ condense into the vacuum such that the system takes a topological phase transition into the topological order $\mathcal{U}$. In theory, such a phase transition occurs in two steps:
\begin{equation}
  \mathcal{P} \to \mathcal{T} \to \mathcal{U},
\end{equation}
where $\mathcal{T}$ is the intermediate theory. The real-space version of the transition is shown in Fig.~\ref{fig:anyoncond}: the left half of the 2D system is the theory $\mathcal{P}$ before condensation, the right half is the theory $\mathcal{U}$ after condensation, and the immediate theory $\mathcal{T}$ lives on the gapped domain wall and characterizes it. Topological excitations (anyons) and their fusion and braiding properties in $\mathcal{P}$, $\mathcal{T}$ and $\mathcal{U}$ are different but closely connected. Anyon condensation theory is the theory that establishes the connection. Below we discuss the connection for bosonic topological orders. When we study fermionic topological orders, they will be turned into bosonic topological orders by gauging the fermion parity. So, the following discussion will be enough for our purpose.

According to Refs.~\onlinecite{BaisPRB2009,kong2014}, the relation between the anyons in $\mathcal{P}$ and $\mathcal{T}$ is described by the restriction map:
\begin{equation}
  r(\alpha) = \sum_{t\in\mathcal{T}} n_{\alpha, t} t,
\end{equation}
where $\alpha\in\mathcal{P}$ and the non-negative integer $n_{\alpha,t}$ is called the restriction coefficient. It means that when $\alpha$ moves to the domain wall, it can turn into $t$ in $n_{\alpha,t}$ distinct ways. If $n_{\alpha,t}=0$, it means $\alpha$ cannot be turned into $t$ on the domain wall. Specifically to which anyon $t$ and in which way it takes to turn into $t$ depend on the detail of how $\alpha$ moves to the domain wall.  An important property to remark is that the restriction map commutes with the fusion of anyons in $\mathcal{P}$. More specifically, for anyons $\alpha, \beta \in \mathcal{P}$,
\begin{equation}
  r(\alpha) \times r(\beta) = r(\alpha \times \beta).
\end{equation}
The explicit expression is
\begin{equation}
  \sum_{r,s \in \mathcal{T}} n_{\alpha,r} n_{\beta,s} N_{rs}^{t} = \sum_{\gamma \in \mathcal{P}} N_{\alpha \beta}^{\gamma} n_{\gamma,t},
  \label{eq: restriction commutes with fusion}
\end{equation}
where $N_{rs}^{t}$ and $N_{\alpha \beta}^{\gamma}$ are respectively the fusion coefficients in $\mathcal{P}$ and $\mathcal{T}$.

Anyons in the intermediate theory $\mathcal{T}$ are divided into two kinds: \emph{confined} and \emph{deconfined} anyons. The confined anyons are not the usual anyons as the energy cost to pull a pair of particle and anti-particle depends on the distance. So, they cannot move freely but are confined on the gapped domain wall. Fusion involving confined anyons may not commute, i.e., $t_i \times t_j \neq t_j \times t_i$. In addition, a confined anyon $t$ does not have a well-defined topological spin. On the other hand, deconfined anyons behave as the usual anyons, i.e., they can move freely out of the gapped domain wall and have commutative fusion and well-defined braiding properties. The set of deconfined anyons are closed under fusion and form the topological order $\mathcal{U}$.

The inverse of the restriction map is called the lifting map and denoted as:
\begin{equation}
  l(t) = \sum_{\alpha \in \mathcal{P}} n_{\alpha,t} \alpha
\end{equation}
where the lifting coefficient $n_{\alpha,t}$ is the same as the restriction coefficient in $r(\alpha)$. It expresses any anyon $t \in \mathcal{T}$ as the superposition of the anyons in $\mathcal{P}$.
We can determine the deconfined anyons in $\mathcal{T}$ via the following criterion that: $t$ is deconfined if and only if all $\alpha \in l(t)$ have the same topological spin. Here, when we say ``$\alpha\in l(t)$'', it means $n_{\alpha,t}\neq 0$. Then, $\theta_t = \theta_\alpha$ for any $\alpha \in l(t)$ if $t$ is deconfined.

As a subset of anyons in $\mathcal{T}$, the deconfined anyons in $\mathcal{U}$ also satisfy  Eq.~\eqref{eq: restriction commutes with fusion},  i.e., when we restrict $r,s$ to be inside $\mathcal{U}$ in Eq.~\eqref{eq: restriction commutes with fusion}. Another important constraint for deconfined anyons is that the matrix $n$ commutes with the modular $S$ and $T$ matrices, in the following sense,
\begin{equation}
  S^{\mathcal{P}} n = n S^{\mathcal{U}},\quad T^{\mathcal{P}} n = n T^{\mathcal{U}}. 
  \label{eq:stu}
\end{equation}
The explicit expression for the former equation is
\begin{equation}
  \sum_{\beta \in \mathcal{P}} S_{\alpha, \beta} n_{\beta. t} = \sum_{s \in \mathcal{U}} n_{\alpha, s} S_{s,t},
  \label{eq:stu2}
\end{equation}
where $\alpha \in \mathcal{P}$ and $t \in \mathcal{U}$.
Since $T_{\alpha,\beta} = \delta_{\alpha,\beta} \theta_\alpha$, the latter equation in \eqref{eq:stu} reduces to that $\theta_\alpha = \theta_t$ as long as $n_{\alpha, t} \neq 0$, for any $\alpha$ and $t$, which has already been mentioned above.

\section{Derivation of $\teta_3$ from anyon condensation}
\label{appd:anycond-boson-teta3}

In this appendix, we use anyon condensation theory to derive the expression of $\teta_3$ in terms of SET quantities. Compared with the approach used in Sec.~\ref{sec:teta3-derive}, it is technically more complicated. However, it is still worth carrying out such an analysis to have an alternative and perhaps better understanding of the physics. We will consider $U(1)\times\Z_2^\calM$ and $U(1)\rtimes\Z_2^\calM$ simultaneously. As discussed in Sec.~\ref{sec:define-eta-u1}, $\teta_3$ is related to the SRE state in the mirror plane which only needs protection from $U(1)$ and its value is determined by the topological spin of $w$-vortices. Therefore, to express $\teta_3$ in terms of SET quantities, it is sufficient to gauge the $\Z_2^w$ subgroup of $U(1)$.  In Appendix \ref{appd:anycond-boson-teta45} where we discuss derivations of $\teta_4$ and $\teta_5$, we will gauge the bigger $\Z_2^w\times\Z_2^x$ subgroup.  Although $\teta_3$ is independent of the mirror symmetry, our following discussions do assume the presence of mirror symmetry and mirror symmetry properties will be extensively studied, which paves the way for deriving $\teta_4$ and $\teta_5$ in the next appendix.

\subsection{Gauging $\Z_2^w$ before folding}
\label{sec: gaugez2p}

We will use $\mathcal{C}_l$ and $\mathcal{C}_r$ to denote the topological orders living on the left and right wings of the T-like junction in Fig.~\ref{fig:folding}(b), respectively, although they are actually the same, namely $\mathcal{C}_l=\mathcal{C}_r=\mathcal{C}$. After gauging $\Z_2^w$, they will be extended to topological orders $\mathcal{B}_l$ and $\mathcal{B}_r$ respectively. Unlike $\mathcal{C}_l$ and $\mathcal{C}_r$, the topological orders $\mathcal{B}_l$ and $\mathcal{B}_r$ are not necessarily the same, as we will discuss below. Later after we establish a few results, we will drop the subscripts ``$l$'' and ``$r$'' for notational simplicity.

First, let us discuss the anyon content of $\mathcal{B}_l$. Anyons in $\mathcal{B}_l$ are divided into two collections $\mathcal{B}_{l,0}$ and $\mathcal{B}_{l,1}$. Without any confusion, we will simply denote this split as $\mathcal{B}_l = \mathcal{B}_{l,0}\oplus\mathcal{B}_{l,1}$.  The first collection is a set of anyons which we denote as 
\begin{align}
\mathcal{B}_{l,0}=\{a_{{}_+}, a_{{}_-}|\ \forall a\in \mathcal{C}_l\}.
\end{align}
Namely, each anyon in $\mathcal{C}_l$ splits into two in $\mathcal{B}_{l,0}$. The special anyon $\mathbbm{1}_{+}$ is the trivial anyon in $\mathcal{B}_l$, while $\mathbbm{1}_{-}$ is the $\Z_2^w$ gauge charge, which is a Abelian boson. For general $a$,  the sign ``$\pm$'' in $a_\pm$  has no absolute meaning of carrying or not carrying a $\Z_2^w$ gauge charge. It only means that $a_+$ and $a_-$ differ \emph{relatively} by a gauge charge $\mathbbm{1}_-$. They satisfy the fusion rule $a_\pm\times \mathbbm{1}_-= a_\mp$.  
The mutual statistics between $a_\pm$ and  $\mathbbm{1}_{-}$ is trivial, for every $a$. The topological spins and quantum dimensions satisfy
\begin{align}
\theta_{a_{\pm}}=\theta_a, \quad d_{a_\pm}=d_{a}.
\end{align}
To fix our notation, we make use of the full $U(1)$ symmetry to distinguish $a_+$ and $a_-$. Before gauging $\Z_2^w$, anyon $a$ may carry a $U(1)$ charge $Q=q_a + n$, where $0 \le q_a<1$ is the fractional charge and $n$ is any integer charge. Then, we define $a_+$ as the one with $Q=q_a \modulo{2}$,  and $a_-$ as the one with $Q=q_a+1 \modulo{2}$. Here, ``$\modulo{2}$'' is taken because we only care about $\Z_2^w$ charges. For later convenience, we define the $\Z_2^w$ charge carried by $a_\lambda$,
\begin{align}
q_{a_\lambda} :=q_a + \frac{1-\lambda}{2}
\label{eq:qalambda}
\end{align} 
where we have interpreted the sign ``$\pm$'' as ``$\pm 1$''. The $\Z_2^w$ charges should be conserved modulo 2, upon fusion of anyons. With this understanding, we obtain the following fusion rule
\begin{align}
a_\lambda \times b_\mu = \sum_c N_{ab}^c c_{\sigma}
\end{align}
where the coefficient $N_{ab}^c$ inherits from $\mathcal{C}$, and the signs $\lambda,\mu,\sigma$ satisfy the following condition
\begin{align}
q_{a_\lambda} + q_{b_\mu} = q_{c_\sigma} \modulo{2}
\label{eq:sign-cond}
\end{align}
Note that the condition \eqref{eq:q-cond} guarantees a unique solution of $\sigma$ in \eqref{eq:sign-cond}, provided that $\lambda$ and $\mu$ are given.  By setting $c_\sigma = \mathbbm{1}_+$, we can obtain the anti-particle $\overline{a_\lambda} = \bar{a}_{\lambda'}$, where $\lambda'$ is fixed  by the condition that $q_{a_\lambda} + q_{\bar{a}_{\lambda'}}=0 \modulo{2}$. Below we will also take the convention that $\bar{\lambda} = -\lambda$ to make some expressions more compact.

The other collection $\mathcal{B}_{l,1}$ contains all the $\Z_2^w$  vortices. First, there exists a special Abelian vortex $V$, whose mutual statistics with anyons $a_{\pm}$ is determined by the fractional charge $q_a$.\cite{wen-book, BarkeshliPRB2019} The mutual statistics between $V$ and $a_\lambda$ is given by
\begin{align}
M_{V, a_\lambda} = e^{i\pi q_{a_\lambda}} =  \lambda e^{i\pi q_{a}}.
\label{eq:Vl-def}
\end{align}
In fact, this vortex is the one obtained by adiabatically inserting a $\pi$ flux of the $U(1)$ symmetry before gauging, and \eqref{eq:Vl-def} is simply the Aharonov-Bohm phase between the charge carried by $a_\lambda$ and the $\pi$ flux. It also means that $V$ itself carries a  $U(1)$ charge $\sigma_H/2$.  Second, all other vortices in $\mathcal{B}_{l,1}$ can be considered as the composite of $V$ and anyons in $\mathcal{B}_{l,0}$. We denote them as $V_{a_{\pm}}$, obtained by the following fusion product 
\begin{align}
V_{a_{\pm}} =V\times a_{\pm} 
\label{eqn:pi-anyon-l}
\end{align}
Since $V$ is Abelian, the fusion outcome on the right-hand side is unique. It is understood that $V_{\mathbbm{1}_+}\equiv V$.  The quantum dimension and the topological spin are 
\begin{align}
d_{V_{a_\lambda}}& =d_{a_\lambda}=d_a, \nonumber\\
\theta_{V_{a_{\lambda}}}& =\lambda \theta_{V}\theta_{a} e^{i\pi q_a}. \label{eqn:ts_flux}
\end{align}
The latter follows from the general relation $\theta_\gamma = \theta_\alpha\theta_\beta M_{\alpha\beta}^\gamma$, where $M_{\alpha\beta}^\gamma$ is the mutual statistics between $\alpha$ and $\beta$ in fusion channel $\gamma$\cite{Kitaev06}. One can check that the total quantum dimension is $D_{\mathcal{B}_l} =2 D_{\mathcal{C}_l}$, and the $S$ matrix elements are given by
\begin{align}
S(V_{a_\lambda}, b_\mu) =\frac{1}{2}M_{V,b_\mu}^*S_{a,b} = \frac{\mu}{2}e^{-i\pi q_b}S_{a,b}.
\label{eq:smatrix-B}
\end{align}
 The fusion product of two $V$'s will turn out to be important, which we denote as
\begin{align}
V\times V=m_\lambda
\label{eqn:fusion_flux}
\end{align}
where $m_\lambda$ is an Abelian anyon in $\mathcal{B}_{l,0}$. Using the $U(1)$ charge carried by $V$, we see that the sign $\lambda$ is determined by the Hall conductance: if $\sigma_H$ is even, $\lambda=+1$; if $\sigma_H$ is odd, $\lambda=-1$.  Note that the anyon  $m\in\mathcal{C}_l$ is the ``avatar" of the $2\pi$ monopole in the surface topological order $\mathcal{C}_l$ before gauging. 

It is worth emphasizing that given $\mathcal{C}_l$ and $\{q_a\}$, the $\Z_2^w$-gauged theory $\mathcal{B}_l$ is not unique. According to the SET classification\cite{BarkeshliPRB2019}, there exist two types of gauged theories, which differ by stacking a $\Z_2^w$ SPT state before gauging. If one theory has $\theta_{V}^2 = \alpha$, the other has $\theta_V^2=-\alpha$, where the value of $\alpha$ depends on $\sigma_H$ or $\{q_a\}$. However, this ambiguity will not affect our derivation of anomaly indicator. Note that the two types of theories have the same topological properties for anyons in $\mathcal{B}_{l,0}$, and the difference appears only for vortices in $\mathcal{B}_{l,1}$.

We now move on to discuss the topological order $\mathcal{B}_r$. Since it is obtained from $\mathcal{C}_r=\mathcal{C}$ with the same fractional charges $\{q_a\}$ by gauging $\Z_2^w$, $\mathcal{B}_r$ should also be one of the two possible $\Z_2^w$ gauge theories, like $\mathcal{B}_l$. Nevertheless, we should keep in mind that the left and right wings of the T-junction are mirror images of each other, as discussed in Sec.~\ref{sec:general}. So, upon gauging  $\Z_2^w$, the gauged theories $\mathcal{B}_l$ and $\mathcal{B}_r$ should also be mirror images of each other. Therefore, the choices of $\mathcal{B}_l$ and $\mathcal{B}_r$ among the two possible $\Z_2^w$-gauged theories are not independent. (However, whether $\mathcal{B}_l$ and $\mathcal{B}_r$ are the same or not is a separate question. In fact, the answer to this question determines whether the mirror plane is a $\Z_2^w$ SPT or not.)

To be more precise, the fact that $\mathcal{B}_l$ and $\mathcal{B}_r$ are mirror images of each other implies that there exists an \emph{anti-equivalence} $\rho_m^\mathcal{B}:\mathcal{B}_l \rightarrow \mathcal{B}_r$. That is, each anyon $\alpha\in\mathcal{B}_l$ is mapped to $\rho_m^\mathcal{B}(\alpha)\in\mathcal{B}_r$, such that all fusion and braiding properties are preserved after complex conjugation. Complex conjugation is needed because mirror symmetry reverses orientation. Specifically, some of the conditions are  
\begin{align}
N_{\rho_m^\mathcal{B}(\alpha),\rho_m^\mathcal{B}(\beta)}^{\rho_m^\mathcal{B}(\gamma)} & = N_{\alpha,\beta}^\gamma\nonumber\\
S_{\rho_m^\mathcal{B}(\alpha),\rho_m^\mathcal{B}(\beta)} & = S_{\alpha,\beta}^*\nonumber\\
 \theta_{\rho_m^\mathcal{B}(\alpha)} & = \theta_{\alpha}^*
\end{align}
The first condition means that fusion commutes with the map $\rho_m^\mathcal{B}$. Mutual statistics such as the one in \eqref{eq:Vl-def} is also preserved after complex conjugation. Therefore, $\mathcal{B}_r$ is determined to $\mathcal{B}_l$ by the map $\rho_m^\mathcal{B}$.

One understands that $\rho_m^\mathcal{B}$ is an extension of $\rho_m^\mathcal{C}:\mathcal{C}_l\rightarrow \mathcal{C}_r$ into the gauged theory. The latter is given as an input of the original SET. There is a difference between the two. Since $\mathcal{C}_l=\mathcal{C}_r=\mathcal{C}$, the map $\rho_m^\mathcal{C}$ is an anti-autoequivalence. It can compose with itself and satisfies $\rho_m^\mathcal{C}\circ\rho_m^\mathcal{C} = \mathbbm{1}$, the identity map. On the other hand,  $\mathcal{B}_l$ is not necessarily the same as $\mathcal{B}_r$. So, $\rho_m^\mathcal{B}$ is an anti-equivalence between two topological orders. It cannot compose with itself. For later convenience, we define its inverse map $\varphi_m^\mathcal{B}: \mathcal{B}_r\rightarrow \mathcal{B}_l$, such that $\varphi_m^\mathcal{B}\circ\rho_m^\mathcal{B}=\rho_m^\mathcal{B}\circ\varphi_m^\mathcal{B} =\mathbbm{1}$.

We now discuss the map $\rho_m^\mathcal{B}$ more specifically. We consider anyons in $\mathcal{B}_{l,0}$ and $\mathcal{B}_{l,1}$ separately. Anyons in $\mathcal{B}_{l,0}$ originate from those in $\mathcal{C}_l$, so $\rho_m^\mathcal{B}$ is mostly determined by $\rho_m^\mathcal{C}$. For anyon $a_\sigma\in \mathcal{B}_{l,0}$, the map $\rho_m^\mathcal{B}$ must be given by 
\begin{align}
\rho_m^\mathcal{B}(a_\sigma)=\rho_m^\mathcal{C}(a)_{\sigma'}
\label{eq:rhob-rhoc}
\end{align}
where the right-hand side is an anyon in $\mathcal{B}_{r,0}$. The sign $\sigma'$ is not arbitrary but determined by the following condition
\begin{align}
q_{a_\sigma} = \zeta q_{\rho_m^\mathcal{C}(a)_{\sigma'}} =\zeta\left(q_{\rho_m^\mathcal{C}(a)} + \frac{1-\sigma'}{2}\right)\modulo{2}
\label{eq:rho-B-sign}
\end{align}
where $\zeta=1$ for $U(1)\times\Z_2^\calM$ and $\zeta=-1$ for $U(1)\rtimes\Z_2^\calM$. It follows from the fact that $U(1)$ charge changes sign under $\calM$ for $U(1)\rtimes\Z_2^\calM$, but does not for $U(1)\times\Z_2^\calM$. The inverse map $\varphi_m^\mathcal{B}$ is very similar: for $a_\sigma\in \mathcal{B}_r$, we have
\begin{align}
\varphi_m^\mathcal{B}(a_\sigma)=\rho_m^\mathcal{C}(a)_{\sigma'}
\end{align}
where $\sigma'$ is again determined by \eqref{eq:rho-B-sign}. Since the two maps looks exactly the same, below we will simply use the convention that $\varphi_m^\mathcal{B}\equiv\rho_m^\mathcal{B}$ when restricted to anyons in $\mathcal{B}_{l,0}$ and $\mathcal{B}_{r,0}$. In the special case $a=\mathbbm{1}$, we obtain $\rho_{m}^\mathcal{B}(\mathbbm{1}_\pm) = \mathbbm{1}_\pm$.

For the vortex anyon $V_{a_{\lambda}} \in \mathcal{B}_{l,1}$, it is mapped to a vortex anyon in $\mathcal{B}_{r,1}$ under $\rho_m^\mathcal{B}$, and vice versa by $\varphi_m^\mathcal{B}$.  Under $\rho_m^\mathcal{B}$, the special vortex $V\in \mathcal{B}_{l}$ must be mapped to an Abelian vortex in $V_{z_\sigma}\in\mathcal{B}_{r}$.  That is, 
\begin{align}
\rho_m^\mathcal{B}(V^l) =V^r_{z_\sigma}
\label{eqn_ar_lr_b}
\end{align}
where to distinguish the special vortices from $\mathcal{B}_l$ and $\mathcal{B}_r$, we have put the superscripts ``$l$'' and ``$r$''. The Abelian anyon $z_\sigma\in \mathcal{B}_r$ can be determined by the requirement that, for any $a_\lambda\in \mathcal{B}_l$, 
\begin{align}
e^{i\pi q_{a_\lambda}}=M_{V^l, a_\lambda} &  = M_{\rho_m^\mathcal{B}(V^l),\rho_m^\mathcal{B}(a_\lambda)}^* \nonumber\\
& = M^*_{V^r,\rho_m^\mathcal{C}(a)_{\lambda'}} M_{z,\rho_m^\mathcal{C}(a)}^* \nonumber\\
& = e^{-i \zeta \pi q_{a_\lambda}} \frac{D_{\mathcal{C}}}{d_a} S_{z,\rho_m^\mathcal{C}(a)}\nonumber
\end{align}
One can check that the $\lambda$ dependence on the two sides cancel. The above equation further simplifies to 
\begin{align}
S_{z,a} = \frac{d_a}{D_\mathcal{C}} e^{i\pi (1+\zeta)q_{a}}, \ \forall a\in\mathcal{C}
\end{align}
where we have used $q_{\rho_m^\mathcal{C}(a)} = \zeta q_a \modulo{1}$. Then, because of modularity in $\mathcal{C}$, we find that $z=\mathbbm{1}$ when $\zeta=-1$; and $z= m$, the ``avatar'' anyon of the bulk monopole defined in \eqref{eq:mma} when $\zeta=1$.  Other mirror image of vortices in $\mathcal{B}_{l,1}$ can be obtained using the property that $\rho_m^\mathcal{B}$ commutes with fusion map.

\subsection{Folding}
\label{sec_fold_b_apd}

Having understood $\mathcal{B}_l$, $\mathcal{B}_r$ and their relation, we now consider folding. We fold the right wing of the T-junction  along the intersection line to the bottom of the left wing, as shown in Fig.~\ref{fig:folding}(c). Since folding reverses orientation, let us use $\mathcal{B}_r^{\text{rev}}$ to denote the topological order in the bottom layer. Also, let $\alpha^\rev $ be the anyon corresponding to $\alpha\in\mathcal{B}_r$. All fusion and braiding quantities in $\mathcal{B}_r^\rev$ are complex conjugates of the counterparts in $\mathcal{B}_r$. In particular, topological spins satisfy $\theta_{\alpha^\text{rev}}=\theta_{\alpha}^*$ and elements of the $S$ matrix satisfy $S_{\alpha^\text{rev}, \beta^\text{rev}}=S_{\alpha, \beta}^*$, for any $\alpha, \beta \in \mathcal{B}_r$. Let us define a folding map $\mathcal{F}$ from $\mathcal{B}_r$ to $\mathcal{B}_r^\rev$:
\begin{align}
\mathcal{F}: \alpha \mapsto \alpha^\text{rev}.
\end{align}
Like $\rho_m^\mathcal{B}$, the map $\mathcal{F}$ is also an anti-equivalence.  Accordingly, the composite map $\mathcal{F}\circ\rho_m^\mathcal{B}$ establishes a usual equivalence between $\mathcal{B}_l$ and $\mathcal{B}_r^\rev$, under which all fusion and braiding properties are preserved (without complex conjugation). That is,  $ \alpha \in \mathcal{B}_l$ is mapped to anyon $ [\rho_m^\mathcal{B}(\alpha)]^\rev$ in $\mathcal{B}_r^\rev$, and the two anyons have exactly the same topological properties.

The folded system is a double-layer topological order $\mathcal{B}_l\boxtimes \mathcal{B}_r^\rev$. Since $\mathcal{B}_r^\rev$ and $\mathcal{B}_l$ are equivalent, we find it convenient to rename anyons in $\mathcal{B}_r^\rev$ using the labels of $\mathcal{B}_l$, i.e., 
\begin{align}
\left[\rho_m^\mathcal{B}(\alpha)\right]^\text{rev} \xrightarrow{\ \ \rm rename \ \ } \alpha
\label{eqn:anyonid}
\end{align}
With this new notation, we can now denote the double-layer topological order as $\mathcal{B}_l\boxtimes \mathcal{B}_l$, whose anyons can be labeled by $(\alpha, \beta)$. The mirror permutation acts simply as exchanging the anyons in the two layers, 
\begin{align}
\mathcal{M}: (\alpha, \beta)\rightarrow (\beta, \alpha).
\label{appeq:layerex}
\end{align}
It becomes an internal symmetry. We observe that the information of $\rho_m^\mathcal{B}$ disappears in the mirror action after renaming \eqref{eqn:anyonid}. It turns out that it is encoded in properties of the 1D gapped domain wall between the double-layer system and the mirror plane in Fig.~\ref{fig:folding}(c), which will be discussed soon.

Without causing confusion, below we will simply omit the subscript $l$ regarding to the notations $\mathcal{C}_l$, $\mathcal{B}_l$, $\mathcal{B}_{l,0}$, $\mathcal{B}_{l,1}$ etc, and denote them as $\mathcal{C}, \mathcal{B}, \mathcal{B}_0, \mathcal{B}_1$ etc.  The double-layer topological order will be denoted as as $\mathcal{B}\boxtimes\mathcal{B}$.

Before proceeding, we make a comment on the ``$\Z_2^w$ gauge charge'' carried by $(a_\lambda,b_\mu)$ in the double-layer topological order $\mathcal{B}\boxtimes\mathcal{B}$ after renaming (\ref{eqn:anyonid}).  The anyon $a_\lambda$ in the top layer carries a $\Z_2^w$ gauge charge $q_{a_\lambda}$ as defined in  \eqref{eq:qalambda}. At the same time, $b_\mu$ originates from the anyon $\rho_m^\mathcal{B}(b_\mu)$ in $\mathcal{B}_r$ before renaming. It carries a $\Z_2^w$ charge $q_{\rho_m^\mathcal{B}(b_\mu)} = \zeta q_{b_\mu}$. Since folding does not modify charge, we have that $(a_\lambda,b_\mu)$ carries a total $\Z_2^w$ charge 
\begin{align}
q_{(a_\lambda,b_\mu)} = q_{a_\lambda} +\zeta q_{b_\mu}.
\label{eq:totalz2charge}
\end{align}
This property will be useful below when we discuss anyon condensation on the gapped domain.

\subsection{Anyon condensation}
\label{sec_anyoncond_apd}

With the above understanding, we now study the 1D gapped domain wall in Fig.~\ref{fig:folding}(c). The left side to the domain wall is the double layer topological order $\mathcal{B}\boxtimes\mathcal{B}$, while the right side is the mirror plane. Since we have gauged $\Z_2^w$, the mirror plane supports a $\Z_2^w$ gauge theory, which we denote as $\mathcal{U}$.  It is either the toric code or double semion topological order.  There are four anyons in $\mathcal{U}$, denoted as $\{\mathbbm{1}, \mathbbm{1}_{-}, w, w_{-}\}$, where $w$ is the $\pi$ flux vortex and $\mathbbm{1}_{-}$ are the $\Z_2^w$ gauge charge. The two kinds of $\Z_2^w$ gauge theories are characterized by $\theta_w^2=\theta_{w_-}^2 = 1$ and $-1$ respectively (see Sec.~\ref{sec:define-eta-u1-b}). Our goal is to determine $\mathcal{U}$, out of the two possibilities, by analyzing properties of the gapped domain wall through anyon condensation theory.

Following Refs.~\onlinecite{folding, Mao2020}, we claim that the symmetric gapped domain wall is described by the following anyon condensate:
\begin{align}
\mathcal{L}=\sum_{a_\lambda\in \mathcal{B}_0} (a_{\lambda}, \bar \rho_m({a_{\lambda}}))
\label{eqn_conds_1}
\end{align}
where $a_\lambda$ goes through all anyons in $\mathcal{B}_0$, and $\bar\rho_m(\alpha)\equiv \overline{\rho_m^\mathcal{B}(\alpha)}$ for notational simplicity. Below, we will also use $\rho_m$ to denote $\rho_m^\mathcal{B}$ or $\rho_m^\mathcal{C}$ for simplicity, where the precise meaning will be self-explaining by the involving anyon. 

A few explanations are in order. First, to see the form of $\mathcal{L}$, we consider an anyon $a_\lambda\in\mathcal{B}_l$ in the left wing and $b_\mu\in\mathcal{B}_r$ in the right wing of Fig.~\ref{fig:folding}(b). When they meet in the intersection line, they can annihilate each other into the vacuum if and only if (1) $b=\bar{a}$ and (2) the total $\Z_2^w$ charge is zero, i.e., $q_{a_\lambda} + q_{b_\mu}=0 \modulo{2}$. That means, $b_\mu = \overline{a_\lambda}$ according to the fusion rule \eqref{eq:sign-cond}. (Strictly speaking, $\overline{a_\lambda}$ is an anyon in $\mathcal{B}_{l,0}$. However, anyons in  $\mathcal{B}_{l,0} $ and  $\mathcal{B}_{r,0}$ are exactly the same, so we abuse the anyon labels.) That means, after folding and renaming \eqref{eqn:anyonid}, the anyon $(a_\lambda, \rho_m(\overline{a_\lambda}))$ condenses into the vacuum on the gapped domain. Indeed, according to \eqref{eq:totalz2charge}, it carries a $\Z_2^w$ charge $q_{a_\lambda}-\zeta q_{\rho_m(a_\lambda)} = q_{a_\lambda}-\zeta^2q_{a_\lambda}=0$. Second, the condensate is mirror symmetric. This is because that both $(a_\lambda, \bar\rho_m(a_\lambda))$ and $(\bar\rho_m(a_\lambda), a_\lambda)$ are contained in $\mathcal{L}$. Mirror properties will be discussed in more details in the next appendix. Third, the anyon $(\mathbbm{1}_-, \mathbbm{1}_-)$ belongs to $\mathcal{L}$. It is very important. Strictly speaking, in the double-layer topological order $\mathcal{B}\boxtimes\mathcal{B}$, we have gauged $\Z_2^w$ in each layer, i.e., a total $\Z_2^w\times\Z_2^w$ group is gauged. The $\Z_2^w$ symmetry in the mirror plane should correspond to the \emph{diagonal} $\Z_2$ subgroup of $\Z_2^w\times\Z_2^w$. Condensing $(\mathbbm{1}_-, \mathbbm{1}_-)$ effectively ``ungauges'' the off-diagonal $\Z_2$ symmetry.

The category $\mathcal{T}$ that lives on the domain wall consists of the following anyons
\begin{align}
\mathcal{T} = \mathcal{B}_0\oplus \{w_{\pm}^1, w_{\pm}^2, \dots\} \oplus \text{others}
\label{eq:T-eta3}
\end{align}
where $w_\pm^i$ are various $\Z_2^w$ vortices and ``others'' are additional confined anyons. To see this constitution, one may first consider the case before gauging. In that case, it is not hard to see that $\mathcal{T}=\mathcal{C}$ on the gapped domain wall, with the braiding information in $\mathcal{C}$ omitted. Then, the actual $\mathcal{T}$ should be a $\Z_2^w$-gauged version of $\mathcal{C}$. That is, $\mathcal{T}$ contains anyons in $\mathcal{B}_0$ and $\Z_2^w$ vortices. Moreover, the gapped domain wall is obtained by condensing $(\mathbbm{1}_-, \mathbbm{1}_-)$ after gauging the $\Z_2^w\times\Z_2^w$ in the double layer, $\mathcal{T}$ also contains defects associated with the off-diagonal $\Z_2$, which are denoted as ``others'' in  \eqref{eq:T-eta3}.

With that, we  claim that the restriction maps of the anyons in $\mathcal{B}\boxtimes\mathcal{B}$ are as follows:
\begin{subequations}
\begin{align}
r\{(a_\lambda,b_\mu)\} & =  \sum_{c\in\mathcal{C}} N_{a\rho_m(b)}^c c_{\sigma} \label{appc-r1}\\
r\{(a_\lambda,V_{b_\mu})\} & =  \text{confined anyons only} \label{appc-r2}\\
r\{(V_{a_\lambda},b_\mu)\} & =  \text{confined anyons only} \label{appc-r3}\\
r\{(V_{a_\lambda},V_{b_\mu})\} & =  \sum_{w_\sigma^i} n_{a_\lambda,b_\mu; w^i_\sigma} w_\sigma^i \label{appc-r4}
\end{align}
\end{subequations}
where  $\sigma$ on the right-hand side of \eqref{appc-r1} is fixed by the charge conservation condition $q_{a_\lambda} + \zeta q_{b_\mu} = q_{c_\sigma} \modulo{2}$.  A few special cases of \eqref{appc-r1} are
\begin{align}
r\{(\mathbbm{1}_+, \mathbbm{1}_+)\} & = r\{(\mathbbm{1}_-, \mathbbm{1}_-)\} = \mathbbm{1}_+, \nonumber\\
r\{(\mathbbm{1}_+, \mathbbm{1}_-)\} & = r\{(\mathbbm{1}_-, \mathbbm{1}_+)\} = \mathbbm{1}_- \label{eq:special-r}
\end{align}
To see \eqref{appc-r2} and \eqref{appc-r3}, we notice that both $(a_\lambda, V_{b_\mu})$ and $(V_{a_\lambda}, b_\mu)$ have the mutual statistics being $-1$ with respect to $(\mathbbm{1}_-,\mathbbm{1}_-)$. Since the latter condenses to the vacuum, $(a_\lambda, V_{b_\mu})$ and $(V_{a_\lambda}, b_\mu)$ must be confined. The restriction map \eqref{appc-r4} is to be determined. We have only included $w^i_\pm$ only  on the right-hand side, as one can show that  $(V_{a_\lambda},V_{b_\mu})$ is a $\Z_2^w$ vortex by checking its mutual statistics with the charges $(\mathbbm{1}_\pm, \mathbbm{1}_\pm)$. Using the commutativity between fusion and restriction and the special restrictions in \eqref{eq:special-r}, one can show that
\begin{align}
r\{(V_{a_\lambda},V_{b_\mu})\} & = r\{(V_{a_{\bar\lambda}},V_{b_{\bar\mu}})\}\nonumber
\end{align}
and
\begin{align}
r\{(V_{a_\lambda},V_{b_\mu})\}\times\mathbbm{1}_- & = r\{(V_{a_{\lambda}},V_{b_{\bar\mu}})\} =r\{(V_{a_{\bar \lambda}},V_{b_{\mu}})\} \nonumber
\end{align} 
These relations imply
\begin{align}
n_{a_\lambda,b_\mu; w^i_\sigma} = n_{a_{\bar\lambda},b_{\bar\mu}; w^i_\sigma}= n_{a_{\bar\lambda,b_\mu}; w^i_{\bar\sigma}}=n_{a_\lambda,{ b_{\bar\mu}}; w^i_{\bar\sigma}}, \nonumber\\
n_{a_{\bar\lambda},b_{\bar\mu}; w^i_{\bar\sigma}} = n_{a_{\bar\lambda},b_{\mu}; w^i_\sigma}= n_{a_{\lambda,b_{\bar\mu}}; w^i_{\sigma}}=n_{a_\lambda,{ b_{\mu}}; w^i_{\bar\sigma}}, 
\end{align}
where ``$\bar\ $'' on the indices $\lambda,\mu,\sigma$ means the opposite.

The lifting maps can be easily read off from the restriction maps. We only list the lifting maps for deconfined anyons: 
\begin{subequations}
\begin{align}
l(\mathbbm{1}) & =\sum_{a_\lambda}  (a_{\lambda}, \bar \rho_m({a_{\lambda}})) \\
l(\mathbbm{1}_-)& =\sum_{a_\lambda}  (a_{\bar\lambda}, \bar \rho_m({a_{\lambda}})) \\
l(w) & =\sum_{a_\lambda,b_\mu} n_{a_\lambda,b_\mu;w}(V_{a_\lambda},V_{b_\mu})\\
l(w_{-})& =\sum_{a_\lambda,b_\mu} n_{a_\lambda,b_\mu;w_-}(V_{a_{\lambda}},V_{b_\mu})
\end{align}
\end{subequations}
where the summations go through all $a_\lambda$ (and $b_\mu$) in $\mathcal{B}_0$, and $n_{a_\lambda,b_\mu}$ is a short-hand notation for $n_{a_\lambda,b_\mu;w}$.

Our goal is to determine $\theta_w^2$, which further determines the topological order $\mathcal{U}$. To do that, we consider the relation $S^{\mathcal{B}\boxtimes\mathcal{B}} n=n S^\mathcal{U}$, where $n$ is the matrix of lifting coefficients. More explicitly, we have
\begin{align}
\sum_{\beta \in \mathcal{B}} S_{\alpha,\beta} n_{\beta,t}=\sum_{s\in \mathcal{U}} n_{\alpha, s} \tilde S_{s, t}
\end{align}
Let us take $t=\mathbbm{1}\in\mathcal{U}$  and $\alpha=(V_{b_\mu},V_{c_\sigma})$. Then, we have the equation
\begin{align*}
\sum_{a_\lambda} S[(V_{b_\mu}, V_{c_\sigma}), (a_\lambda, \bar\rho_m(a_\lambda))] = \frac{n_{b_\mu,c_\sigma;w}+n_{b_\mu,c_\sigma;w_-}}{2}
\end{align*}
The left-hand side can be further simplified as follows:
\begin{align}
\text{l.h.s}&= \sum_{a_\lambda} S_{V_{b_\mu},a_\lambda} S_{V_{c_\sigma},\bar{\rho}_m(a_\lambda)}\nonumber \\
& = \sum_{a_\lambda} \frac{1}{2}e^{-i\pi q_{a_\lambda}}  S_{b,a} \times \frac{1}{2} e^{-i\pi q_{\bar{\rho}_m(a_\lambda)}} S_{c,\bar\rho_m(a)} \nonumber\\
& = \frac{1}{4}\sum_{a_\lambda} e^{-i\pi (1-\zeta) q_{a_\lambda}} S_{b,a} S_{c,\bar\rho_m(a)}\nonumber\\
& = \frac{1}{2}\sum_{a} e^{-i\pi (1-\zeta) q_{a}} S_{b,a} S^*_{\bar\rho_m(c),a}\nonumber\\
&=\begin{cases}\frac{1}{2} \delta_{b, \rho_m(\bar c)}, & \text{  for } U(1)\times \Z_2^\calM\\ \frac{1}{2}\delta_{mb,\rho_m(\bar c) }, & \text{  for } U(1)\rtimes \Z_2^\calM\end{cases}
\end{align}
In the second line, we used the expression \eqref{eq:smatrix-B} of $S$ matrix in $\mathcal{B}$. From the third to fourth line, we used the fact that the expression is independent of $\lambda$. To obtain the last line, we use the definition \eqref{eq:mma} of the Abelian anyon $m$.  Therefore, we have 
\begin{align}
n_{b_\mu,c_\sigma;w}+n_{b_\mu,c_\sigma;w_-}=\begin{cases} \delta_{b, \rho_m(\bar c)} & \text{  if } U(1)\times \Z_2^\calM\\ \delta_{mb,\rho_m(\bar c) } & \text{  if } U(1)\rtimes \Z_2^\calM\end{cases}
\label{eqn:b3wbc}
\end{align}
The $\delta$ function on the right-hand side is 1, when one and only one between $n_{b_\mu,c_\sigma;w}$ and $ n_{b_\mu,c_\sigma;w_-}$ is nonzero. Precisely which one is non-zero can be further determined by matching the topological spins of $(V_{b_\mu},V_{c_\sigma})$ and $w_\pm$. However, it is not important for our later discussions.

\subsection{Anomaly indicator $\teta_3$}
\label{eta3_bbc}

The anomaly indicator is given by $\teta_3 = \theta_w^2 = \theta_{w_-}^2$. The topological spin $\theta_w^2$ is equal to that of $(V_{b_\mu},V_{c_\sigma})$ in the lifting map $l(w)$ with a non-zero lifting coefficient $n_{b_\mu,c_\sigma;w}$, and similarly for $\theta_{w_-}^2$. When the $\delta$ function is equal to 1 in \eqref{eqn:b3wbc}, we observe that $(V_{b_\mu},V_{c_\sigma})$ must have a non-zero lifting coefficient either in $l(w)$ or $l(w_-)$. Accordingly,  we have
\begin{align}
\teta_3 &=\theta^2_{(V_{b_\mu}, V_{c_\sigma})} \nonumber \\
&= \theta_{V_{b_\mu}}^2 \theta^2_{Vc_\sigma} \nonumber \\
&= \theta_V^4 \theta_b^2 \theta_c^2 e^{i2\pi (q_{b}+q_{c})}
\label{eq:et3-bc}
\end{align}
where \eqref{eqn:ts_flux} is used. The anyons $b$ and $c$ satisfy $\delta_{b,\rho_m(\bar c)}=1$ for  $U(1)\times \Z_2^\calM$, or $\delta_{mb,\rho_m(\bar c)}=1$ for $U(1)\rtimes \Z_2^\calM$. In the former case, 
\begin{align}
\theta_b^2 & = \theta_{\rho_m(\bar c)}^2 = (\theta_c^*)^2\nonumber\\
q_b & = q_{\rho_m(\bar{c})} = -q_c \modulo{1}
\end{align}
Then, \eqref{eq:et3-bc} simplifies to $\teta_3 = \theta_V^4$. In the latter case, 
\begin{align}
\theta_b^2 & = \theta_{\bar m\rho_m(\bar c)}^2 = (\theta_c^*)^2  e^{-i4\pi q_c}\nonumber\\
q_b & = q_{\bar m \rho_m(\bar{c})} = q_c \modulo{1}
\end{align}
where we used the facts that $\theta_{\bar{m}}^2 =1$ and $q_{\bar{m}}=0\modulo{1}$. Inserting them into \eqref{eq:et3-bc}, we again obtain $\teta_3 = \theta_V^4$. Therefore, we obtain
\begin{align}
\teta_3=(\theta_V)^4=\theta_{V\times V} =\theta_{m_\sigma}=\theta_m
\end{align}
where $V\times V=m_\sigma$ is used. This is exactly the relation obtained in Sec.~\ref{sec:teta3-derive} using Hall conductance argument. Then, expressing $\theta_m$ in terms of surface quantities, we obtain $\teta=\eta_1\eta_3$, which is discussed in the main text.

\section{Derivation of $\teta_4$ and $\teta_5$ from anyon condensation}
\label{appd:anycond-boson-teta45}

In this appendix, we continue to use the anyon condensation theory to derive the expressions of $\tilde \eta_4$ and $\tilde \eta_5$. Both indicators are related to the SRE state that needs joint protection from $\mathcal{M}$ and $U(1)$. Therefore,  we need to gauge both the $\Z_2^x$ and $\Z_2^w$ symmetries. Again, we will discuss the cases of $U(1)\times\Z_2^\calM$ and $U(1)\rtimes\Z_2^\calM$ simultaneously. The $\Z_2^w$ gauged theory is discussed in Appendix \ref{appd:anycond-boson-teta3}. So,  we only need to further gauge $\Z_2^x$ in this section.
 
Before proceeding, we make a remark first. The group $\Z_2^w\times \Z_2^x$ is not a normal subgroup of  $U(1)\rtimes \Z_2^x$. The consequence is that the gauged system does not preserve the quotient group $[U(1)\rtimes \Z_2^x]/[\Z_2^w\times \Z_2^x]$. That is, gauging breaks the remaining symmetries in the case of $U(1)\rtimes \Z_2^\mathcal{M}$.  On the contrary, the remaining symmetries are preserved in the case of $U(1)\times\Z_2^x$. Nevertheless, as discussed in Sec.~\ref{sec:define-eta-u1}, all information about the anomaly indicators $\teta_4$ and $\teta_5$ are encoded in the vortices of the $\Z_2^w\times \Z_2^x$ subgroup. Accordingly, breaking the remaining symmetries in the $U(1)\rtimes \Z_2^x$ case does not lose information for our purpose.

\subsection{More on $\Z_2^w$-gauged theory}
\label{app_z2w_review}

In Appendix \ref{appd:anycond-boson-teta3}, we have already discussed many topological and mirror properties of the $\Z_2^w$-gauged theory $\mathcal{B}$. Here, we specialize to properties of the anyons that satisfy $\rho_m^\mathcal{C}(a)=\bar{a}$, which will be useful later for studying anyon condensation.

Recall from Sec.\ref{sec:STO-b} that the mirror fractionalization $\mu_a=\pm 1$ is defined as the mirror eigenvalue of a two-anyon state $|a,\bar{a}\rangle$, where the two anyons are located symmetrically on two sides of the mirror axis and  they satisfy $\rho_m^\mathcal{C}(a)=\bar{a}$. Also, the two anyons are in the vacuum fusion channel. When $a$ and $\bar{a}$ move towards the mirror axis and annihilate each other, it results in a local mirror charge $\mu_a$ on the axis. 

We would like to see the total $U(1)$ charge $Q_{\rm tot}$ of the two-anyon state $|a,\bar{a}\rangle$. The presence of $\calM$ symmetry constrains the possible total charge. Let $Q_a$ be the absolute $U(1)$ charge carried by $a$ in this state, which may differ by states. Then, $\bar{a}$ must carry  a charge $\zeta Q_a$ to fulfil the mirror symmetry, where $\zeta=1$ for $U(1)\times \Z_2$ and $\zeta=-1$ for $U(1)\rtimes \Z_2$. Accordingly, the total charge is 
\begin{align}
Q_{\rm tot} = (1+\zeta)Q_a
\end{align}
The total $\Z_2^w$ charge $q_{\rm tot} = Q_{\rm tot} \modulo{2}$ is then given by
\begin{align}
q_{\rm tot} = (1+\zeta)q_a = \left\{
\begin{array}{ll}
0, & \zeta=-1\\
0, & \zeta=1 \ \& \ q_a=0\\
1, & \zeta=1 \ \& \ q_a=1/2
\end{array}
\right.
\end{align}
where we have used the fact that $q_a$ can only be 0 or $1/2$ for $\rho_m^\mathcal{C}(a)=\bar{a}$ in the case of $U(1)\times \Z_2^\calM$.

When $\Z_2^w$ is gauged, $a$ turns into $a_\lambda\in\mathcal{B}_0$ with the sign $\lambda$ determined by requiring $q_{a_\lambda} = Q_a \modulo{2}$ (see Appendix \ref{sec: gaugez2p} for our convention). Similarly, $\bar{a}$ turns into $\bar{a}_{\lambda'}$ with $q_{\bar{a}_{\lambda'}}=\zeta Q_{a} \modulo{2}$. By definition \eqref{eq:rhob-rhoc} of $\rho_m^\mathcal{B}$, we have $\bar{a}_{\lambda'} = \rho_m^\mathcal{B}(a_\lambda)$. Also, we know $\bar a_{\lambda'}$ is either $\overline{a_\lambda}$ or $\overline{a_{\bar\lambda}}$, where $\bar{\lambda} = -\lambda$. Then, depending on $q_{\rm tot}$, we have the following relation:
\begin{align} 
{\rho}_m^\mathcal{B}({a_\lambda})=
\begin{cases}
\overline{a_\lambda},\quad \text{$\zeta=-1$, or $\zeta=1\ \& \ q_a=0$  }\\[3pt]
\overline{a_{\bar\lambda}}, \quad \text{$\zeta=1\ \& \ q_a=1/2$}
\end{cases}
\label{eqn_app_inverse}
\end{align}
where the sign $\lambda$ varies as the charge $Q_a$ varies. In other words, when $a_\lambda$ and $\bar a_{\lambda'}$ move towards the mirror axis, they fuse into either the vacuum $\mathbbm{1}$ or the $\Z_2^w$ charge $\mathbbm{1}_-$, depending on the values of $\zeta$ and $q_a$.

For later convenience, we introduce the set $\mathcal{I}$ to collect all the anyons in $\mathcal{C}$ that satisfy $\rho_m(a) =\bar{a}$, i.e.,
\begin{align}
\mathcal{I}=\{a\in \mathcal{C}|\rho_m^\mathcal{C}({ a})=\bar a\}.
\end{align}
Furthermore, $\mathcal{I}$ splits into two subsets $\mathcal{I}_0$ and $\bar{\mathcal{I}}_0$, defined as
\begin{align}
&\mathcal{I}_0=\{a\in \mathcal{I}|\overline{a_\lambda} =\rho_m^\mathcal{B}({ a_{\lambda}})\}\label{eqn_app_charge_set1}\\
&\bar{\mathcal{I}}_0=\{a\in \mathcal{I} |\overline{a_{\bar\lambda}}=\rho_m^\mathcal{B}( {a_{\lambda}})\}\label{eqn_app_charge_set2}
\end{align}
From \eqref{eqn_app_inverse}, we see that for $U(1)\rtimes \Z_2^\mathcal{M}$,  $\mathcal{I}_0=\mathcal{I}$ while $\bar{\mathcal{I}}_0$ is empty. On the contrary, both $\mathcal{I}_0$ and $\bar{\mathcal{I}}_0$ might not be empty for $U(1)\times \Z_2^\mathcal{M}$.

\subsection{Further gauging $\Z_2^x$ symmetry}
\label{sec_eta45_dp}

As discussed in Appendix \ref{appd:anycond-boson-teta3}, after gauging $\Z_2^{w}$, folding and renaming, the left side of Fig.~\ref{fig:folding}(c) is described by the double-layer topological order  $\mathcal{B}\boxtimes \mathcal{B}$.  The mirror symmetry $\mathcal{M}$ acts just as the layer exchange in \eqref{appeq:layerex}. In this section, we further gauge $\Z_2^\calM$, i.e., $\Z_2^x$ after renaming. We will denote the $\Z_2^x$-gauged theory as $\mathcal{E}$. It has already been obtained in Ref.~\cite{folding}. Below, we review some of the results that are useful to us. We also discuss the topological order  $\tilde{\mathcal{U}}$   of the right-hand side of Fig.~\ref{fig:folding}(c), obtained by gauging both $\Z_2^w$ and $\Z_2^x$.

First  we discuss the anyon content of the  topological order $\mathcal{E}$ and some useful relations  that were obtained in Ref.\cite{folding}.  (Here we only briefly illustrate the results. For those who are interested in the derivations, we refer them to Ref.\cite{folding}.)  Anyons in $\mathcal{E}$
are divided into two collections $\mathcal{E}_0$ and $\mathcal{E}_1$. The first collection $\mathcal{E}_0$ contains the set of anyons that are inherited from $\mathcal{B}\boxtimes \mathcal{B}$. Following the notation in Ref.\cite{folding}, they are denoted as $(\alpha, \alpha)^{\pm} \text{ for } {\forall \alpha \in \mathcal{B}}$,  and  $[\alpha,\beta] \text{ for } \alpha\neq\beta \in \mathcal{B}$. 
The superscript of $(\alpha, \alpha)^{\pm}$ denotes the $\Z_2^\mathcal{M}$ gauge charge.  In particular, $(\mathbbm{1},\mathbbm{1})^-$ is the pure $\Z_2^\mathcal{M}$ gauge charge. The anyon $[\alpha,\beta]$ is the symmetrization of two anyons $(\alpha,\beta)$ and $(\beta,\alpha)$ in $\mathcal{B}\boxtimes \mathcal{B}$ that are interchanged under mirror symmetry. The quantum dimension of $(\alpha, \alpha)^{\pm}$ is $d_\alpha^2$, while that of $[\alpha, \beta]$ is $2 d_\alpha d_\beta$. The topological spin of these anyons can be related to the ones  of anyons in $\mathcal{B}$ by $\theta_{(\alpha, \alpha)^{\pm}}^\mathcal{E}=(\theta_\alpha^{\mathcal{B}})^2$ while $\theta_{[\alpha, \beta]}^\mathcal{E}=\theta_\alpha^{\mathcal{B}}\theta_\beta^{\mathcal{B}}$ where we use the superscript $\mathcal{E,B}$ to reflect in what topological order the  topological data  are defined. As for the $S$ matrix, there are relations that are $S_{(\alpha, \alpha)^{\pm}, (\beta, \beta)^{\pm}}^\mathcal{E}=\frac{1}{2}(S^\mathcal{B}_{\alpha,\beta})^2$, $S_{(\alpha, \alpha)^{\pm}, [\beta, \gamma]}^\mathcal{E}=S^\mathcal{B}_{\alpha,\beta}S^\mathcal{B}_{\alpha,\gamma}$, and also $S_{[\alpha, \rho], [\beta, \gamma]}^\mathcal{E}=S^\mathcal{B}_{\alpha,\beta}S^\mathcal{B}_{\rho,\gamma}+S^\mathcal{B}_{\alpha,\gamma}S^\mathcal{B}_{\rho,\beta}$.

The second collection $\mathcal{E}_1$ is the set of $\Z_2^x$ vortices, denoted as
$X_\alpha^{\pm} , \text{ } {\forall \alpha\in\mathcal{B}}$.
The sign of $X_\alpha^\pm$ does not have absolute meaning but relfects that
$X_\alpha^+$ and $X_\alpha^-$ differ by a $\Z_2^x$ gauge charge.
 The quantum dimension of $X_\alpha^{\pm}$ is $d_\alpha D_\mathcal{B}$ and $D_{\mathcal{B}}$ is the total quantum dimension of $\mathcal{B}$. For the topological spin, we have the relation 
\begin{align}
\theta_{X_\alpha^\pm}=\pm e^{ic\pi/8} \sqrt{\theta_{\alpha}}
\label{eq:X-topospin}
\end{align}
where $c$ is the chiral central charge of the topological order. For the $S$ matrix, we have the following relations
\begin{align}
&S_{X_{\alpha}^\pm, [\beta, \gamma]}^\mathcal{E}=0,\label{eqn_S_1}\\
&S_{X_{\alpha}^\pm, (\beta, \beta)^{\mu}}^\mathcal{E}= \frac{1}{2}\mu S_{\alpha, \beta}^\mathcal{B}.\label{eqn_S_2}
\end{align}
where $\mu=\pm$.

A few useful fusion rules are listed as follows
\begin{align}
(\alpha,\alpha)^\pm \times (\mathbbm{1}_+,\mathbbm{1}_+)^- & = (\alpha,\alpha)^\mp \nonumber\\
[\alpha,\beta]\times (\mathbbm{1}_+,\mathbbm{1}_+)^- & = [\alpha,\beta]\nonumber\\
X_\alpha^\pm\times (\mathbbm{1}_+,\mathbbm{1}_+)^- & = X_\alpha^\mp\nonumber\\
(\alpha,\alpha)^\pm \times (\mathbbm{1}_-,\mathbbm{1}_-)^+ & = (\alpha\times\mathbbm{1}_-,\alpha\times\mathbbm{1}_-)^\pm \nonumber\\
[\alpha,\beta]\times (\mathbbm{1}_-,\mathbbm{1}_-)^+ & = [\alpha\times\mathbbm{1}_-,\beta\times\mathbbm{1}_-]\nonumber\\
X_\alpha^\pm\times [\mathbbm{1}_+,\mathbbm{1}_-] & = X_{\alpha\times\mathbbm{1}_-}^++X_{\alpha\times\mathbbm{1}_-}^-
\label{eq:E-fusion}
\end{align}
where $\alpha\times\mathbbm{1}_-$ is an anyon determined by the fusion rules of $\mathcal{B}$.

Secondly, we discuss the anyons content of the topological order $\tilde{\mathcal{U}}$ on the mirror plane. It is  a $\Z_2^w\times \Z_2^x$ gauge theory, obtained by further gauging $\Z_2^x$ in $\mathcal{U}$. In Appendix \ref{appd:anycond-boson-teta3}, we have used the subscript $\pm$ to denote that $\Z_2^w$ gauge charge in $\mathcal{U}$. Here, we use the  superscript $\pm$ to denote the gauge charge of $\Z_2^x$. We use $w$ to label $\Z_2^w$ vortices, $x$ to label the pure $\Z_2^x$ vortices, and $y$ to label the composite of $w$ and $x$ vortices. The anyons in $\tilde{U}$ are 
\begin{align}
\tilde{\mathcal{U}}=\{\mathbbm{1}_{\pm}^{\pm}, w_{\pm}^{{\pm}},  x_{\pm}^{{\pm}},  y_{\pm}^{{\pm}}      \}\nonumber
\end{align}
where  $\mathbbm{1}^+_+ \equiv \mathbbm{1}$ is the vacuum. We also take the short-hand notations that $x_+^+ \equiv x$, $y_+^+\equiv y$ and $w_+^+\equiv w$. All anyons in $\tilde{\mathcal{U}}$ are Abelian. Some of the fusion rules are 
\begin{align}
x\times\mathbbm{1}_-^+ &=x_-^+ \nonumber\\
x\times\mathbbm{1}_+^- &=x_+^-\nonumber\\
x\times w&=y.
\end{align}
The topological spins of $w$-vortices take two possible kinds of values $\pm i$ or $\pm1$,  where the $\pm$ signs come from the different gauge charge contents of the vortices. Similarly, $x$- and $y$-vortices also can take two possibilities. The total eight combinations reflect the $(\Z_2)^3$ classification of $\Z_2\times \Z_2$ gauge theories in bosonic systems. 

As discussed in Sec.~\ref{sec:define-eta-u1-b}, the indicators are $\teta_4 = M_{w,x}^2$  and $\teta_5 = \theta_y^2$. To determine their values, we need to know $\tilde{U}$ is which of the eight possible $\Z_2\times\Z_2$ gauge theories. The latter is determined by the nature of $\mathcal{E}$ and properties of the gapped domain wall in Fig.~\ref{fig:folding}(c). In the next subsection, we study properties of the gapped domain wall using anyon condensation theory.

\subsection{Anyon condensation}
\label{app_eta4_boson_anyon_cond}
Here we discuss the anyon condensation on the gapped domain wall, which will help us to establish the boundary-bulk correspondence and obtain expressions for the anomaly indicators.

Following Refs.~\onlinecite{folding, Mao2020}, we claim that the gapped domain wall is associated with the following condensate
\begin{align}
\mathcal{L}=&\sum_{a_{\lambda}=\bar \rho_m(a_{\lambda})} (a_{\lambda}, a_{\lambda})^{\mu_a} +\sum_{a_{\lambda}\neq\bar \rho_m(a_{\lambda})} [a_{\lambda}, \bar \rho_m(a_{\lambda})], 
\label{eqn_cdz2pm}
\end{align}
For notational simplicity, we use $\rho_m$ to denote either $\rho_m^\mathcal{B}$ or $\rho_m^\mathcal{C}$ if no confusion is caused. That is, anyons in $\mathcal{L}$ can be annihilated by local operators at the domain wall. A few remarks are listed as follows. First,  the first summation is over all $a_\lambda\in\mathcal{B}_0$ that satisfy $a_{\lambda}=\bar \rho_m(a_{\lambda})$, while the second summation is over all pairs $\{a_{\lambda}, \bar \rho_m(a_{\lambda})\}$ that satisfy  $a_{\lambda}\neq\bar \rho_m(a_{\lambda})$, or equivalently, it is over only one of the two $a_{\lambda}$ and  $\rho_m(a_{\lambda})$.  Second, compared to (\ref{eqn_conds_1}),
the condensate (\ref{eqn_cdz2pm}) can be understood as the $\Z_2^x$-gauged version of (\ref{eqn_conds_1}).  If $a_{\lambda}\neq\bar \rho_m(a_{\lambda})$, the pair $(a_{\lambda},\bar \rho_m(a_{\lambda}))$ and its mirror image $(\bar \rho_m(a_{\lambda}), a_{\lambda})$ in  (\ref{eqn_conds_1}) are combined into $[a_{\lambda},\bar \rho_m(a_{\lambda})]$ in (\ref{eqn_cdz2pm}).  If $a_{\lambda}=\bar \rho_m(a_{\lambda})$,  only one between $(a_\lambda,a_\lambda)^+$ and $(a_\lambda,a_\lambda)^-$ is condensed; the other turns into the $\Z_2^x$ charge $\mathbbm{1}_+^-$ instead. Our claim is $(a_\lambda,a_\lambda)^{\mu_a}$ is the condensed one. To see the claim, we recall that before gauging $\Z_2^w\times\Z_2^x$, the pair $(a,a)$ corresponds to the two-anyon state $|a, \bar\rho_m(a)\rangle$ (before renaming in \eqref{eqn:anyonid}), which carries mirror eigenvalue by $\mu_a$. Since the condensate should be mirror neutral, the pair  $(a, a)$ shall be condensed on the domain wall together with a mirror charge $\mu_a$ to respect the mirror symmetry. Accordingly, it is  $(a_\lambda, a_\lambda)^{\mu_a}$ appears in the condensate (\ref{eqn_cdz2pm}). Thirdly,   the special anyon $(\mathbbm{1}_-,\mathbbm{1}_-)^+$ is in $\mathcal{L}$ as  $\mu(\mathbbm{1})=1$. 

To better describe the anyon condensation, we study more details on the fusion category $\tilde{\mathcal{T}}$ that lives the domain wall. We claim that it contains anyons as follows
\begin{align}
\tilde{\mathcal{T}}=\tilde{\mathcal{B}}_0& \oplus \{w_\pm^{1\pm},w_\pm^{2\pm},...\}\oplus\{ x_\pm^{1\pm},x_\pm^{2\pm},...\}\nonumber\\
& \oplus\{y_\pm^{1\pm},y_\pm^{2\pm},... \}\oplus \text{others}
\label{eq:T-eta4}
\end{align}
where 
\begin{align}
\tilde{\mathcal{B}}_0=\{\mathbbm{1}_\pm^\pm, a_\pm^\pm,...\}.
\end{align}
Several remarks  are as follows.  First of all, $\tilde{\mathcal{T}}$ is the $\Z_2^x$-gauged version of $\mathcal{T}$ in (\ref{eq:T-eta3}). Specifically, $\tilde{\mathcal{B}}_0$ is a $\Z_2^x$-gauged version of  $\mathcal{B}_0$ in  (\ref{eq:T-eta3}) such that every anyon $a_\pm\in\mathcal{B}_0$ is further decorated with a superscript $\pm$, representing the $\Z_2^x$ charge. Similarly, $\Z_2^w$ vortices in $\mathcal{T}$ are further decorated by a superscript $\pm$, representing the $\Z_2^x$ charge. In addition, new vortices are introduced, including $\Z_2^x$ vortices and composites of $\Z_2^w$ and $\Z_2^x$ vortices, which are labelled as $x_\pm^{i\pm}$ and $y_\pm^{i\pm}$, respectively. Most of these $w$-, $x$- and $y$-vortices will be confined, except those in $\tilde{\mathcal{U}}$. We will take the convention that $w_\pm^{1\pm} = w_\pm^\pm$, $x_\pm^{1\pm}=x_\pm^\pm$, and $y_\pm^{1\pm}=y_\pm^\pm$, which are discussed in Appendix \ref{sec_eta45_dp}.  The ``others'' are extra confined vortices associated with the off-diagonal $\Z_2$ of the $\Z_2^w\times\Z_2^w$ group and their composites with other vortices. Since all of them are confined, they are not important for our following discussions.

Next, we study more details of the anyon condensation. We claim that the restriction maps are given as follows:
\begin{subequations}
\begin{align}
r\{[a_{\lambda}, b_\gamma]\}&=\sum_{c}N_{a,\rho_m(b)}^c (c_\sigma^++c_\sigma^-)\label{eqn_app_mp_rest_1}\\
r\{(a_\lambda, a_\lambda)^+\}&=\sum_c N_{a,\rho_m(a)}^c c_\sigma^{\mu[a,\rho_m(a);c]}\label{eqn_app_mp_rest_2}\\
r\{(a_\lambda, a_\lambda)^-\}&=\sum_c N_{a,\rho_m(a)}^c c_\sigma^{-\mu[a,\rho_m(a);c]}\label{eqn_app_mp_rest_3}\\
r\{[a_\lambda, V_{b_\gamma}]\}&=\text{confined anyons only} \label{eqn_app_mp_rest_4}\\
r\{[V_{a_\lambda}, V_{b_\gamma}]\}&=\text{$w$-vortices  only}\label{eqn_app_mp_rest_5}\\
r\{(V_{a_\lambda}, V_{a_\lambda})^\pm\}&=\text{$w$-vortices  only}\label{eqn_app_mp_rest_6}\\
r\{X_{V_{a_\lambda}}^\pm\}&=\text{confined anyons only}\label{eqn_app_mp_rest_7}\\
r\{X_{a_\lambda}^\pm\}&=\text{$x$ and $y$ anyons only}\label{eqn_app_mp_rest_8}
\end{align}
\end{subequations}
Some explainations are listed as follows. First of all,
the restriction map (\ref{eqn_app_mp_rest_1})-(\ref{eqn_app_mp_rest_6}) are closely related to (\ref{appc-r1})-(\ref{appc-r4}).  The sign $\sigma$ of $c_\sigma^\pm$ in (\ref{eqn_app_mp_rest_1})-(\ref{eqn_app_mp_rest_3}) are determined in the same way as that in (\ref{appc-r1}), i.e., by the condition  $q_{a_\lambda} + \zeta q_{b_\gamma} = q_{c_\sigma} \modulo{2}$.  The summation over $c$ in (\ref{eqn_app_mp_rest_1})-(\ref{eqn_app_mp_rest_3}) are over the anyons in $\mathcal{C}$. In general, we do not know the expression of the funcion $\mu[a,\rho_m(a);c]$. However, in the special case that $a=\bar \rho_m(a)$, the sign $\mu(a,\bar{a};\mathbbm{1}) = \mu_a$, the mirror eigenvalue defined in Sec.~\ref{sec:STO-b}. This can be understood following the discussion below \eqref{eqn_cdz2pm}. The sign difference in (\ref{eqn_app_mp_rest_2}) and (\ref{eqn_app_mp_rest_3}) can be understood from the fact that $(a_\lambda, a_\lambda)^-=(a_\lambda, a_\lambda)^+\times (\mathbbm{1},\mathbbm{1})^-$ and the property \eqref{eq: restriction commutes with fusion}, i.e., the restriction map commutes with fusion. Secondly, to understand (\ref{eqn_app_mp_rest_7}) and (\ref{eqn_app_mp_rest_8}), we consider the mutual statistics between  $(\mathbbm{1}_-,\mathbbm{1}_-)^+$ and $X_{\alpha}^\pm$. According to \eqref{eqn_S_2}, the mutual statistics is given by
\begin{align}
M_{(\mathbbm{1}_-,\mathbbm{1}_-)^+, X^\pm_{\alpha}}=M_{\mathbbm{1}_-,\alpha}=\begin{cases}-1, \quad \text{ if $\alpha\in \mathcal{B}_0$}\\ 1,\,\,\,\,\, \quad \text{ if $\alpha\in \mathcal{B}_1$} \end{cases}
\end{align}
The anyon $(\mathbbm{1}_-,\mathbbm{1}_-)^+$ is condensed at the domain wall. Accordingly, any anyon that has non-trivial mutual statistics with $(\mathbbm{1}_-,\mathbbm{1}_-)^+$  will be confined, which gives rises to \eqref{eqn_app_mp_rest_7}.

The restriction map (\ref{eqn_app_mp_rest_8}) is a very important one, so we expand it as follows
\begin{align}
r(X_{a_\lambda}^+)=n_{a_\lambda} x^+_++p_{a_\lambda} y^+_++n'_{a_\lambda} x^+_-+p'_{a_\lambda} y^+_-+\cdots
\end{align}
where  $n_\alpha$, $ p_\alpha$ and  $n'_\alpha$, $p'_\alpha$ are unknown non-negative integer, and ``$\dots$'' denotes the confined $x$- and $y$-vortices. The superscript signs on the two sides  are chosen to be the same as our convention for the mirror charges on the $x$- and $y$-vortices in $\tilde{\mathcal{U}}$. The restriction map $r(X_{a_\lambda}^-)$ is determined by $r(X_{a_\lambda}^-) = r(X_{a_\lambda}^+)\times r((\mathbbm{1}_+,\mathbbm{1}_+)^-)$. 
Finally, we consider $r(X^+_{a_\lambda}\times [\mathbbm{1}_+,\mathbbm{1}_-])=r(X^+_{a_\lambda})\times r([\mathbbm{1}_+,\mathbbm{1}_-])$. It results in the relations  
\begin{align} 
n_{a_\lambda}&=n_{a_{\bar \lambda}}'\label{eqn:acs5}\\
p_{a_\lambda}&=p_{a_{\bar \lambda}}'.\label{eqn:acs6}
\end{align}
Therefore,  among the two sets of data $\{ n_{a_\lambda}, p_{a_\lambda}\}$ and $\{n'_{a_\lambda}, p'_{a_\lambda}\}$, only one is independent.

With these restriction maps, we obtain the following useful lifting maps:
\begin{align}
l(\mathbbm{1}) &= \!\!\sum_{a_{\lambda}=\bar \rho_m(a_{\lambda})} (a_{\lambda}, a_{\lambda})^{\mu_a}
+\!\!\sum_{a_{\lambda}\neq\bar \rho_m(a_{\lambda}) } [a_{\lambda}, \bar \rho_m(a_{\lambda})] 
\label{eqn_lfm_z2pm_1}\\
l(\mathbbm{1}^+_-) &=\!\!\sum_{a_{\lambda}=\bar \rho_m(a_{\bar\lambda})} (a_{\lambda}, a_{\lambda})^{\mu_a} +\sum_{a_{\lambda}\neq\bar \rho_m(a_{\bar \lambda})} [a_{\lambda}, \bar \rho_m(a_{\bar \lambda})] \label{eqn_lfm_z2pm_1_}\\
l({x}_+^\pm) &= \sum_{a_\lambda\in B_0} n_{a_\lambda} X_{a_\lambda}^\pm \label{eqn_lfm_z2pm_x} \\
l(y^\pm_+) &= \sum_{a_\lambda\in B_0} p_{a_\lambda} X_{a_\lambda}^\pm  \label{eqn_lfm_z2pm_xv}\\
l({x}_-^\pm) &= \sum_{a_\lambda\in B_0} n'_{a_\lambda} X_{a_\lambda}^\pm \label{eqn_lfm_z2pm_x1} \\
l(y^\pm_-) &= \sum_{a_\lambda\in B_0} p'_{a_\lambda} X_{a_\lambda}^\pm  \label{eqn_lfm_z2pm_xv1}
\end{align}
where the $\bar \lambda$ denote the opposite of $\lambda$. The second summation in the lifting map $l(\mathbbm{1})$ is only  over one out of the pair $a_{\lambda}$ and $\bar\rho_m(a_{ \lambda})$, and similarly for $l(\mathbbm{1}_-^+)$.  An important remark is that for $U(1)\rtimes \Z_2^\mathcal{M}$ the lifting map $l(\mathbbm{1}^+_-)$ contains only the second summation, since there is no anyon $a_\lambda \in \mathcal{B}_0$ such  that $a_\lambda=\bar \rho_m(a_{\bar\lambda})$, which has been discussed in detail in Appendix \ref{app_z2w_review}.

So far, the integers $n_{a_\lambda}$ and $p_{a_\lambda}$ are completely unknown. Below we show that they are related to the symmetry  and topological data of the surface topological order through the following expressions,
\begin{align}
&n_{b_\lambda}+n_{b_{\bar\lambda}}=\sum_{a\in \mathcal{C}} \mu_a S_{b,a}^{\mathcal{C}} \label{eqn:acs1} \\
& p_{b_\lambda}+p_{b_{\bar \lambda}}=\sum_{a\in \mathcal{C}} e^{i\pi (1+\zeta) q_a}\mu_a S_{b,a}^{\mathcal{C}} \label{eqn:acs2}  
\end{align}
where $\mu_a$ is set to 0 for $a\neq \bar\rho_m(a)$ according to our convention Eq.~\eqref{eq:mu=0} in Sec.~\ref{sec:STO-b}. 
We further define
\begin{align}
n_b&=n_{b_+}+n_{b_-}\label{eqn:acs3}\\
p_b&=p_{b_+}+p_{b_-},\label{eqn:acs4}
\end{align}
Then,  (\ref{eqn:acs1}) and (\ref{eqn:acs2}) become 
\begin{align}
&n_{b}=\sum_{a\in \mathcal{C}} \mu_a S_{b,a}^{\mathcal{C}} \label{eqn:acs7} \\
&p_{b}=\sum_{a\in \mathcal{C}} e^{i(1+\zeta)\pi q_a}\mu_a S_{b,a}^{\mathcal{C}}. \label{eqn:acs8}  
\end{align}

Below we prove the two relations (\ref{eqn:acs1}) and (\ref{eqn:acs2}) in  two steps. 
First, we show that
\begin{align}
&\frac{1}{2}(n_{b_{\lambda}}+p_{b_{\lambda}}+n'_{b_{\lambda}}+p'_{b_{\lambda}})=\sum_{a\in\mathcal{I}_0} \mu_a  S_{b, a   }^{\mathcal{C}} \label{eqn:deconfined1}\\
&\frac{1}{2}(n_{b_{\lambda}}-p_{b_{\lambda}}+n'_{b_{\lambda}}- p'_{b_{\lambda}})=\sum_{a\in\bar{\mathcal{I}}_0}\mu_a S_{b, a }^{\mathcal{C}}.
\label{eqn:deconfined2}
\end{align}
We note that $\mathcal{I}_0$ and $\bar{\mathcal{I}}_0$ are defined in (\ref{eqn_app_charge_set1}) and (\ref{eqn_app_charge_set2}).
To derive (\ref{eqn:deconfined1}),  we make use of the property that restriction maps commute with $S$ matrices, i.e., Eq.~\eqref{eq:stu2}. In the current context, it is
\begin{align}
\sum_{\beta \in \mathcal{E}} S^\mathcal{E}_{\alpha,\beta} n_{\beta,t}=\sum_{s\in \tilde{\mathcal{U}}} n_{\alpha, s}  S^{\tilde{\mathcal{U}}}_{s, t}
\label{eq:stu3}
\end{align}
By setting $\alpha=X_{a_\lambda}^+$ and $t=\mathbbm{1}$, we have
\begin{align}
\text{r.h.s.}&=\sum_{s\in \tilde{\mathcal{U}}} n_{X_{a_\lambda}^+, s}  S^{\tilde{\mathcal{U}}}_{s,\mathbbm{1}} \nonumber \\
&=\frac{1}{4} (n_{X_{a_\lambda}^+, x_+^+}+n_{X_{a_\lambda}^+, y_+^+}+n_{X_{a_\lambda}^+, x_-^+}+n_{X_{a_\lambda}^+, y_-^+})\nonumber \\
&=\frac{1}{4} (n_{a_\lambda}+p_{a_\lambda}+n'_{a_\lambda}+p'_{a_\lambda})
\end{align}
where we have used $S_{s,\mathbbm{1}}^{\tilde{\mathcal{U}}}=1/4$ for any $s$. Meanwhile,
\begin{align}
\text{l.h.s.}&=\sum_{\beta \in \mathcal{E}} S^{{\mathcal{E}}}_{X_{a_\lambda}^+,\beta} n_{\beta,\mathbbm{1}}\nonumber \\
&=\sum_{\substack{b_{\gamma} \in \mathcal{B}_0\\ \& b_\gamma=\bar \rho_m(b_{\gamma})}} S^{{\mathcal{E}}}_{X_{a_\lambda}^+, (b_{\gamma}, b_{\gamma})^{\mu_b}}  \nonumber \\
&=\sum_{\substack{b_{\gamma} \in \mathcal{B}_0\\ \& b_\gamma=\bar \rho_m(b_{\gamma})}} \frac{1}{2}\mu_b S_{a_\lambda, b_{\gamma}   }^{\mathcal{B}} \nonumber \\
&=\sum_{b \in \mathcal{I}_0}\frac{1}{2}\mu_b S_{a, b   }^{\mathcal{C}} 
\end{align}
where, from the first line to the second line, we have used the restriction map (\ref{eqn_lfm_z2pm_1_}) and the relations (\ref{eqn_S_1}); from the second to the third line, we have used the relation (\ref{eqn_S_2}); from the third to the fourth, we have used the relation
\begin{align}
&\sum_{\gamma =\pm} S_{{ a_{\lambda},b_{\gamma}}}^{\mathcal{B}} =\sum_{\gamma =\pm} \frac{1}{2} S_{a, b}^{\mathcal{C}} = S_{a, b}^{\mathcal{C}}
\end{align}
and the definition \eqref{eqn_app_charge_set1} of $\mathcal{I}_0$. Combining expressions of l.h.s and r.h.s gives rise to Eq.~\eqref{eqn:deconfined1}.

To  derive  (\ref{eqn:deconfined2}), we apply \eqref{eq:stu3} by setting $\alpha=X_{a_\lambda}^+$, $t=\mathbbm{1}_-^+$. The calculation is very similar to the derivation of (\ref{eqn:deconfined1}), as follows:
\begin{align}
\text{r.h.s.}&=\sum_{s\in \tilde{\mathcal{U}}} n_{X_{a_\lambda}^+, s}  S^{\tilde{\mathcal{U}}}_{s,\mathbbm{1}_-} \nonumber \\
&=\frac{1}{4} (n_{X_{a_\lambda}^+, x_+^+}-n_{X_{a_\lambda}^+, y_+^+}+n_{X_{a_\lambda}^+, x_-^+}-n_{X_{a_\lambda}^+, y_-^+})\nonumber \\
&=\frac{1}{4} (n_{a_\lambda}-p_{a_\lambda}+n'_{a_\lambda}-p'_{a_\lambda})
\end{align}
and
\begin{align}
\text{l.h.s.}&=\sum_{\beta \in \mathcal{E}} S^{{\mathcal{E}}}_{X_{a_\lambda}^+,\beta} n_{\beta,\mathbbm{1}_-}\nonumber \\
&=\sum_{\substack{b_{\gamma} \in \mathcal{B}_0\\ \& b_\gamma=\bar \rho_m(b_{\bar\gamma})}} S^{{\mathcal{E}}}_{X_{a_\lambda}^+, (b_{\gamma}, b_{\gamma})^{\mu_b}}  \nonumber \\
&=\sum_{\substack{b_{\gamma} \in \mathcal{B}_0\\b_\gamma=\bar \rho_m(b_{\bar\gamma})}} \frac{1}{2}\mu_bS_{a_\lambda, b_{\gamma}   }^{\mathcal{B}} \nonumber \\
&=\sum_{b \in \bar{\mathcal{I}}_0}\frac{1}{2}\mu_b S_{a, b   }^{\mathcal{C}} 
\end{align}
Combining the two expressions, we obtain \eqref{eqn:deconfined2}.

Secondly, by adding the equations (\ref{eqn:deconfined1}) and (\ref{eqn:deconfined2}) and using the relation (\ref{eqn:acs5}), we immediately obtain  (\ref{eqn:acs1}). To obtain (\ref{eqn:acs2}), we  consider subtracting  (\ref{eqn:deconfined1}) with (\ref{eqn:deconfined2}) and using (\ref{eqn:acs6}), and obtain
\begin{align}
p_{b_\lambda}+p_{b_{\bar\lambda}}
&=\sum_{a\in \mathcal{I}_0}\mu_a S_{b, a }^{\mathcal{C}}-\sum_{a\in \bar{\mathcal{I}}_0}\mu_a S_{b, a }^{\mathcal{C}} \nonumber \\
&=\sum_{a\in\mathcal{I}} e^{i(1+\zeta)\pi q_a}\mu_a S_{b, a }^{\mathcal{C}}
\end{align}
where  we have used the property \eqref{eqn_app_inverse} of anyons in $\mathcal{I}$.  In fact, $p_b$ can be further expressed as
\begin{align}
p_b=\sum_{\substack{a\in \mathcal{C}}} \mu_a' S_{b, a }^{\mathcal{C}} 
\label{eq:pb}
\end{align}
where $ \mu_a'=e^{i(1+\zeta)\pi q_a}\mu_a$ is given in \eqref{eq:muprime} in Sec.~\ref{sec:STO-b} and the convention that $\mu_a=0$ for $a\notin \mathcal{I}$.

Finally, we make two remarks on properties of $n_b$ and $p_b$. First, for $U(1)\rtimes \Z_2^\mathcal{M}$, we have
\begin{align}
p_b=n_b=n_{\bar{b}}.\label{eqn:wam1}
\end{align}
The first equality is easily seen from Eqs.~\eqref{eqn:acs7} and \eqref{eqn:acs8}, and the second equality can be obtained by combining the fact that $\mu_{a}=\mu_{\bar{a}}$ and  Eq.~\eqref{eqn:acs7}. On the other hand, for $U(1)\times \Z_2^\mathcal{M}$,  we have
\begin{align}
p_{b}=n_{mb} = n_{\bar{m}b} =n_{\bar{m}\bar{b}}.\label{eqn:wam}
\end{align}
where  $m$ is the monopole anyon defined with the relation \eqref{eq:mma}. These equalities can be easily obtained by combining the relations $S_{b m, a}^{\mathcal{C}}= S_{b,a}^{\mathcal{C}}M_{m,a}^*$, $M_{m,a}=e^{i2\pi q_a}$, $M_{\bar m,a}=e^{-i2\pi q_a}$, and $e^{i(1+\zeta)\pi q_a}\mu_a=e^{-i(1+\zeta)\pi q_a}\mu_a $ for all anyons in $\mathcal{C}$ for $U(1)\times\Z_2^\calM$, Eq.~\eqref{eqn:acs7} and Eq.~\eqref{eqn:acs8}.

Second, we claim that
\begin{align}
\sum_{a\in \mathcal{C}} n_a d_a & =D_\mathcal{C} 
\label{eqn:qd1}\\
\sum_{a\in \mathcal{C}} p_a d_a & =D_\mathcal{C}
\label{eqn:qd2}
\end{align}
where $D_\mathcal{C}$ is the total quantum dimension of $\mathcal{C}$. These properties follow from a general result \cite{NeupertPRB2016} of anyon condensation
\begin{align}
\sum_{\alpha\in\mathcal{P}} n_{\alpha,\mathbbm{1}}d_\alpha =  \frac{1}{d_t} \sum_{\alpha\in\mathcal{P}} n_{\alpha,t}d_\alpha
\label{eqn:qd3}
\end{align}
where $\mathcal{P}$ is the parent topological order, $t$ is any deconfined anyon and $n_{\alpha,t}$ is the restriction or lifting coefficient. Taking $\mathcal{P}=\mathcal{E}$ and lifting map \eqref{eqn_lfm_z2pm_1}, we have the left-hand side equal to $2D_\mathcal{C}^2$. Then, it is straightforward to check that \eqref{eqn:qd1}  and \eqref{eqn:qd2} can be obtained from \eqref{eqn:qd3} by taking $t=x_+^+$ and $y_+^+$, respectively. 

\subsection{Anomaly indicator $\teta_4$ and $\teta_5$}
\label{eta4_indc2}

We now derive the expressions of the anomaly indicators $\teta_4$ and $\teta_5$ in terms of SET data. First, we define  
\begin{align}
\eta_4&=\frac{1}{D_{\mathcal{C}}}\sum_{a\in \mathcal{C}} d_a\theta_a \mu_a e^{i(1+\zeta)\pi q_a }\label{eqn:eta_4_exp}
\end{align}
which reduces to $\eta_4$ in \eqref{eq:eta_b4} in the main text for $\zeta=1$. For $\zeta=-1$, it reduces to $\eta_2$. With this definition, we will show
\begin{align}
\teta_4 & =\eta_1\eta_2\eta_3\eta_4 \label{eq:app_teta4}\\
\teta_5 & =\eta_4 \label{eq:app_teta5}
\end{align}
where the expressions of $\eta_1,\eta_2,\eta_3$  are given in Eqs.~\eqref{eq:eta_b1}, \eqref{eq:eta_b2} and \eqref{eq:eta_b3}. These results have been derived in Sec.~\ref{sec:boson-U(1)} and here we will derive them  from the anyon condensation theory described above.  In addition, we derive an alternative expression for $\teta_4$:
\begin{align}
\teta_4 &=\frac{1}{D_{\mathcal{C}}} \sum_{a,b} \mu_a S_{a,b} d_be^{i2\pi q_b} \label{eqn:teta_4_exp} 
\end{align}
which is certainly equivalent to \eqref{eq:app_teta4}. This expression is also given in Ref.~\onlinecite{LapaPRB2019} the time-reversal systems.

We start by relating $\teta_4$ and $\teta_5$ to properties of anyons in $\mathcal{C}$. By definition, we have $\teta_5=\theta_{y}^2$, where $y\equiv y_+^+$ is the deconfined anyon in $\tilde{U}$ whose the lifting map is \eqref{eqn_lfm_z2pm_xv}. Using the fact that topological spin remains the same for deconfined anyons before and after anyon condensation, we have that, for any $a$ with $p_a\neq 0$, i.e., either $p_{a_+}\neq 0$ or $p_{a_-}\neq 0$, 
\begin{align}
\teta_5 = \theta_{X_{a_\lambda}^+}^2 =e^{i c\pi/4} \theta_{a_\lambda} =e^{ic\pi/4}\theta_{a}
\label{eq_app:dev1}
\end{align}
where $\lambda$ is either ``$+$'' or ``$-$'', and \eqref{eq:X-topospin} for the topological spin of $X_{a_\lambda}^+$ is used to obtain the second equality. Meanwhile, the relations \eqref{eqn:wam1} and \eqref{eqn:wam} imply that $n_b \neq 0$ if $p_a\neq 0$, where $b=a\bar m$ for $U(1)\times \Z_2^{\mathcal{M}}$ and $b=a$ for $U(1)\rtimes \Z_2^\mathcal{M}$. By the definition of $\teta_2$ and the lifting map \eqref{eqn_lfm_z2pm_x},  we then have
\begin{align}
\teta_2 = \theta_{X_{b_\lambda}^+}^2 = e^{i c\pi/4} \theta_{b_\lambda} =e^{ic\pi/4}\theta_{b}
\label{eq_app:dev2}
\end{align} 
We note that since $b=a$ for $U(1)\rtimes\Z_2^\calM$, the equality $\teta_5=\teta_2$ is justified from anyon condensation. Then, we recall the relation $\teta_5 = \teta_2\teta_3\teta_4$ from Sec.~\ref{sec:U(1)xZ2-b} and the result $\teta_3 = \theta_m$ from Appendix \ref{eta3_bbc}. Putting them together with \eqref{eq_app:dev1} and \eqref{eq_app:dev2}, we obtain
\begin{align}
\teta_4 = \frac{\theta_a}{\theta_b\theta_m} =\left\{
\begin{array}{ll}
\theta_m^*, & U(1)\rtimes\Z_2^\calM\\[3pt]
M_{b,m}, & U(1)\times\Z_2^\calM
\end{array}
\right.
\label{eq_app:dev3}
\end{align}
That means, we establish $\teta_4=\teta_3^* = \teta_3$ for $U(1)\rtimes\Z_2^\calM$ from anyon condensation theory.

Next, we prove \eqref{eq:app_teta5}. Starting with \eqref{eq_app:dev1}, we have
\begin{align}
\tilde \eta_5 &=e^{ic\pi/4}\theta_a\nonumber \\
&=e^{ic\pi/4}\frac{1}{D_{\mathcal{C}}} \sum_{b\in \mathcal{C}} p_b d_b \theta_b \nonumber\\ 
&=e^{ic\pi/4}\frac{1}{D_{\mathcal{C}}} \sum_{b\in \mathcal{C}} \left(\sum_{a\in \mathcal{C}} \mu'_a S_{b,a}^\mathcal{C}\right) d_b \theta_b \nonumber\\ 
&=e^{ic\pi/4}\frac{1}{D_{\mathcal{C}}^2} \sum_{a,b} \mu'_a d_b \theta_b  \sum_{c\in \mathcal{C}}N^c_{a\bar{b}} \frac{{\theta_c}}{ \theta_a \theta_b}d_c\nonumber\\ 
&=e^{ic\pi/4}\frac{1}{D_{\mathcal{C}}^2} \sum_{a,c}  \mu'_a \theta_a^*    d_c{\theta_c}  \sum_{b}  d_b  N^b_{a\bar c}\nonumber\\ 
&=e^{ic\pi/4}\frac{1}{D_{\mathcal{C}}^2} \sum_{a,c}  \mu'_a \theta_a^*    d_c{\theta_c} d_c d_a\nonumber\\ 
&=e^{ic\pi/4}\left(\frac{1}{D_{\mathcal{C}}} \sum_{c}d_c^2{\theta_c^*}\right)\left(\frac{1}{D_{\mathcal{C}}} \sum_{a}  \mu_a'\theta_a  d_a \right) \nonumber\\
&=\eta_4 \label{eqn_teta5_deriv}
\end{align}
In the second line, we use the relation (\ref{eqn:qd2}) and the fact that $\theta_a$ is the same for any $a$ with $ p_a\neq 0$. In the third line, we insert the expression \eqref{eq:pb}. In the fourth line, the definition of $S$ matrix is inserted. In the fifth line, we use the fact that $N_{a\bar{b}}^c = N_{a \bar{c}}^b$. In the sixth line, we use the property that $\sum_{b} N_{a\bar{c}}^b d_b = d_ad_c$.  In the seventh line, we use \eqref{eq:centralcharge} to cancel the factor $e^{ic\pi/4}$, and finally with the definition of $\mu'_a$, we obtain the relation that $\teta_5 = \eta_4$. With this,  Eq.~\eqref{eq:app_teta4} follows directly from the relation $\teta_4=\teta_2\teta_3\teta_5$ and the facts that $\teta_2=\eta_2$ and $\teta_3=\eta_1\eta_3$.  

The expression (\ref{eqn:teta_4_exp}) can be obtained from \eqref{eq_app:dev3}. In the case of $U(1)\times\Z_2^\mathcal{M}$,  for any $b$ with $n_b\neq 0$, we have 
\begin{align}
\teta_4&=M_{b,m}\nonumber\\
&=\frac{1}{D_{\mathcal{C}}} \sum_{b} n_b d_b M_{m,b}  \nonumber\\
&=\frac{1}{D_{\mathcal{C}}} \sum_{a,b} \mu_a S_{a,b} d_be^{i2\pi q_b} 
\label{eqappd-teta4}
\end{align}
For the first to the second equality, we used the relation \eqref{eqn:qd1}. From the second to third equility, we have used the relations \eqref{eqn:acs7} and $M_{m,b}=e^{i2\pi q_b}$. A corollary is
\begin{equation}
\teta_4 = \mu_m. \label{eq:monople-teta4}
\end{equation}
To see that, from \eqref{eqn:acs7}, we have
\begin{align}
\mu_m =\sum_b S_{m,b}^{\mathcal{C}\dag} n_b = \frac{1}{D_\mathcal{C}}\sum_b M_{m,b} n_b d_b
\end{align}
which is exactly the second line of \eqref{eqappd-teta4}. Hence, \eqref{eq:monople-teta4} holds. 

\bibliographystyle{apsrev4-2}
\bibliography{lie}

\end{document}